\documentclass[]{aa}  

\usepackage{graphicx}
\usepackage{txfonts}
\usepackage{txfonts}
\usepackage{array}
\usepackage{amssymb}
\usepackage{natbib}
\usepackage{lscape}
\usepackage{rotating}
\usepackage{longtable}
\newcommand{\zp}{$\zeta$\,Pup}

\newcommand{\xmm}{{\sc{XMM}}\emph{-Newton}}
\newcommand{\swift}{{\em Swift}}
\begin{document} 

\title{A detailed X-ray investigation of \zp\ \\ IV. Further characterization of the variability}

\author{Ya\"el~Naz\'e\inst{1}\fnmsep\thanks{F.R.S.-FNRS Research Associate.}, Tahina Ramiaramanantsoa\inst{2}, Ian R. Stevens\inst{3}, Ian D. Howarth\inst{4}, Anthony F.J. Moffat\inst{2}}

\institute{Groupe d'Astrophysique des Hautes Energies, STAR, Universit\'e de Li\`ege, Quartier Agora (B5c, Institut d'Astrophysique et de G\'eophysique), All\'ee du 6 Ao\^ut 19c, B-4000 Sart Tilman, Li\`ege, Belgium\\
\email{naze@astro.ulg.ac.be}
\and
Centre de Recherche en Astrophysique du Qu\'ebec (CRAQ) and D\'epartement de physique, Universit\'e de Montr\'eal, C.P. 6128, Succ. Centre-Ville, Montr\'eal, QC H3C 3J7, Canada
\and
School of Physics and Astronomy, University of Birmingham, Edgbaston, Birmingham B15 2TT, UK
\and
Department of Physics \& Astronomy, University College London, Gower St, London WC1E 6BT, UK 
}
\authorrunning{Naz\'e et al.}
\titlerunning{X-rays from \zp - IV}

 
  \abstract
   {One of the optically brightest and closest massive stars, \zp, is also a bright X-ray source. Previously, its X-ray emission was found to be variable with light curves harbouring ``trends'' with a typical timescale longer than the exposure length, i.e. $>$1\,d. The origin of these changes was proposed to be linked to large-scale structures in the wind of \zp, but further characterization of the variability at high energies was needed to investigate this scenario.}
   {Since the previous papers of this series, a number of new X-ray observations have become available. Furthermore, a cyclic behaviour with a 1.78\,d period was identified in long optical photometric runs, which is thought to be associated with the launching mechanism of large-scale wind structures.}
   {We analysed these new X-ray data, revisited the old data, and compared the X-ray light curves with the optical data, notably those taken simultaneously.}
   {The behaviour of \zp\ in X-rays cannot be explained in terms of a perfect clock because the amplitude and shape of its variations change with time. For example, \zp\ was much more strongly variable between 2007 and 2011 than before and after this interval. Comparing the X-ray spectra of the star at maximum and minimum brightness yields no compelling difference beyond the overall flux change: the temperatures, absorptions, and line shapes seem to remain constant, well within errors. The only common feature between X-ray datasets is that the variation amplitudes appear maximum in the medium (0.6--1.2\,keV) energy band. Finally, no clear and coherent correlation can be found between simultaneous X-ray and optical data. Only a subgroup of observations may be combined coherently with the optical period of 1.78\,d, although the simultaneous optical behaviour is unknown. }
   {The currently available data do not reveal any obvious, permanent, and direct correlation between X-ray and optical variations. The origin of the X-ray variability therefore still needs to be ascertained, highlighting the need for long-term monitoring in multiwavelengths, i.e. X-ray, UV, and optical.}



   \maketitle
%

\section{Introduction}

Massive stars are well known for their ability to launch dense and fast stellar winds with important consequences for their evolution and their feedback on the environment. However, numerous uncertainties remain on their properties, notably regarding their structure. The first question that arose regarded clumping. The winds of massive stars are very different from those from solar-like stars since the winds of massive stars are propelled by intense UV radiation \citep{luc70}. Photons can be absorbed by the many transitions from metallic ions;  since photons initially flow radially while scattering occurs in all directions, this leads to a net radial acceleration of those ions and, subsequently, of the entire atmospheric material after ion interactions. However, this line driving is an unstable process. If we take Doppler shifting into account, a slightly faster ion can absorb more photons and accelerate further, while slightly slower ions are deprived of photons and decelerate further since the needed photons were already absorbed by the layers just below the ion position. The line-driven winds are thus intrinsically unstable \citep[see e.g.][]{luc70,car80} with  numerous small-scale stochastic perturbations pervading the wind flow \citep[e.g.][]{eve98}. The presence of such perturbations alters the interpretation of mass-loss diagnostics, and this impact varies with the chosen diagnostic since some of the diagnostics depend on the square of the density while others vary linearly with density. After considering those effects, actual mass-loss rates are now believed to be lower than previous used unclumped values by a factor $\sim3$ \citep[e.g.][]{bou05}. These wind instabilities are also responsible for the generation of X-rays \citep[and references therein]{luc82,owo88,fel97,osk04}. Stochastic variability should then arise on hourly timescales (e.g. wind flow time $R_*/v_{\infty}$ is $\sim$1.6h for \zp) with an amplitude linked to the number of involved clumps. However, dedicated analyses of sensitive X-ray observations failed to detect such changes, implying a very large number of clumps ($>10^5$; \citealt{naz13}). 

The structure of stellar winds is not restricted to small-scale clumping, however, because large-scale features are also inferred to be ubiquitous in massive hot-star winds. Indeed, variable discrete absorption components (DACs) observed in unsaturated UV P-Cygni profiles of O stars \citep{mas95,how95} have been interpreted as a manifestation of co-rotating interaction regions (CIRs; see e.g. \citealt{cra96,lob08}). Related variability was detected for line profiles in the optical range as well \citep[e.g.][]{pri96,kap97}. These large-scale ($>R_*$) wind disturbances are believed to originate at the base of the wind from bright spots on the photosphere. The presence of such spots has also been inferred from optical variations, for example modelling of the photometry of $\xi$\,Per and \zp, \citep{ram14,ram17} and modelling of the He\,{\sc ii}\,$\lambda$4686 profile of $\lambda$\,Cep, (\citealt{sud16}). But, until recently, CIRs were not considered to be a potential source of X-rays since models predicted only a velocity kink and  not a large jump \citep[e.g.][]{lob08}.

\begin{sidewaystable*}
\centering
\footnotesize
\caption{Journal of X-ray observations.}
\label{journal}
\begin{tabular}{lcccccccccccc}
\hline\hline
\multicolumn{5}{l}{\it \swift\ observations}\\
ID & ObsID & EL & HJD && Count rate (cts\,s$^{-1}$) & $HR=H/M$ & \multicolumn{2}{c}{$VI$} & \multicolumn{2}{c}{$F_{var}$}\\
   &       & (ks)&    && 0.5--10.\,keV             &           & M-band & H-band & M-band & H-band \\
\hline
S-01  & 00032727001 &1.4& 2456350.796 &&  &  \\
S-2/3 &00032727002+3&1.1& 2457664.147 && 0.59$\pm$0.03 &0.44$\pm$0.05 &  \\
S-04  & 00032727004 &1.0& 2457748.540 && 0.66$\pm$0.03 &0.56$\pm$0.05 &  0.22$\pm$0.05 & 0.43$\pm$0.07 & 0.14$\pm$0.04 & 0.21$\pm$0.06 \\
S-05  & 00032727005 &1.0& 2457748.760 && 0.47$\pm$0.02 &0.42$\pm$0.05 &   \\
S-06  & 00032727006 &1.0& 2457748.929 && 0.36$\pm$0.04 &0.34$\pm$0.06 &   \\
S-07  & 00032727007 &1.0& 2457749.129 && 0.38$\pm$0.04 &0.49$\pm$0.07 &   \\
S-08  & 00032727008 &1.1& 2457749.262 && 0.49$\pm$0.02 &0.42$\pm$0.05 &   \\
S-09  & 00032727009 &1.0& 2457749.461 && 0.46$\pm$0.03 &0.46$\pm$0.05 &   \\
S-10  & 00032727010 &1.0& 2457749.540 && 0.52$\pm$0.03 &0.37$\pm$0.04 &   \\
S-11  & 00032727011 &0.5& 2457749.794 && 0.55$\pm$0.03 &0.39$\pm$0.06 &   \\
S-12  & 00032727012 &1.0& 2457749.992 && 0.50$\pm$0.02 &0.44$\pm$0.05 &   \\
S-13  & 00032727013 &1.0& 2457750.126 && 0.43$\pm$0.04 &0.57$\pm$0.07 &   \\
S-14  & 00032727014 &1.0& 2457765.553 && 0.64$\pm$0.03 &0.46$\pm$0.05 &  0.15$\pm$0.04 & 0.22$\pm$0.06 & 0.08$\pm$0.03 & 0.14$\pm$0.05\\
S-15  & 00032727015 &0.9& 2457765.802 && 0.49$\pm$0.02 &0.37$\pm$0.04 &  \\
S-16  & 00032727016 &1.0& 2457765.936 && 0.62$\pm$0.03 &0.35$\pm$0.04 &  \\
S-17  & 00032727017 &1.0& 2457766.084 && 0.65$\pm$0.03 &0.42$\pm$0.04 &  \\
S-18  & 00032727018 &0.9& 2457766.271 && 0.53$\pm$0.03 &0.34$\pm$0.04 &  \\
S-19  & 00032727019 &1.0& 2457766.480 && 0.54$\pm$0.05 &0.34$\pm$0.04 &  \\
S-20  & 00032727020 &0.9& 2457766.631 && 0.60$\pm$0.03 &0.43$\pm$0.04 &  \\
S-21  & 00032727021 &1.0& 2457766.799 && 0.52$\pm$0.03 &0.43$\pm$0.05 &  \\
S-23  & 00032727023 &1.0& 2457767.070 && 0.68$\pm$0.03 &0.39$\pm$0.04 &  \\
S-24/1&00032727024+1&1.2& 2457847.727 && 0.48$\pm$0.02 &0.38$\pm$0.04 & 0.14$\pm$0.04 & 0.19$\pm$0.06 & 0.10$\pm$0.04 & 0.14$\pm$ 0.06\\
S-25/2&00032727025+2&1.3& 2457847.794 && 0.54$\pm$0.02 &0.37$\pm$0.04 &  \\
S-26/3&00032727026+3&1.3& 2457847.994 && 0.65$\pm$0.03 &0.42$\pm$0.04 &  \\
S-27/4&00032727027+4&1.2& 2457848.193 && 0.54$\pm$0.02 &0.51$\pm$0.05 &  \\
S-28/5&00032727028+5&1.2& 2457848.259 && 0.55$\pm$0.02 &0.39$\pm$0.04 &  \\
\hline
\multicolumn{5}{l}{\it \xmm\ observations}\\
Rev & ObsID & EL & HJD & Bgd & \multicolumn{4}{c}{$VI$}& \multicolumn{4}{c}{$F_{var}$}\\
    &         &(ks)          &     &     & T-band & S-band & M-band & H-band & T-band& S-band & M-band & H-band\\
\hline
0636 & 0159360101  &21.5 & 2452790.119 & 2 & 0.067$\pm$0.005 & 0.019$\pm$0.009 & 0.013$\pm$0.007 & 0.026$\pm$0.014 &                 & 0.012$\pm$0.008 &                 &                 \\
0795 & 0159360301  &17.4 & 2453108.018 & 2 & 0.009$\pm$0.005 & 0.009$\pm$0.009 & 0.015$\pm$0.006 & 0.035$\pm$0.014 & 0.003$\pm$0.007 &                 & 0.006$\pm$0.006 & 0.022$\pm$0.012 \\
0903 & 0159360401  &20.9 & 2453323.292 & 3 & 0.020$\pm$0.005 & 0.018$\pm$0.009 & 0.029$\pm$0.007 & 0.047$\pm$0.014 & 0.012$\pm$0.005 & 0.004$\pm$0.014 & 0.018$\pm$0.007 & 0.027$\pm$0.012 \\
0980 & 0159360501  &29.3 & 2453476.954 & 2 & 0.009$\pm$0.005 & 0.020$\pm$0.009 & 0.015$\pm$0.007 & 0.023$\pm$0.013 &                 & 0.002$\pm$0.002 & 0.005$\pm$0.006 &                 \\
1071 & 0159360701  &14.8 & 2453658.852 & 3 & 0.014$\pm$0.005 & 0.022$\pm$0.009 & 0.011$\pm$0.007 & 0.022$\pm$0.014 & 0.009$\pm$0.005 & 0.015$\pm$0.009 &                 &                 \\
1096 & 0159360901  &35.6 & 2453708.477 & 3 & 0.007$\pm$0.005 & 0.014$\pm$0.009 & 0.014$\pm$0.007 & 0.042$\pm$0.014 &                 &                 &                 & 0.019$\pm$0.009 \\
1343 & 0414400101  &38.1 & 2454200.470 & 2 & 0.029$\pm$0.005 & 0.025$\pm$0.009 & 0.036$\pm$0.007 & 0.042$\pm$0.014 & 0.019$\pm$0.005 & 0.002$\pm$0.018 & 0.026$\pm$0.007 & 0.021$\pm$0.009 \\
1620 & 0159361301  &42.7 & 2454753.553 & 4 & 0.039$\pm$0.005 & 0.048$\pm$0.009 & 0.040$\pm$0.006 & 0.040$\pm$0.013 & 0.025$\pm$0.006 & 0.027$\pm$0.007 & 0.026$\pm$0.006 & 0.016$\pm$0.008 \\
1814 & 0561380101  &44.7 & 2455139.776 & 4 & 0.060$\pm$0.005 & 0.061$\pm$0.010 & 0.077$\pm$0.007 & 0.046$\pm$0.014 & 0.042$\pm$0.008 & 0.036$\pm$0.008 & 0.053$\pm$0.011 & 0.015$\pm$0.008 \\
1983 & 0561380201  &53.5 & 2455477.463 & 4 & 0.032$\pm$0.005 & 0.040$\pm$0.009 & 0.037$\pm$0.007 & 0.063$\pm$0.014 & 0.017$\pm$0.004 & 0.018$\pm$0.005 & 0.021$\pm$0.005 & 0.030$\pm$0.008 \\
2183 & 0561380301  &44.4 & 2455875.706 & 3 & 0.039$\pm$0.005 & 0.021$\pm$0.009 & 0.058$\pm$0.007 & 0.064$\pm$0.013 & 0.025$\pm$0.005 & 0.005$\pm$0.007 & 0.037$\pm$0.008 & 0.040$\pm$0.010 \\
2363 & 0561380501  &41.9 & 2456234.613 & 4 & 0.028$\pm$0.005 & 0.028$\pm$0.009 & 0.036$\pm$0.007 & 0.035$\pm$0.014 & 0.018$\pm$0.004 & 0.011$\pm$0.005 & 0.023$\pm$0.005 &                 \\
2533 & 0561380601  &45.9 & 2456573.969 & 4 & 0.025$\pm$0.005 & 0.028$\pm$0.009 & 0.033$\pm$0.007 & 0.043$\pm$0.014 & 0.015$\pm$0.004 & 0.011$\pm$0.005 & 0.020$\pm$0.005 & 0.010$\pm$0.010 \\
2540 & 0159361601  &35.2 & 2456588.474 & 3 & 0.010$\pm$0.005 & 0.018$\pm$0.010 & 0.019$\pm$0.007 & 0.018$\pm$0.014 &                 &                 & 0.004$\pm$0.006 &                 \\
2817 & 0561380701  &35.4 & 2457140.795 & 2 & 0.012$\pm$0.005 & 0.025$\pm$0.009 & 0.015$\pm$0.007 & 0.051$\pm$0.014 & 0.002$\pm$0.007 & 0.010$\pm$0.006 &                 & 0.024$\pm$0.009 \\
2911 & 0561380801  &30.5 & 2457328.107 & 3 & 0.012$\pm$0.005 & 0.029$\pm$0.009 & 0.010$\pm$0.007 & 0.044$\pm$0.014 & 0.004$\pm$0.004 & 0.012$\pm$0.006 &                 & 0.020$\pm$0.010 \\
2989 & 0561380901  &44.2 & 2457483.757 & 2 & 0.017$\pm$0.005 & 0.033$\pm$0.009 & 0.026$\pm$0.007 & 0.048$\pm$0.013 & 0.009$\pm$0.003 & 0.017$\pm$0.005 & 0.010$\pm$0.004 & 0.027$\pm$0.008 \\
3172 & 0561381001  &41.3 & 2457848.028 & 2 & 0.016$\pm$0.005 & 0.029$\pm$0.009 & 0.019$\pm$0.007 & 0.042$\pm$0.014 & 0.008$\pm$0.003 & 0.013$\pm$0.005 & 0.007$\pm$0.004 & 0.021$\pm$0.008 \\
\hline
\end{tabular}
\\
\vspace*{-0.25cm}
\tablefoot{\tiny \swift\ observations 00032727002 and 00032727003, 00032727024 and 00010039001, 00032727025 and 00010039002, 00032727026 and 00010039003, 00032727027 and 00010039004, and 00032727028 and 00010039005 were taken during the same spacecraft orbit, hence their data were combined (which is identified by a ``+'' sign in the ObsID column). No count rate is provided for \swift\ dataset 00032727001 because of centroiding problems during the light curve generation; however, its associated spectrum could be extracted (see Table \ref{swiftfit}). EL and HR stand for actual exposure length (i.e. after flare filtering and considering a live time of 71\%) and hardness ratio, respectively, while S, M, and H correspond to soft (0.3--0.6\,keV), medium (0.6--1.2\,keV), and hard (1.2--4.0\,keV) energy bands. $VI$ is a variability index, expressed as $(max-min)/(max+min)$ - it thus compares the amplitude of the count rate variation (i.e. half the peak-to-peak variation amplitude) to the mean. To correct for measurement noise, the fractional variability amplitudes $F_{var}$, defined as $\sqrt((S^2-\sigma^2_{err})/X_m),$ where $X_m=\sum X_i/N$, $S^2=\sum (X_i-X_m)^2/(N-1)$, and $\sigma_{err}^2=\sum sigma_{err,i}^2/N$, are also shown. These amplitudes are provided for each \xmm\ exposure and for each cycle observed by \swift, except if the measurement noise exceeds the variance ($S^2<\sigma^2_{err}$) for $F_{var}$. }
\end{sidewaystable*}

The situation changed with the detection of trends in the X-ray light curves of several massive stars. The analysis of an extensive \xmm\ dataset of \zp, while failing to unveil the expected short-term stochastic variability associated with clumping, clearly revealed shallow monotonic increases, monotonic decreases, or modulations of the X-ray light curves with a relative peak-to-peak amplitude of $\sim$20\% \citep{naz13}. No full cycle could be observed over the exposure lengths, which are usually less than a day, but time analysis tools point towards potential timescales between a day and a few days. A few other (relatively bright) single massive stars hinted at a similar variability. Variations with 15\% amplitude were detected in $\xi$\,Per \citep{mas14}. The {\it Chandra} light curve could be fitted using a sinusoidal function with the period detected in UV P-Cygni profiles, although less than one cycle was observed and a linear trend was able to fit the data as well as this sinusoidal modulation. Changes by 15\% were also detected in the four \xmm\ observations of $\lambda$\,Cep \citep{rau15} and they appear to be compatible with the 4.1\,d variation of the star observed in H$\alpha$ at the same epoch, although each exposure covers only about 1\,d and no single cycle was fully covered. In the past, a variation of 20\% amplitude was also reported for $\zeta$\,Oph using the less sensitive {\it ASCA} observatory, and it appeared possibly related to the rotation period detected in the UV \citep{osk01}. As these X-ray changes presented strong similarities (in pattern or timescale) to the optical/UV variability associated with CIRs, these features were naturally suggested to explain the X-ray variations. A definite link is not proven yet, however, as there is a clear misfit between observing cadence at X-ray energies and the possible periodicities: no massive star has ever been monitored over a full --or to be more conclusive, several consecutive-- variability cycle(s) in X-rays. 

To better characterize these X-ray trends, additional data are thus needed. This paper aims at such a goal, thanks to new observations of \zp\ and revisiting old data. The choice of the target comes from its brightness: the signal to noise achieved for the other massive stars with such variability is many times below that which can be achieved for \zp. Re-examining this star requires a good knowledge of its cycle length, however. In the past, several variability timescales were reported for this star. In particular, a value of $\sim$5.1\,d was proposed as the rotation period for this star by \citet{mof81} from H$\alpha$ observations, but this period was not recovered in the last 20 years. The presence of a 15--19\,h period was found in UV \citep{pri92,how95}, H$\alpha$ line \citep{rei96} and even X-ray observations \citep[{\it ROSAT}; ][]{ber96} although no sign of this period was later found in {\it ASCA} or \xmm\ data \citep{osk01,naz13}. Recent extensive photometry now shows that timescale of 15--19\,h remains coherent for a few days at best, with very frequent complete disappearances \citep{ram17}. Recently, \citet{how14} unveiled a much more stable periodicity in \zp\ of  $P=1.780938\pm0.000093$\,d,  from long, 40 to 236\,d, {\it Coriolis/SMEI} (Solar-Mass Ejection Imager) observing runs spread over three years. The presence of this signal is confirmed in {\it BRITE} (BRIght Target Explorer) observations of \zp\ taken in 2014--2015 as well as in observations of the He\,{\sc ii}\,$\lambda$4686 line profile taken simultaneously \citep{ram17}. While \citet{how14} proposed this signal to be pulsational in nature, \citet{ram17} interpreted this behaviour as a rotational modulation associated with bright surface spots. Both cases could however launch CIRs in the wind, which should thus present the same recurrence timescale as detected for the variations of the He\,{\sc ii}\,$\lambda$4686 wind emission line. Armed with this precise period, this paper revisits the X-ray variability of \zp. Section 2 presents the X-ray data and their reduction, while Section 3 presents and discusses the observational results, and Section 4 summarizes these results.

\section{Observations and data reduction} 

\subsection{\swift }
At our request, \zp\ (O4I) was observed by \swift\ in December 2016 and January 2017. Each time, the X-ray emission of \zp\ was followed with \swift -XRT (X-Ray Telescope) over one entire 1.78\,d cycle with a set of 1\,ks observations taken every 3--6 h. Three older datasets and a set of calibration exposures taken simultaneously with the last \xmm\ observation were also available and are thus also analysed here; see Table~\ref{journal} for the full list of observations. 

Corrected (full PSF) count rates in the same energy bands as previous \xmm\ analyses \citep[see also next subsection]{naz12} were obtained, considering only grade 0, for each observation from the UK online tool\footnote{http://www.swift.ac.uk/user\_objects/}. Heliocentric corrections were added to mid-observation times. To get spectra, XRT data were also locally reprocessed using {\it xrtpipeline} of HEASOFT v6.18 with calibrations available in winter 2016 and an explicit input of the Simbad coordinates of \zp. 

\zp\ is a bright target both in the optical ($V$=2.25) and at X-ray energies, hence we could not use the photon-counting mode of \swift -XRT, which would be heavily affected by optical loading for such a bright source. However, even the alternative (windowed timing) mode has some problems. Indeed, below 0.5\,keV, the count rates sometimes increased erratically (up to a factor of 3) and the spectra showed a varying, unrealistic bump. Therefore, to avoid contamination by optical/UV photons as much as possible, only the best quality data ($grade=0$, $E>0.5$\,keV) were considered. Furthermore, as recommended by the XRT team, the source spectra were extracted within a circular aperture\footnote{Using an annular region to remove the core only degrades the signal-to-noise ratio without removing the low-energy contamination nor the energy scale offset (see Sect. 3.2).} centred on the Simbad coordinates of the target\footnote{However, the use of  a fixed extraction region such as here can lead to PSF correction uncertainties of up to 20\%, especially for snapshots in which the target appears near the bad columns (as e.g. for ObsID 00032727026 and 00032727027).} with radius 10px to minimize the background contribution, while an annular background region with radii 70 and 130px was used\footnote{see http://www.swift.ac.uk/analysis/xrt/backscal.php}. Spectra were then binned (using {\it grppha}) to reach 10 counts per bin (or a signal-to-noise ratio of 3 per bin). The redistribution matrix file (RMF), to be used for calibrating the energy axis of spectra at the observing date, was retrieved from the calibration database, after checking that the general RMF matrix or the position-dependent RMF matrices\footnote{http://www.swift.ac.uk/analysis/xrt/posrmf.php} yielded the same results in our case. Finally, specific ancillary response files (ARFs), used to calibrate the flux axes of spectra, were calculated for each dataset (using {\it xrtmkarf}) considering the associated exposure map. 

\subsection{\xmm }
\zp\ was used as a calibration target for \xmm, resulting in an exceptional number of 26 exposures taken since 2000. These observations were taken in a variety of modes and with various exposure lengths \citep[see][for details]{naz12}. For this new analysis, these datasets were re-reduced using the most recent calibration available (SAS v15.0.0 with calibration files available in winter 2016) and following the recommendations of the \xmm\ team\footnote{SAS threads, see \\ http://xmm.esac.esa.int/sas/current/documentation/threads/ }. 

This paper primarily deals with variability issues, hence it is focussed on light curves. The most sensitive data in this regard are those taken by the pn camera, but they are affected by pile-up and/or optical loading in all but one mode (small window + thick filter). This mode has been used since 2003, but the first datasets were rather short and very affected by soft proton flares, mostly due to solar activity; the pn camera was even not switched on in one case (Rev. 1164 in 2006). The pn data have the additional advantage that they are the least affected by sensivity degradation; this detector is in fact considered stable by the \xmm\ calibration team, to within a few percents across the mission lifetime, which facilitates inter-pointing comparisons.

After the initial pipeline ({\it epproc}) processing, the pn data were filtered to keep only those with the best quality, using {\sc pattern}=0+{\sc flag}=0 and eliminating the flares whenever needed; time intervals with count rates above 10\,keV beyond 0.045\,cts\,s$^{-1}$ were discarded. The barycentric correction was calculated and applied to the recorded times (SAS task {\it barycen}). Light curves were then extracted in the same regions as in \citet{naz12}: the source region is a circle centred on the Simbad coordinates of \zp\ and with a radius of 850\,px (42.5''); the background region is also a circle centred on a nearby position (an identification number given in Table \ref{journal} refers to the background region positions listed in Table 2 of \citealt{naz12}); these circular regions have a radius of 700\,px except for Revs. 2363 and 2817, where it had to be reduced to 600\,px and 450\,px, respectively. We considered time bins of 5\,ks and four energy bands, i.e. soft band S=0.3--0.6\,keV, medium band M=0.6--1.2\,keV, hard band H=1.2--4.0\,keV, and total band T=0.3--4.0\,keV (see \citealt{naz12}). The light curves were further processed by the task {\it epiclccorr}, which corrects for loss of photons due to  vignetting, off-axis angle, or other problems such as bad pixels. In addition, to avoid very large errors and bad estimates of the count rates, we discarded bins displaying effective exposure time lower than 50\% of the time bin length. For these light curves, the background contribution is less than half the $1\sigma$ error bars, hence background plays no significant role in the variability studied here.

\setcounter{figure}{5}
\begin{figure*}
\includegraphics[height=6.5cm, angle=-90, bb=568 5 74 699, clip]{pnspeccompa.ps}
\includegraphics[width=6.cm]{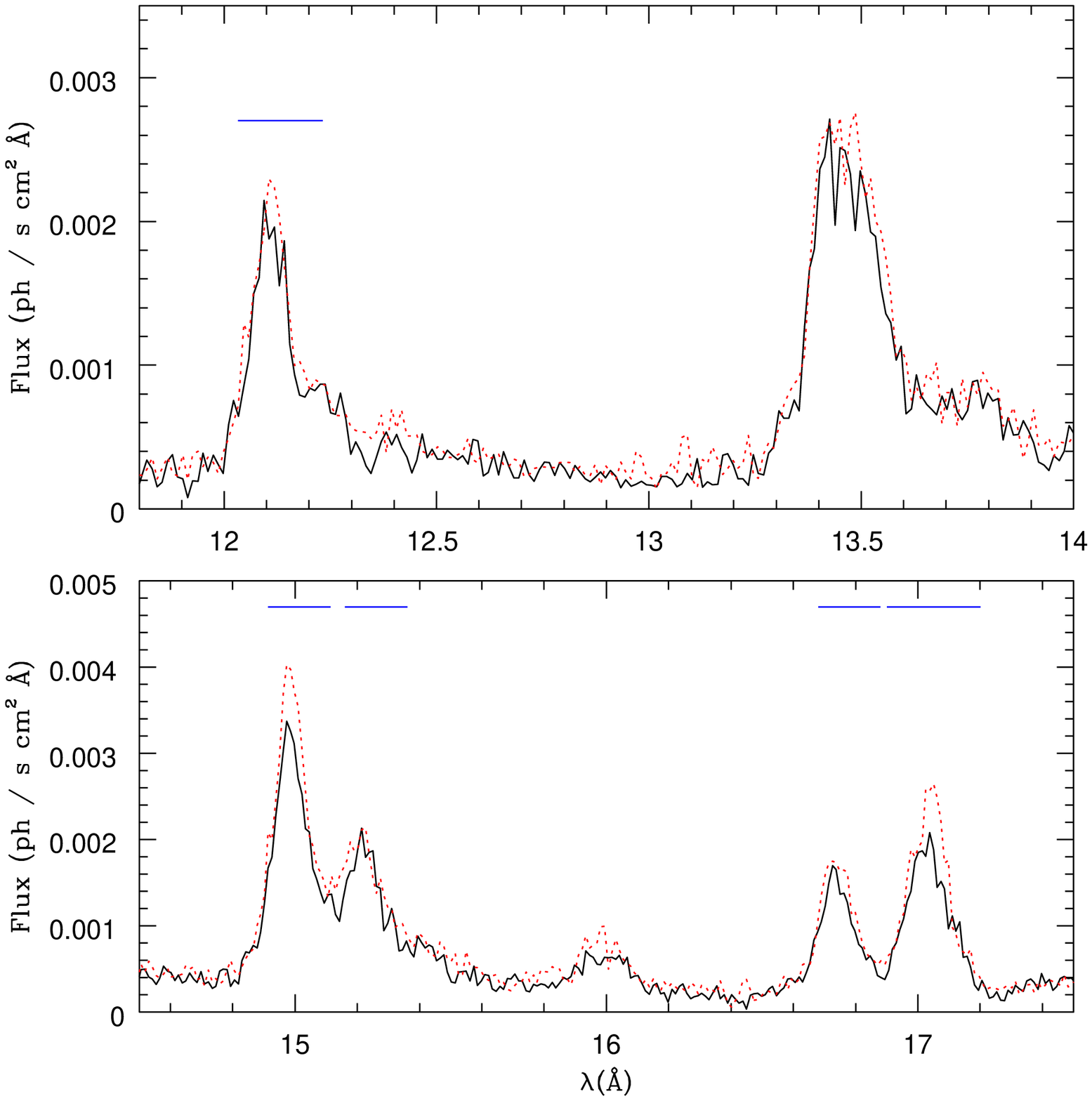}
\includegraphics[width=6.cm]{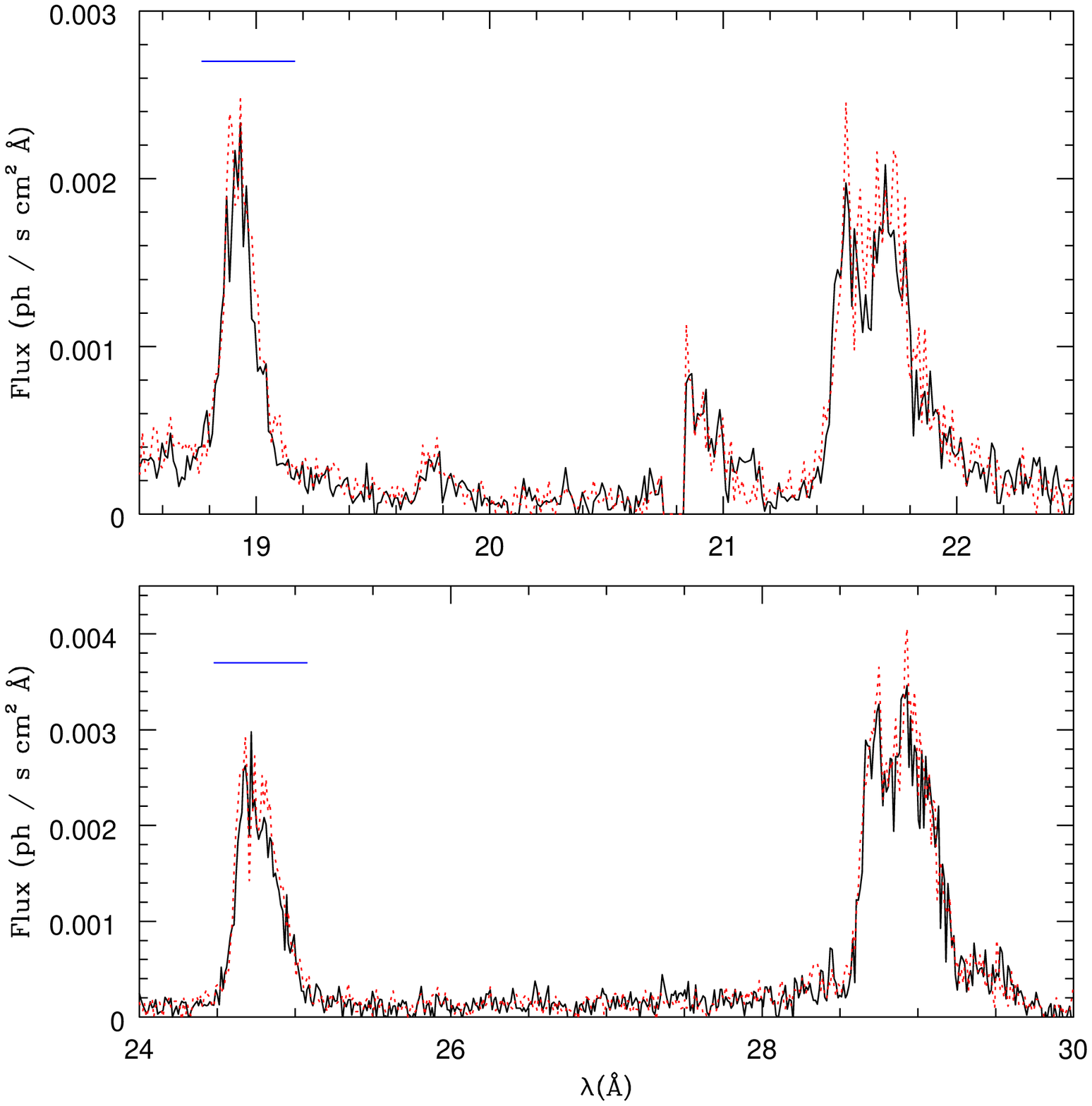}
\caption{Comparison of the  pn (left) and RGS (middle and right) spectra observed at minimum brightness (first part of Rev. 1814; solid black line) and maximum brightness (first part of Rev. 2183; dotted red line). Blue segments above lines of RGS spectra indicate the intervals chosen for the moment calculations.}
\label{comp}
\end{figure*}

As explained in the next section, \xmm\ spectra of two extreme cases were specifically analysed. In those cases, pn spectra were extracted for the chosen time intervals via the task {\it{especget}} in the same regions as the light curves. Dedicated ARF and RMF response matrices were also derived for each observation. The pn spectra were finally grouped, with {\it{specgroup}}, to obtain an oversampling factor of five and to ensure that a minimum signal-to-noise ratio of 3 (i.e.\ a minimum of 10 counts) was reached in each spectral bin of the background-corrected spectra. After the initial pipeline ({\it rgsproc}) processing, Reflection Grating Spectrometer (RGS) data were also filtered to keep only events in the chosen time intervals and the associated spectra and their response matrices were then calculated via the appropriate SAS tasks ({\it rgsspectrum, rgsrmfgen}). This was carried out for the default regions since \zp\ has no neighbour of similar X-ray brightness. Finally, the RGS spectra were individually grouped with {\it{specgroup}} to obtain an oversampling factor of five; calibrated combinations of both RGS and both orders were also derived via {\it rgsfluxer}.

\section{Results}

\subsection{Variations in light curve amplitudes}

Figures \ref{xmm3}, \ref{xmm1}, \ref{xmm2}, and \ref{swlc} show the observed light curves, in several energy bands, phased with the ephemerides of \citet[$P=1.780938$\,d, $T_0$=2\,450\,000]{how14} so that several light curves can be represented in the same graph using the same x-axis, facilitating the (direct) comparisons. The amplitude of the variations is immediately seen to change from one observation to the next. To quantify it, a variability index $VI$ was defined as $(max-min)/(max+min)$, i.e. the ratio between half the peak-to-peak amplitude and the mean. If the variations were sinusoidal in shape, this would yield the fractional semi-amplitude of the sine-wave. The values of this index are provided in Table \ref{journal}. In the \xmm\ datasets, the exposure lengths are similar, but $VI$ varies between 1 and 6\% in the total band, and the most recent datasets are the least variable. Indeed, the count rate in the total band varies from slightly below 6\,cts\,$^{-1}$ to slightly more than 7\,cts\,$^{-1}$ in 2007--2011 (Fig. \ref{xmm1}), but only between 6.2 and 6.7\,cts\,$^{-1}$ in 2003--2006 and 2012--2016 (Fig. \ref{xmm2}). Moreover, some light curves clearly appear above or below others\footnote{This does not seem to be linked to a sensitivity degradation with time, as the faintest count rates (e.g. beginning of Rev. 1814) are not necessarily the most recent and the brightest ones (e.g. beginning of Rev. 2183) the oldest. Therefore, such variations reflect intrinsic variations of \zp.}. Analysing all \xmm\ datasets together therefore yields larger overall $VI$: 0.083$\pm$0.005, 0.079$\pm$0.009, 0.099$\pm$0.007, and 0.095$\pm$0.013 for the T, S, M, H energy bands, respectively, confirming the value of 20\% peak-to-peak variations mentioned in \citet{naz13}. This 20\% variation amplitude is also much higher than the possibly remaining calibration uncertainties (\xmm\ help desk, private communication). The $VI$ for various energy bands agree, usually well within 3$\sigma$, but the maximum variability seems to be reached in the medium band. 

To correct these estimates from the variability due to noise, we also determined the fractional variability amplitude $F_{var}$ of the data \citep[see Appendix A of][]{ede02}. $F_{var}$ is the square root of the normalized excess variance which is calculated by subtracting the average variance due to noise from the observed variance of the data and then normalizing the result by the mean. This thus gives an idea of the intrinsic variability of the source. The fractional variability amplitude (Table \ref{journal}) varies from 0 to 4.2\% in the T band (0.3--4.0\,keV), reaching the highest values ($>$1\%) for data in 2007--2013. There is negligible variability for the other years (i.e. $F_{var}\sim0$ within 3$\sigma$), confirming the $VI$ results. Considering all data together, we found $F_{var}$ of 0.034$\pm$0.002, indicating that about half of the observed variability is due to measurement noise. The fractional variability amplitude amounts to 0.033$\pm$0.002, 0.040$\pm$0.002, and 0.027$\pm$0.002 for the S, M, H energy bands, respectively; again an agreement within 3$\sigma$ is found between $F_{var}(S)$ and $F_{var}(M)$ or between $F_{var}(S)$ and $F_{var}(H)$, but the larger value of the variability in the medium band appears here more clearly. 

The hardness ratios indicate only slightly smaller variability indices than the band-by-band values, with $VI(M/S) =0.058\pm0.011$ and $VI(H/M)=0.081\pm 0.015$ or $F_{var}(M/S)=0.019\pm0.002$ and  $F_{var}(H/M)=0.003\pm0.002$. Therefore, we examined correlations between energy bands. The correlation between soft and hard count rates is poor ($r=0.24$), but those between soft and medium count rates and between medium and hard count rates are better ($r$ of 0.79 and 0.49, respectively). Furthermore, there seems to be a correlation between hardness and brightness (Fig. \ref{xmmhr}): as the source brightens, $M/S$ increases (correlation coefficient $r$ of 0.35) while $H/M$ decreases ($r=-0.59$), i.e. the flux of \zp\ principally changes in the medium band.

The \swift\ light curves display variations of larger amplitudes, although the variability indicators ($VI$ and $F_{var}$) can be reconciled with those measured for \xmm\ within 3$\sigma$ considering their larger errors. While the instrumental sensitivities of both observatories are different, which  explains part of these differences, \swift\ light curves may also remain somewhat affected by optical loading. It is however encouraging to find the same brightness--hardness correlation in these data (except for two outliers, see rightmost panel of Fig. \ref{xmmhr}), demonstrating that the X-ray signal dominates over contamination. We come back to \swift\ data in the next section.

\subsection{Spectral differences}

In the \xmm\ light curves, the minimum count rate is reached for the beginning of Rev. 1814\footnote{Corresponding to pn count rate $<$6.1\,cts\,s$^{-1}$ in total band, or \xmm\ $TIME$ in 373670893--373710893\,s (from 2009-Nov-03 at 21h27m to 2009-Nov-04 at 8h34m).}, while the maximum is reached during the first part of Rev. 2183\footnote{Corresponding to pn count rate $>$6.7\,cts\,s$^{-1}$ in total band, or \xmm\ $TIME$ in 437255175--437300175 (from 2011-Nov-09 at 19h45m to 2011-Nov-10 at 8h15m)}. The difference in total band count rate between these two cases amounts to at least 0.6\,cts\,s$^{-1}$, which corresponds to $>$12 times the error bars on the 5\,ks bins. Thus, the beginning of Rev. 1814 and Rev. 2183 are well differentiated in terms of brightness. As these two cases represent the most extreme brightness variations of \zp, their pn and RGS spectra were examined in more detail to determine the nature of the changes. Figure \ref{comp} compares the spectra taken at minimum and maximum brightness. While a change in intensity is readily detected in both low- and high-resolution spectra, no other variation, such as shift of spectrum towards lower or higher energies or variation in line profile shapes, appears obvious by eye.

\begin{table*}
\centering
\caption{Comparison of the fitting results between minimum and maximum brightness cases.  }
\label{pnfit}
\begin{tabular}{l|cc|cc}
\hline\hline
Parameter & Min & Max & Min & Max \\
\hline
$N_{\rm H}(1)$ (10$^{22}$\,cm$^{-2}$)&  0.203$\pm$0.041 & 0.258$\pm$0.025 & 0.230$\pm$0.006 & 0.240$\pm$0.005 \\      
$kT(1)$ (keV)                      & 0.0808$\pm$0.0345&0.0808$\pm$0.0212& \multicolumn{2}{c}{0.0808 (fixed)} \\
$norm(1)$ (cm$^{-5}$)               & 0.111$\pm$0.150 &  0.295$\pm$0.238 & 0.162$\pm$0.012 & 0.229$\pm$0.014 \\   
$N_{\rm H}(2)$ (10$^{22}$\,cm$^{-2}$)&  0.283$\pm$0.022 & 0.372$\pm$0.016 & 0.331$\pm$0.016 & 0.342$\pm$0.012 \\     
$kT(2)$ (keV)                      &  0.259$\pm$0.006 & 0.288$\pm$0.002 & \multicolumn{2}{c}{0.273 (fixed)} \\
$norm(2)$ ($10^{-2}$\,cm$^{-5}$)    &  1.487$\pm$0.162 & 3.064$\pm$0.245 & 1.814$\pm$0.146 & 2.732$\pm$0.162 \\    
$N_{\rm H}(3)$ (10$^{22}$\,cm$^{-2}$)&  0.674$\pm$0.023 & 0.852$\pm$0.038 & 0.742$\pm$0.021 & 0.740$\pm$0.019 \\    
$kT(3)$ (keV)                      &  0.609$\pm$0.006 & 0.679$\pm$0.016 & \multicolumn{2}{c}{0.644 (fixed)} \\
$norm(3)$ ($10^{-2}$\,cm$^{-5}$)    &  1.152$\pm$0.027 & 1.089$\pm$0.059 & 1.028$\pm$0.016 & 1.249$\pm$0.017 \\         
$\chi_{\nu}^2$ (dof)               & 4.2 (109)        & 5.0 (114)       & 4.3 (112)       & 5.2 (117) \\
$F^{\rm obs}_{\rm X}$(T) ($10^{-12}$\,erg\,cm$^{-2}$\,s$^{-1}$)&12.00$\pm$1.30 &15.95$\pm$1.60 & 12.01$\pm$0.04 &15.94$\pm$0.04 \\ 
$F^{\rm obs}_{\rm X}$(S) ($10^{-12}$\,erg\,cm$^{-2}$\,s$^{-1}$)& 3.28$\pm$1.33 & 4.24$\pm$1.35 &  3.28$\pm$0.02 & 4.23$\pm$0.02 \\ 
$F^{\rm obs}_{\rm X}$(M) ($10^{-12}$\,erg\,cm$^{-2}$\,s$^{-1}$)& 6.29$\pm$0.15 & 8.58$\pm$0.20 &  6.26$\pm$0.02 & 8.60$\pm$0.02 \\ 
$F^{\rm obs}_{\rm X}$(H) ($10^{-12}$\,erg\,cm$^{-2}$\,s$^{-1}$)& 2.43$\pm$0.02 & 3.13$\pm$0.02 &  2.43$\pm$0.02 & 3.11$\pm$0.02 \\ 
\hline
\end{tabular}
\\
\tablefoot{For the last two columns, the flux ratios max/min are 1.327 $\pm$ 0.006, 1.289 $\pm$ 0.010, 1.375 $\pm$ 0.005, and 1.282 $\pm$ 0.013 for the total (0.3--4.\,keV), soft (0.3--0.6\,keV), medium (0.6--1.2\,keV), and hard (1.2--4.\,keV) bands, respectively. As explained in \citet{naz12}, the reduced $\chi^2$ values are large mostly because of remaining calibration uncertainties (both in the atomic and instrumental sides), which become particularly visible for such a bright object for which Poissonian errors are very small. Flux errors correspond to 1$\sigma$ and were derived using the ``flux err'' command under Xspec - if asymmetric, the highest value is reported here.}
\end{table*}

\begin{table*}
\centering
\caption{Comparison of line moments between minimum and maximum brightness cases.  }
\label{moment}
\begin{tabular}{lcccccc}
\hline\hline
Line & \multicolumn{3}{c}{Min}     & \multicolumn{3}{c}{Max} \\
     & position & width & skewness & position & width & skewness\\ 
     & \multicolumn{2}{c}{(km\,s$^{-1}$)} & & \multicolumn{2}{c}{(km\,s$^{-1}$)} & \\
\hline
Ne\,{\sc x}\,$\lambda$12.132   & $-213\pm38$ & 1253$\pm$20 & 0.318$\pm$0.064 & $-233\pm35$ & 1258$\pm$18 & 0.284$\pm$0.058 \\
Fe\,{\sc xvii}\,$\lambda$15.014& $-184\pm20$ & 991$\pm$10  & 0.339$\pm$0.039 & $-197\pm17$ & 968$\pm$9   & 0.363$\pm$0.035 \\
Fe\,{\sc xvii}\,$\lambda$15.261& $-280\pm30$ & 1059$\pm$15 & 0.375$\pm$0.053 & $-259\pm28$ & 1095$\pm$13 & 0.336$\pm$0.045 \\
Fe\,{\sc xvii}\,$\lambda$16.780& $-302\pm29$ & 882$\pm$15  & 0.448$\pm$0.061 & $-318\pm25$ & 881$\pm$13  & 0.438$\pm$0.055 \\
Fe\,{\sc xvii}\,$\lambda$17.051,17.096& $-238\pm34$ & 1202$\pm$18 & 0.178$\pm$0.066 & $-206\pm27$ & 1169$\pm$16 & 0.163$\pm$0.060 \\
O\,{\sc viii}\,$\lambda$18.967 & $-447\pm36$ & 1307$\pm$26 & 0.435$\pm$0.085 & $-362\pm32$ & 1321$\pm$22 & 0.426$\pm$0.071 \\
N\,{\sc vii}\,$\lambda$24.779  & $-132\pm37$ & 1500$\pm$23 & 0.259$\pm$0.068 & $-89\pm34$  & 1523$\pm$20 & 0.246$\pm$0.060 \\
Averages                       & $-257\pm12$ & 1171$\pm$7  & 0.336$\pm$0.024 & $-238\pm11$ & 1174$\pm$6  & 0.322$\pm$0.021 \\
\hline
\end{tabular}
\\
\tablefoot{Columns provide position $=M_1=\sum F_i\times v_i /\sum F_i$, width $=\sqrt{M_2}$, and skewness $=M_3/(M_0\times M_2^{1.5}),$ where $M_0=\sum F_i$, $M_2=\sum F_i\times (v_i-M_1)^2 /\sum F_i$, and $M_3=\sum F_i\times (v_i-M_1)^3 /\sum F_i$. The intervals in which these parameters were evaluated are shown on Figure \ref{comp}. The errors were derived from error propagation, hence systematic uncertainties are not included. For example, the RGS  wavelength scale display variations with $\sigma\sim5$\,m\AA\ \citep{dev15}, corresponding to an additional error of 60--125\,km\,s$^{-1}$ on the positions of the chosen lines.  }
\end{table*}

Fits were then performed for the pn spectra with Xspec 12.9.0i, using \citet{asp09} for the reference solar abundances. A model $wabs*(vphabs*vapec+vphabs*vapec+vphabs*vapec)$ was used, where the first absorption was fixed to the interstellar value of $8.9\times10^{19}$\,cm$^{-2}$ \citep{dip94} and CNO abundances of the other components were fixed to those derived in \citet{her13}\footnote{\citet{her13} derived mass fractions of $6.6\times10^{-4}$, $7.7\times10^{-3}$, and $3.05\times10^{-3}$ for C, N, and O, respectively (with a mass fraction of hydrogen equal to 0.70643 since that paper used solar abundances from \citealt{and89}). The Xspec abundances are in number, relative to hydrogen and relative to solar, hence the C, N, and O abundances used here are 0.26, 11.5, and 0.55 times solar, using \citet{asp09} as reference.}. This model is similar to that used in \citet{naz12}, although that model had a fourth thermal component. In our case, adding this hottest component did not significantly improve the fit, so it was discarded for simplicity. Fitting results are provided in Table \ref{pnfit}. In a first trial, the absorptions, temperatures, and normalizations of all thermal components were free to vary. Higher values of these components were derived for the maximum brightness case. However, a second trial was then made, fixing the temperatures to their average values. The absorptions were then found to be consistent within errors, letting only normalization factors significantly vary between minimum and maximum brightness cases. In particular, the smallest variation is detected for the hottest component (0.644\,keV) and the largest variation is detected for the intermediate temperature (0.273\,keV). The largest flux change is recorded in the medium band, while the flux ratios in the soft and hard bands are similar, which confirms the light curve results. The variations in \zp\ thus seem to result from a global increase in the strength of the emission, particularly that of the warm plasma ($kT\sim0.3$\,keV).

To get a more detailed look at variations, X-ray lines recorded by RGS can be studied. However, the X-ray lines of \zp\ are well known to be non-Gaussian \citep[e.g.][]{kah01}. Therefore, to best compare the RGS spectra, line moments (orders 0 to 3) were calculated in the velocity space for strong and relatively isolated lines (a method used e.g. by \citealt{coh06}): Ne\,{\sc x}\,$\lambda$12.132, Fe\,{\sc xvii}\,$\lambda$15.014,15.261,16.780,17.051/96, O\,{\sc viii}\,$\lambda$18.967, and N\,{\sc vii}\,$\lambda$24.779. Table \ref{moment} provides the results, showing that there are no significant differences between position, width, and skewness of lines at minimum or maximum brightness. Comparing the line profile shapes rather than their overall properties does not reveal any obvious feature either, considering the errors (Fig. \ref{comp}). The changes thus seem limited to the line intensities, as found before for low-resolution pn spectra. 

\begin{figure*}
\includegraphics[width=6.cm]{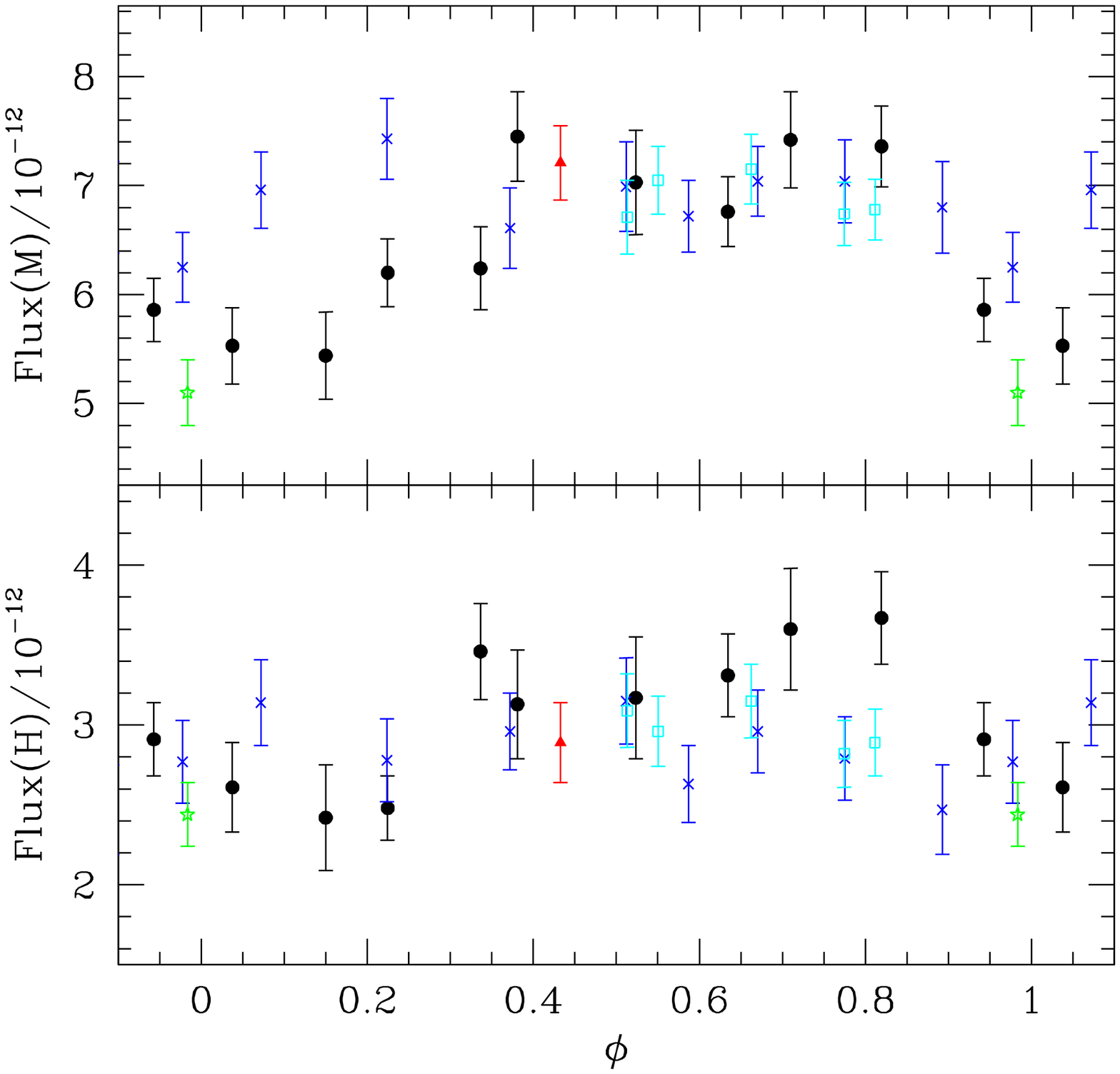}
\includegraphics[width=6.cm]{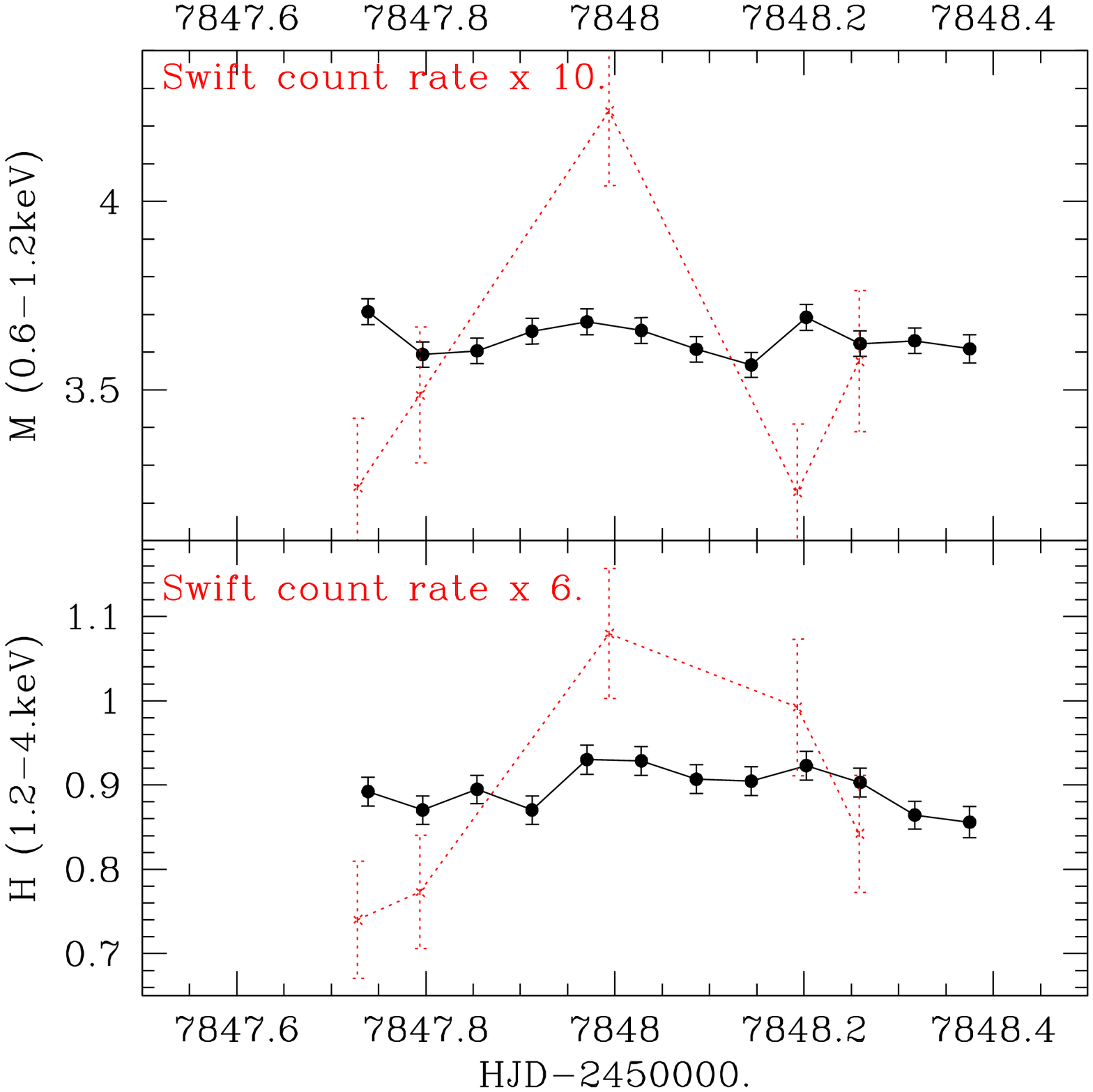}
\includegraphics[width=6.cm]{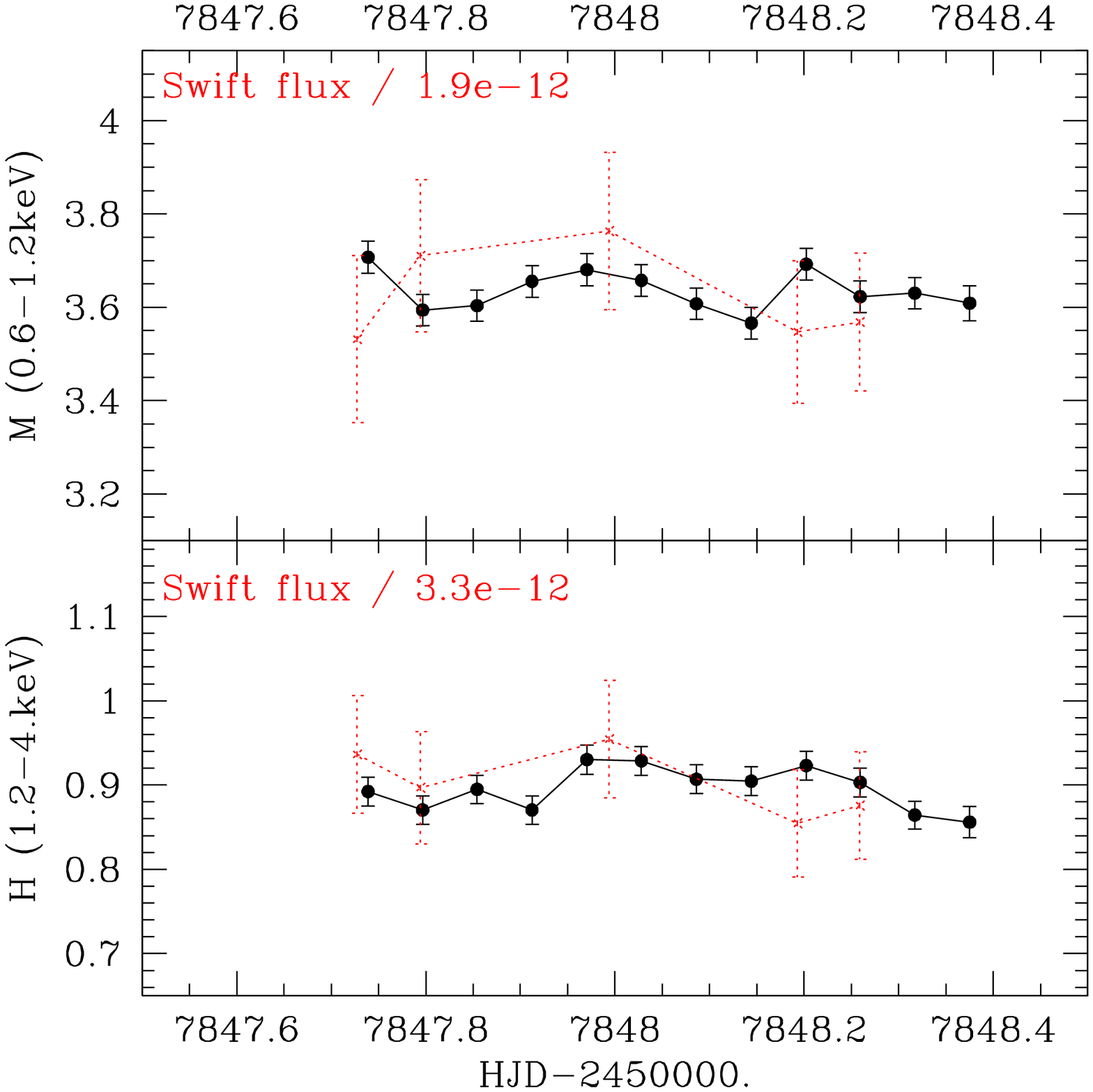}
\caption{{\it Left:} Same as Figure \ref{swlc} but for the fluxes (Table \ref{swiftfit}). The green stars correspond to data from February 2013. {\it Middle and right:} Comparison between the \xmm\ light curve of April 2017 (Rev. 3172, black dots and solid lines) and the (arbitrarily scaled) simultaneous \swift\ light curve (red crosses and dotted lines) derived from count rates (middle) or fluxes (right). }
\label{swiftflux}
\end{figure*}

\begin{table*}
\centering
\caption{Results of the spectral fits to \swift\ data.  }
\label{swiftfit}
\begin{tabular}{lcccccc}
\hline\hline
ID  & $norm(2)$ & $norm(3)$ & $\chi_{\nu}^2$ (dof) & $F^{\rm obs}_{\rm X}$(M) & $F^{\rm obs}_{\rm X}$(H) & $gain\,offset$\\
    & ($10^{-2}$\,cm$^{-5}$) & ($10^{-2}$\,cm$^{-5}$) & & \multicolumn{2}{c}{($10^{-12}$\,erg\,cm$^{-2}$\,s$^{-1}$)} & \\
\hline
S-01   & 1.22$\pm$0.16 & 1.10$\pm$0.13 & 0.78 (34) & 5.10$\pm$0.30 & 2.44$\pm$0.20 & 0.114$\pm$0.006 \\
S-2-3  & 2.07$\pm$0.23 & 1.23$\pm$0.19 & 1.08 (45) & 7.21$\pm$0.34 & 2.89$\pm$0.25 & 0.136$\pm$0.015 \\
S-04   & 1.69$\pm$0.26 & 1.69$\pm$0.27 & 0.92 (43) & 7.36$\pm$0.37 & 3.67$\pm$0.29 & 0.109$\pm$0.008 \\
S-05   & 1.33$\pm$0.18 & 1.34$\pm$0.14 & 1.06 (37) & 5.86$\pm$0.29 & 2.91$\pm$0.23 & 0.100$\pm$0.005 \\
S-06   & 1.34$\pm$0.20 & 1.18$\pm$0.16 & 1.24 (25) & 5.53$\pm$0.35 & 2.61$\pm$0.28 & 0.071$\pm$0.010 \\ 
S-07   & 1.41$\pm$0.25 & 1.07$\pm$0.30 & 0.57 (25) & 5.44$\pm$0.40 & 2.42$\pm$0.33 & 0.048$\pm$0.008 \\
S-08   & 1.78$\pm$0.20 & 1.05$\pm$0.16 & 1.46 (40) & 6.20$\pm$0.31 & 2.48$\pm$0.20 & 0.111$\pm$0.011 \\
S-09   & 1.20$\pm$0.23 & 1.63$\pm$0.19 & 1.18 (36) & 6.24$\pm$0.38 & 3.46$\pm$0.30 & 0.071$\pm$0.013 \\
S-10   & 2.05$\pm$0.25 & 1.35$\pm$0.22 & 0.55 (38) & 7.45$\pm$0.41 & 3.13$\pm$0.34 & 0.123$\pm$0.009 \\
S-11   & 1.80$\pm$0.27 & 1.41$\pm$0.33 & 1.15 (21) & 7.03$\pm$0.48 & 3.17$\pm$0.38 & 0.034$\pm$0.006 \\
S-12   & 1.56$\pm$0.19 & 1.50$\pm$0.15 & 0.94 (42) & 6.76$\pm$0.32 & 3.31$\pm$0.26 & 0.071$\pm$0.009 \\
S-13   & 1.75$\pm$0.24 & 1.65$\pm$0.28 & 1.09 (32) & 7.42$\pm$0.44 & 3.60$\pm$0.38 & 0.108$\pm$0.009 \\
S-14   & 1.70$\pm$0.24 & 1.31$\pm$0.25 & 1.17 (40) & 6.61$\pm$0.37 & 2.96$\pm$0.24 & 0.145$\pm$0.009 \\
S-15   & 1.78$\pm$0.22 & 1.39$\pm$0.17 & 1.23 (40) & 6.99$\pm$0.41 & 3.15$\pm$0.27 & 0.049$\pm$0.004 \\
S-16   & 1.96$\pm$0.30 & 1.10$\pm$0.14 & 0.73 (40) & 6.72$\pm$0.33 & 2.63$\pm$0.24 & 0.099$\pm$0.004 \\
S-17   & 1.94$\pm$0.22 & 1.28$\pm$0.17 & 0.68 (42) & 7.04$\pm$0.32 & 2.96$\pm$0.26 & 0.096$\pm$0.009 \\
S-18   & 2.03$\pm$0.20 & 1.17$\pm$0.15 & 1.05 (37) & 7.04$\pm$0.38 & 2.79$\pm$0.26 & 0.071$\pm$0.009 \\
S-19   & 2.09$\pm$0.20 & 0.99$\pm$0.16 & 1.12 (34) & 6.80$\pm$0.42 & 2.47$\pm$0.28 & 0.154$\pm$0.008 \\
S-20   & 1.63$\pm$0.22 & 1.22$\pm$0.18 & 0.62 (36) & 6.25$\pm$0.32 & 2.77$\pm$0.26 & 0.149$\pm$0.007 \\
S-21   & 1.77$\pm$0.21 & 1.40$\pm$0.21 & 1.12 (41) & 6.96$\pm$0.35 & 3.14$\pm$0.27 & 0.117$\pm$0.009 \\
S-23   & 2.25$\pm$0.20 & 1.14$\pm$0.16 & 0.79 (40) & 7.43$\pm$0.37 & 2.78$\pm$0.26 & 0.105$\pm$0.004 \\
S-24/1 & 1.65$\pm$0.25 & 1.36$\pm$0.18 & 1.28 (49) & 6.71$\pm$0.34 & 3.09$\pm$0.23 & 0.055$\pm$0.003 \\
S-25/2 & 1.95$\pm$0.22 & 1.28$\pm$0.17 & 1.16 (52) & 7.05$\pm$0.31 & 2.96$\pm$0.22 & 0.061$\pm$0.007 \\
S-26/3 & 1.88$\pm$0.19 & 1.39$\pm$0.16 & 0.86 (57) & 7.15$\pm$0.32 & 3.15$\pm$0.23 & 0.106$\pm$0.009 \\
S-27/4 & 1.86$\pm$0.20 & 1.22$\pm$0.19 & 0.98 (51) & 6.74$\pm$0.29 & 2.82$\pm$0.21 & 0.111$\pm$0.013 \\
S-28/5 & 1.84$\pm$0.20 & 1.26$\pm$0.15 & 1.07 (53) & 6.78$\pm$0.28 & 2.89$\pm$0.21 & 0.086$\pm$0.007 \\
\hline
\end{tabular}
\\
\tablefoot{Model is that used for \xmm, i.e. $wabs*(vphabs*vapec+vphabs*vapec+vphabs*vapec)$, with $N_{\rm H}(ISM)=8.9\times10^{19}$\,cm$^{-2}$, $kT$ fixed to 0.0808, 0.273, and 0.644\,keV (see rightmost columns of Table \ref{pnfit}), and CNO abundances fixed to 0.26, 11.5, and 0.55 times the solar abundances \citep{asp09}, respectively. Since the \swift\ spectra are uncertain at low energies, bins with energies $<$0.5\,keV were discarded, the additional absorptions were fixed to the \xmm\ mean $N_{\rm H}$ (2.35, 3.36, and 7.41$\times10^{21}$\,cm$^{-2}$; see rightmost columns of Table \ref{pnfit}), and the ratio $norm(1)/norm(2)$ was fixed to the \xmm\ mean (8.65; see right column of Table \ref{pnfit}). Errors correspond to 1$\sigma$ and were derived using the ``flux err'' command under Xspec for fluxes or the ``error'' command for nomalizations factors - if asymmetric, the largest value is reported here.}
\end{table*}

As a last investigation, the \swift\ spectra were also analysed. The chosen model is the same as used for \xmm, but with some parameters fixed. The \xmm\ fits showed that, if temperatures are fixed, absorptions are similar whatever the brightness, so both parameters were fixed here. Furthermore, since the \swift\ spectra are only examined above 0.5\,keV and the first thermal component dominates at energies below 0.5\,keV, the $norm(1)/norm(2)$ ratio was also fixed. Following recommendations of the XRT team, the spectral fitting was performed considering the possibility of energy scale offsets (command ``gain fit'' under Xspec, with slope fixed to 1). Results are provided in Table \ref{swiftfit} and the left panel of Fig. \ref{swiftflux}. The fluxes present lower variability indicators than the light curves, leading to a better agreement with \xmm\ data. A direct comparison of the simultaneous \xmm/\swift\ observations of April 2017 is shown in the middle and right panels of Fig. \ref{swiftflux}, and it further confirms that fluxes derived from \swift\ spectra better correlate with the \xmm\ data than the \swift\ count rates. Problems due to optical loading seem negligible; the \swift\ spectra truly reflect the actual brightness variations of \zp\ and can thus be trusted. But we need to be aware of their larger Poisson noise, due to shorter exposure times and lower effective area of \swift ,\  and of the potentially larger systematic uncertainties (up to $\sim20$\% in the worst cases).  

\subsection{Shapes of the variations}

There is a second striking feature in Figures \ref{xmm3} to \ref{swlc}: the shape of the light curves. Indeed, one can observe near constancy (e.g. Rev. 2540), linear trends (e.g. Rev. 1343), or parabolic trends (e.g. Revs. 1983 and 2183), and there is even one case in which the X-ray luminosity experiences a slow increase followed by a steeper increase (Rev. 1814). Not only are the mean levels and amplitudes different, well beyond the calibration uncertainties, but the light curve shapes are sometimes incompatible with each other. Indeed, the steep increase at the end of Rev. 1814 cannot be reconciled with the shallower increases observed in, for example Revs. 1620, 1983, or 2183 (Fig. \ref{xmm1}). 

\begin{figure}
\center
\includegraphics[width=8.5cm]{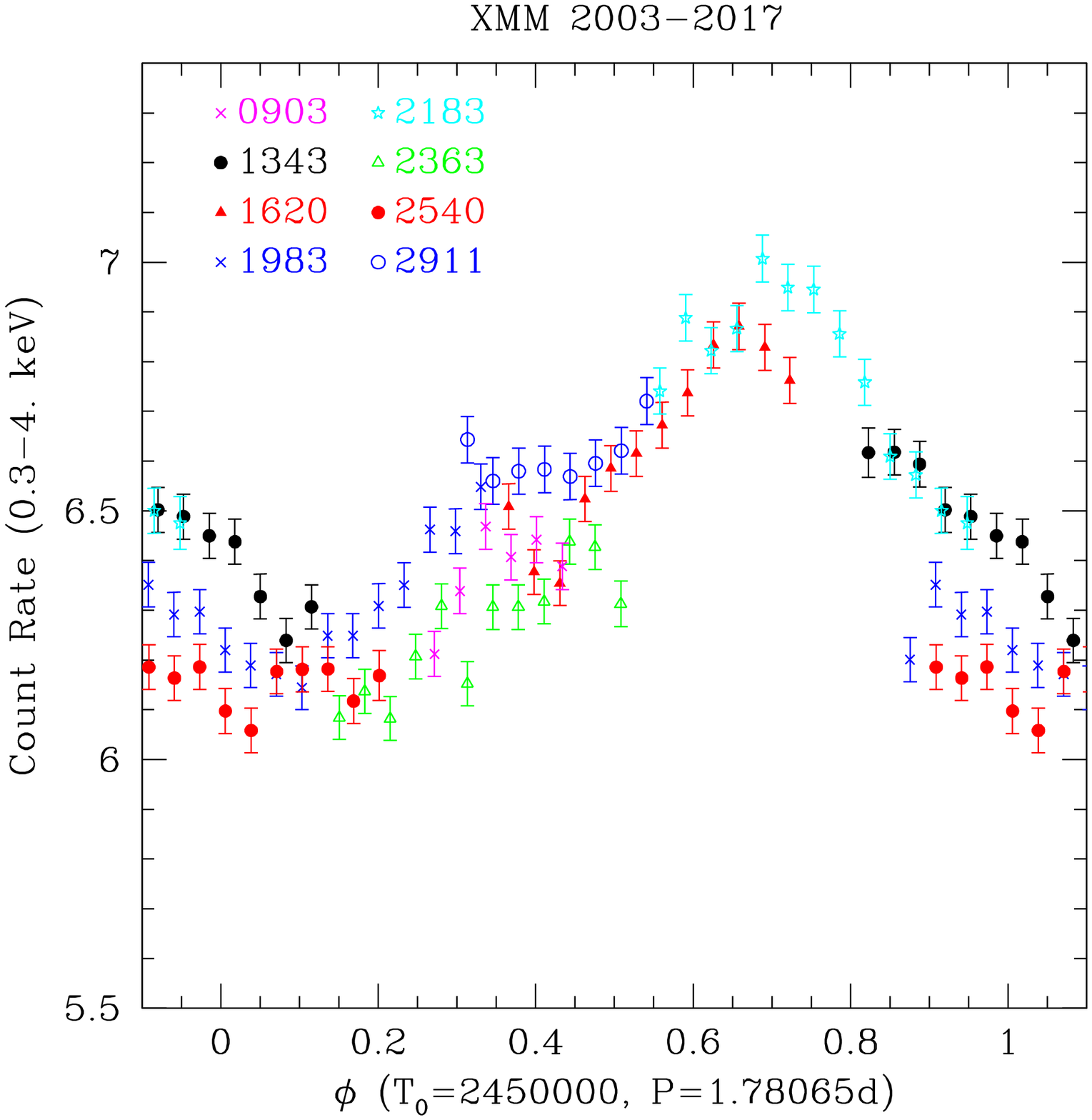}
\includegraphics[width=8.5cm]{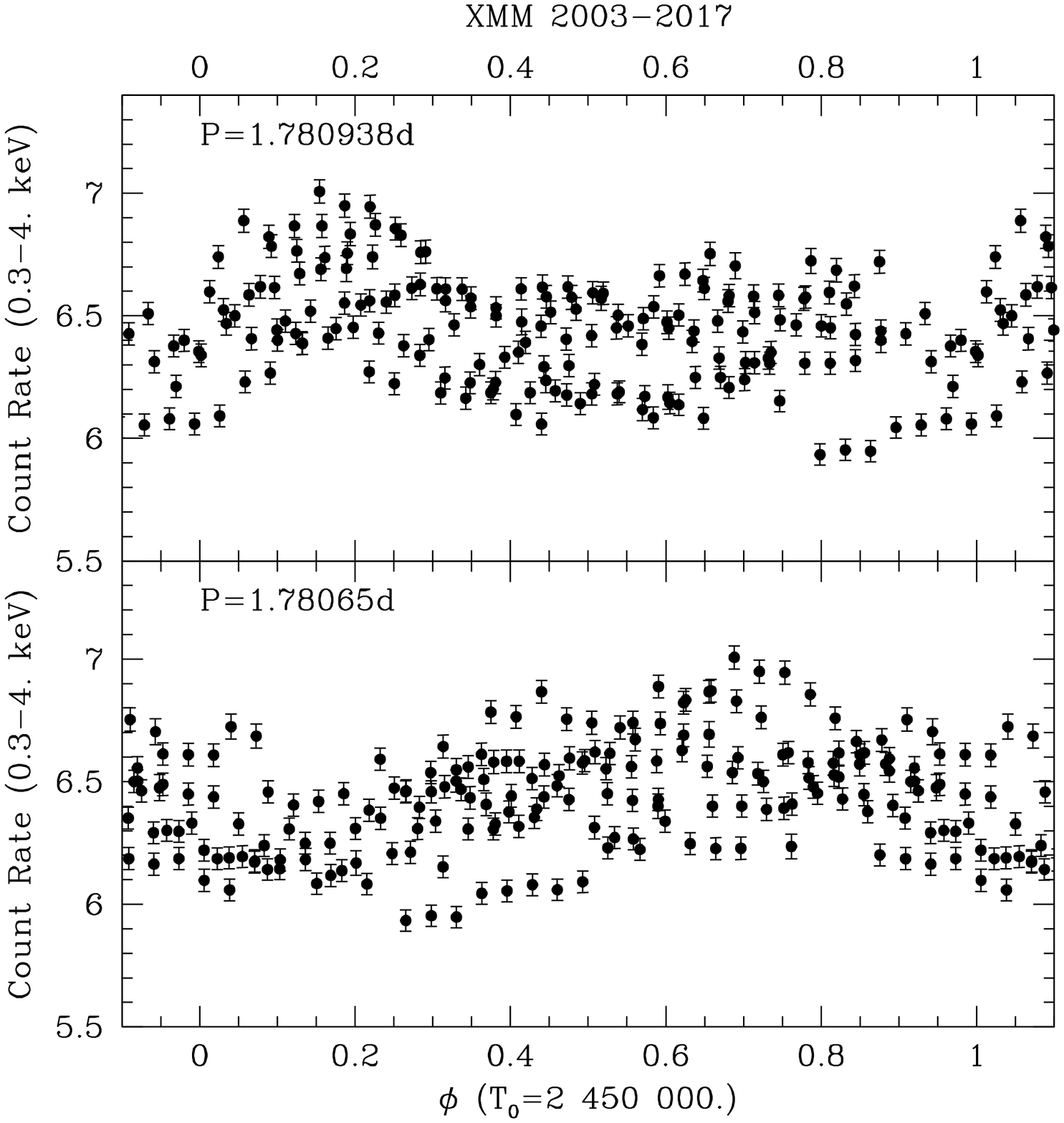}
\caption{Subgroup of \xmm\ light curves can be phased with a period of 1.78065\,d to form a relatively coherent behaviour (top), but considering all available light curves only lead to scatter plots, whatever the value of the 1.78\,d period considered (bottom). }
\label{xmmp}
\end{figure}

A last interesting feature is the phasing of these X-ray light curves, performed using the very precise value for the periodic variability in optical light of \zp\ derived by \citet{how14} from $SMEI$ data, i.e. $P=1.780938\pm0.000093$\,d. Assuming the period to remain stable and considering its error bar, a 1$\sigma$ error on the period would yield a phase shift of only 0.02 after one year, or 0.1 after five years. There should thus be little phase shift between individual X-ray light curves in each of the figures \ref{xmm3} to \ref{swlc}, although phase shifts as large as $\sim0.3$ (for a 1$\sigma$ error) can be expected between the beginning and the end of the 14\,yrs observing campaign in X-rays. However, the observed light curves do not form a coherent behaviour (Fig. \ref{xmm3} to \ref{swlc} and \ref{xmmp}, bottom). Even data taken the same year -- an interval for which phasing errors and any instrumental sensitivity changes should be negligible -- cannot be reconciled with each other. This is the case of Revs. 0795 and 0903, both taken in 2004, of Revs. 0980, 1071, and 1096, taken in 2005, (see Fig. \ref{xmm3}) and, to a lesser extent (only some bins do not match within 3$\sigma$ for these, see Fig. \ref{xmm2}), of Revs. 2533 and 2540, both taken in 2013, and of Revs. 2817 and 2911, both taken in 2015. The light curves of a subgroup of observations, however, can be combined to form a coherent, sinusoid-like overall shape. This is best achieved by slightly modifying the period to, for example 1.78065\,d (Fig. \ref{xmmp}, top), a value similar to that recently found by \citet{ram17} - these authors have found $P=1.78063\pm0.00025$\,d for the weighted average of the period values derived from the analysis of the individual datasets ($SMEI$ 2003--2004, $SMEI$ 2004--2005, $SMEI$ 2005--2006, and $BRITE$ 2014--2015). This provides marginal support for the presence of this period in the X-ray data, but only at some epochs; it should be noted that some exposures taken in between those of this subgroup (e.g. Rev. 1814) do not show the same variations (Fig. \ref{xmm1}).  The X-ray variations of \zp\ definitely appear to differ widely from a perfectly stable modulation, both in terms of phasing and overall light curve shape. Retrospectively, this irregular behaviour may explain the variability detection obtained with {\it ROSAT} \citep{ber96} and the subsequent non-detection in {\it ASCA} data \citep{osk01}. The amplitude of variation found by Berghoefer is similar to what we detect with \xmm; simply, as in 2007--2011, such strong variability was transient.

Such a behaviour is not totally unprecedented for this star. Indeed, \citet{how14} mentioned that the amplitude of the photometric cycle varied between 5 and 12 mmag. \citet{ram17} further showed that the light curve shape changes as well. This could perhaps be understood if the photometric changes arise in a non-constant phenomenon, for example  stellar spots. The behaviour remains coherent but only for a while during the lifetime of the spot, and as new spots arise, a phase shift and a change in the light curve shape can be observed. In any case, this somewhat erratic behaviour is usual for Oef stars such as \zp, where cycles with similar but not perfectly identical periodicities are often found in spectroscopic time series \citep{rau03,deb04,rau15,sud16}. However, at a given date, the star should display one single behaviour. If a CIR scenario in which structures at the stellar surface launch strong wind pertubations is correct, a tight correlation is then expected between optical and X-ray light curves. In fact, in our case, this correlation between the behaviour of \zp\ at optical wavelengths and that at high energies can be directly evaluated as some photometric campaigns were obtained at the same time as the X-ray observations of \zp, for example $SMEI$ data in 2004--2005 \citep{how14} and $BRITE$ data in 2015--2017 (\citealt{ram17} and in preparation).  

For $BRITE$ data, the noise on individual photometric points is small, so that two comparisons are possible: (1) simultaneous X-ray and optical data (Fig. \ref{britea}) and (2) the X-ray light curve and the average 1.78\,d optical cycle around the same date (Fig. \ref{briteb}). The latter solution compares the X-ray light curves only to the 1.78\,d cycle while the former compares these light curves to the full optical photometry, which includes both the 1.78\,d cycle and a stochastic component of similar amplitude \citep{ram17}. Because the origin of the X-ray variations is uncertain, both comparisons should be considered. For $SMEI$ data, the noise on individual photometric points is rather large, so that only averages could be used, erasing the short-term stochastic component. In fact, the optical data taken within $\pm15$\,d of the X-ray observing dates (Table \ref{journal}) were combined and those averages are then compared with the associated X-ray data in Figure \ref{smei}. 

In 2004, X-ray observations were taken in April and November (Fig. \ref{smei}). The former data were taken during an optical photometric minimum and the X-ray light curve is relatively stable, and hence there may be a weak correlation between the two;  the latter data were taken on the descending branch between the maximum and minimum optical brightnesses, but X-rays show an increase in strength (anti-correlation). In 2005, there were three X-ray observations in April, October, and December (Fig. \ref{smei}). All of these observations were taken at and after the minimum optical brightness phase, but the first and third X-ray dataset are stable so there is no correlation between X-ray and optical for these datasets;  the second X-ray light curve reveals an increase in X-ray brightness so a correlation is detected. In April 2015, the X-ray observations were taken as \zp\ displayed a marked increase in optical brightness (Figs. \ref{britea} and \ref{briteb}) but the X-ray light curve remained stable: there is no correlation between the two. In December 2016, \zp\ was observed by \swift\ as it slowly brightened at optical wavelengths but the X-ray light curve displays a dip and then remained relatively stable (Fig. \ref{britea}): the last five points of the X-ray light curve are at a similar level, while their simultaneous $BRITE$ data show a clear increase. In January 2017, \zp\ remained stable both in optical and X-rays. Finally, in April 2017, the X-ray light curve shows a marked decrease while the optical photometry shows both an increase and a decrease over the same interval (Figs. \ref{britea}  and \ref{briteb}). There is thus no obvious correlation between X-ray and optical emissions in the currently available data\footnote{This also confirms that the \swift\ spectra are relatively free from optical contamination, since we would expect a correlation in that case.}. 

Thus, in general, there does not seem to be a clear, unique correlation between the behaviour of \zp\ in optical and X-rays, even when we allow for short, fraction of day, time lags. Furthermore, the only potential evidence for a 1.78\,d cycle in the X-ray domain comes from a subset of observations (mostly from 2007--2013, see top of Fig. \ref{xmmp}), but no optical monitoring is available these years. It is thus difficult to assess whether the brightness of \zp\ in the optical showed a similar variability at that time, or if on the contrary the optical changes had completely disappeared, in a scenario in which the variability source had become more energetic. In summary, we cannot directly and systematically relate the X-ray variability and optical changes, and therefore we find no support for the hypothesis of a simple, direct connection. The infrequent presence of a 1.78\,d timescale at X-ray energies is possible but requires further checking, using additional monitoring, to be fully demonstrated. 

\section{Conclusion}

Because   \zp \ is a bright and nearby object, this star\ has been the target of numerous investigations, including at high energies. In particular, it is known to be variable in X-rays since at least \citet{ber96}. Using X-ray datasets obtained by \xmm\ before 2010, \citet{naz12,naz13} found a lack of the short-term stochastic changes expected to be associated with the clumpy nature of the massive-star winds, but they detected longer term trends. Similar features were then found in other massive stars and they are currently thought to be associated with large-scale structures in the wind, i.e. wind perturbations somehow triggered at the stellar surface (perhaps from spots or pulsations). As the latter can be detected at optical wavelengths, a correlation is expected between the X-ray and optical behaviours. 

Half a decade later, new X-ray observations of \zp\ were obtained by \xmm\ and \swift, and a clear, stable periodicity of 1.78\,d was identified in long optical photometric runs of the star (\citealt{how14} and \citealt{ram17}). We thus revisited the X-ray variability of \zp. 

The full dataset of X-ray light curves now reveals their variety: both the amplitude and the shape of the variations change from one exposure to the next. A subgroup of observations displays very large variations that can be combined to form a sinusoid-like shape for a period of 1.78\,d.  However, the other X-ray datasets behave in a different way, often with rather constant light curves of different mean levels that are difficult to reconcile with one another or with the subgroup light curves. 

Hoping to gather more detailed information on the nature of the X-ray changes, we then compared \xmm\ spectra (both EPIC at low resolution and RGS at high resolution) taken at extreme X-ray brightnesses since changing velocities or temperatures routinely help constrain the origin of high-energy variations. No significant changes were found for the temperatures, absorptions, and line shapes (centroid, width, and skewness). This notably rejects the scenario in which variability arises from changing absorption as denser wind structures come and go into the line of sight. Only changes in flux, slightly stronger in the medium (0.6--1.2\,keV) energy band than at softer or harder energies, are detected. The impossibility to constrain the hot plasma properties more precisely  precludes us from clarifying its nature.

The ephemerides of \citet{how14} are precise enough that datasets separated by a few years can be compared in phase. However, apart maybe from a subgroup, the X-ray light curves cannot be phased coherently because of incompatibilities arising even for datasets taken the same year where calibration and ephemeris uncertainties play little role. Since the 1.78\,d cycle of \zp\ varies in amplitude as well as shape, and since additional optical variations (stochastic changes of similar amplitude) are also present, a more meaningful comparison is achieved if X-ray data and optical photometry obtained simultaneously are examined. This comparison can be made thanks to $SMEI$ in 2004--2005 and $BRITE$ in 2015--2017 but again, no clear, single result is obtained. Sometimes a correlation between optical and X-ray data is detected, sometimes there is an anti-correlation, and there is a total absence of correlation at still other times. Therefore, there is no clear X-ray/optical link and the current data do not validate the scenario of X-rays arising in large-scale structures triggered at the surface of \zp, traced by the optical data. 

To solve the riddle of \zp\ X-ray variations in the future, performing repeated, long-term, high-cadence, and multiwavelength monitoring will be needed. Only a multiwavelength study, which is best performed by combining X-rays to UV and optical data, can unveil the wind properties of the star at a given time. Such a comparison also needs to be long enough (several days/weeks, not a single snapshot) so that a clear view of the long-term (weeks) behaviour is acquired. Finally, since the variability properties of \zp\ are not constant with epoch ($>$weeks), repeated observations are required, so that conclusions drawn at one time can be checked and fully validated.

\begin{acknowledgements}
YN acknowledges support from the Fonds National de la Recherche Scientifique (Belgium), the Communaut\'e Fran\c caise de Belgique, the PRODEX \xmm\ and Integral contracts, and an ARC grant for concerted research actions financed by the French community of Belgium (Wallonia-Brussels Federation). TR acknowledges support from the Canadian Space Agency grant FAST. AFJM is grateful for financial support from NSERC (Canada) and FQRNT (Quebec). We thank Kim Page, from the \swift\ team, for her kind assistance, the \xmm\ team for their help, the referees for valuable comments that improved the paper, and the whole XRT team for their dedication towards making these \swift\ observations possible and usable. ADS and CDS were used for preparing this document.  Based on data collected with \swift, the ESA science mission \xmm\ (an ESA Science Mission with instruments and contributions directly funded by ESA Member States and the USA), the \emph{SMEI}/Coriolis instrument, and the \emph{BRITE-Constellation} satellite mission (designed, built, launched, operated, and supported by the Austrian Research Promotion Agency - FFG, the University of Vienna, the Technical University of Graz, the Canadian Space Agency - CSA, the University of Toronto Institute for Aerospace Studies - UTIAS, the Foundation for Polish Science \& Technology - FNiTP MNiSW, and National Science Centre - NCN).
\end{acknowledgements}

\setcounter{figure}{0}

\begin{figure*}
\includegraphics[width=5.5cm]{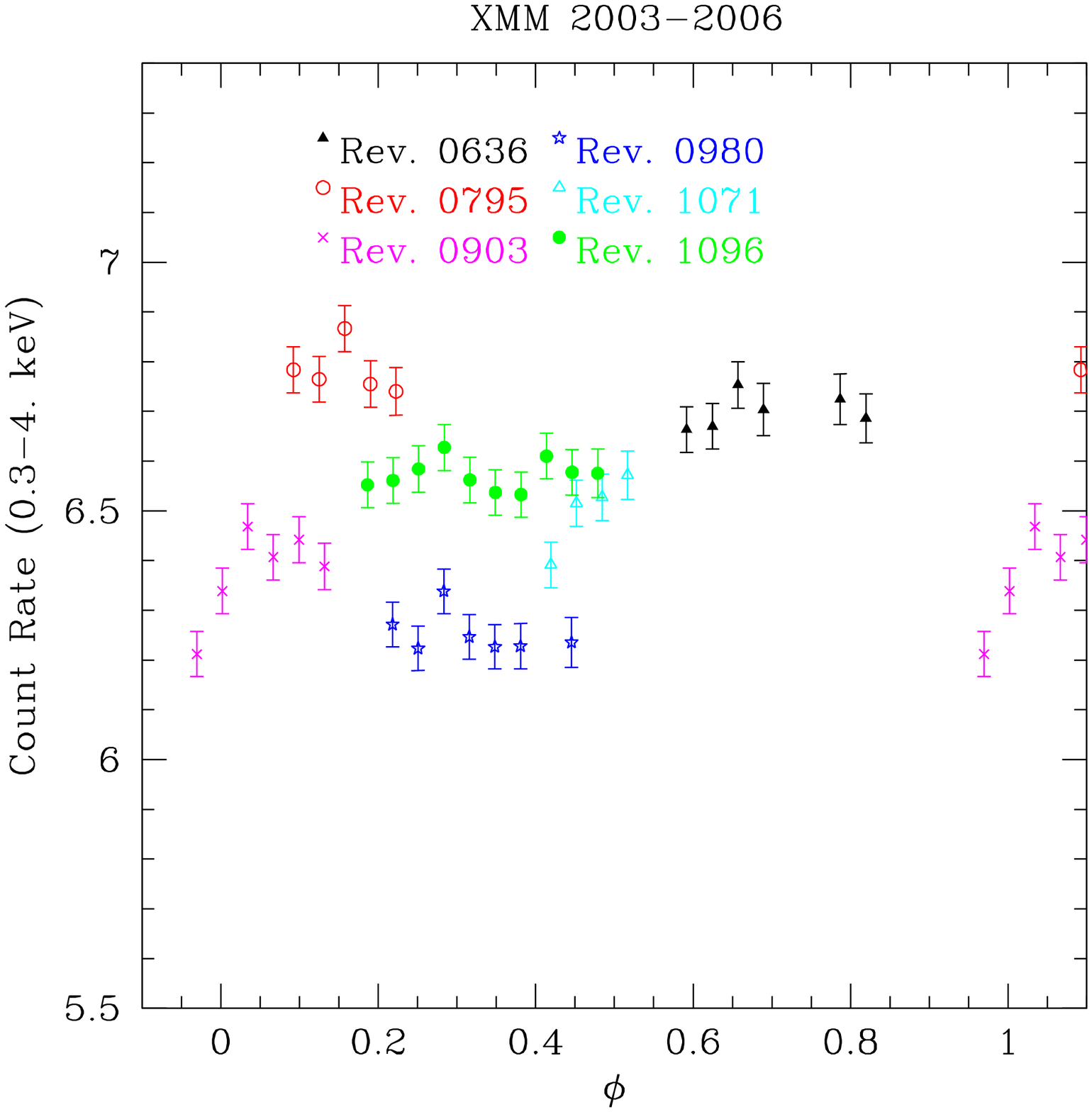}
\includegraphics[width=5.5cm]{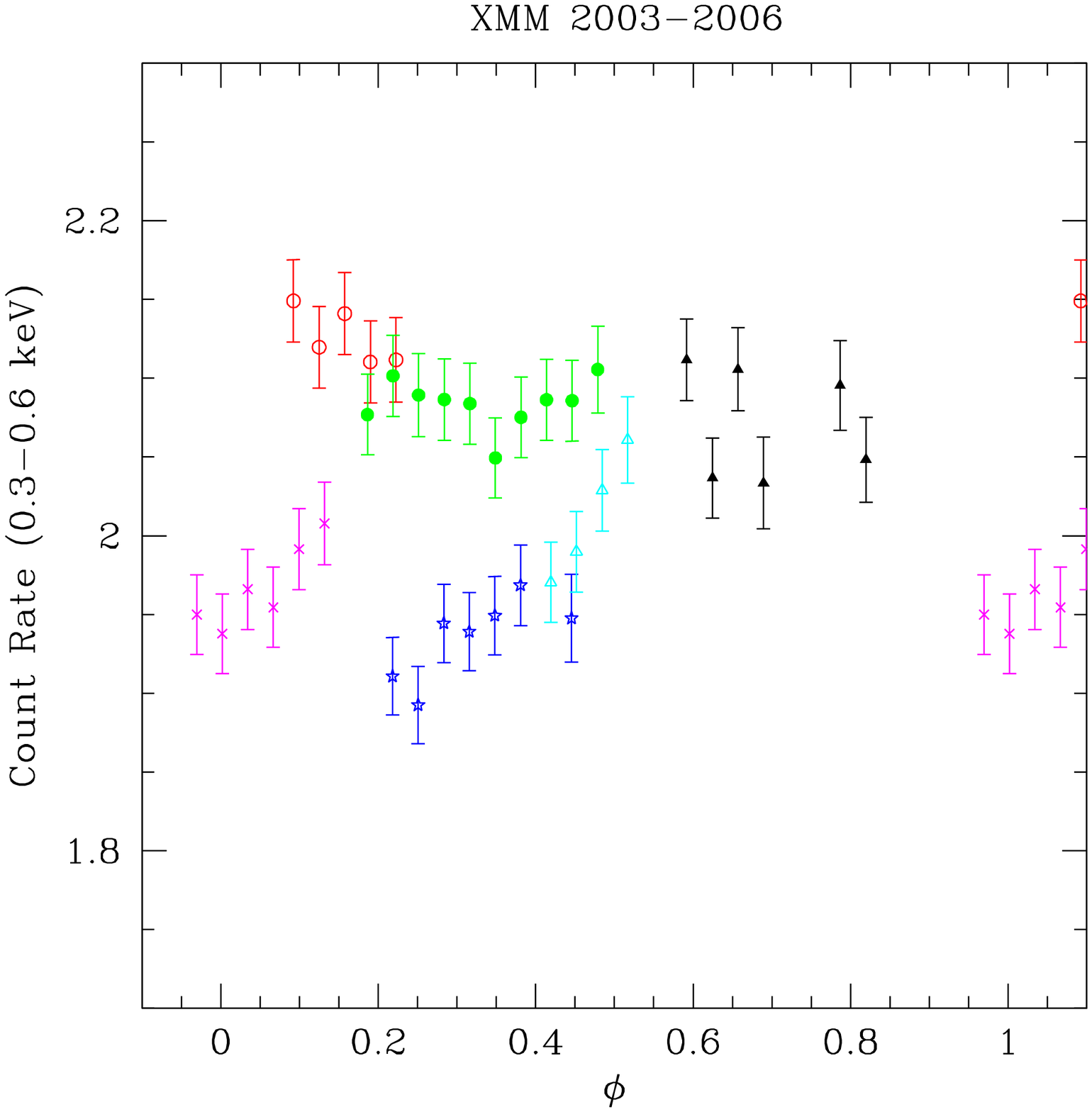}
\includegraphics[width=5.5cm]{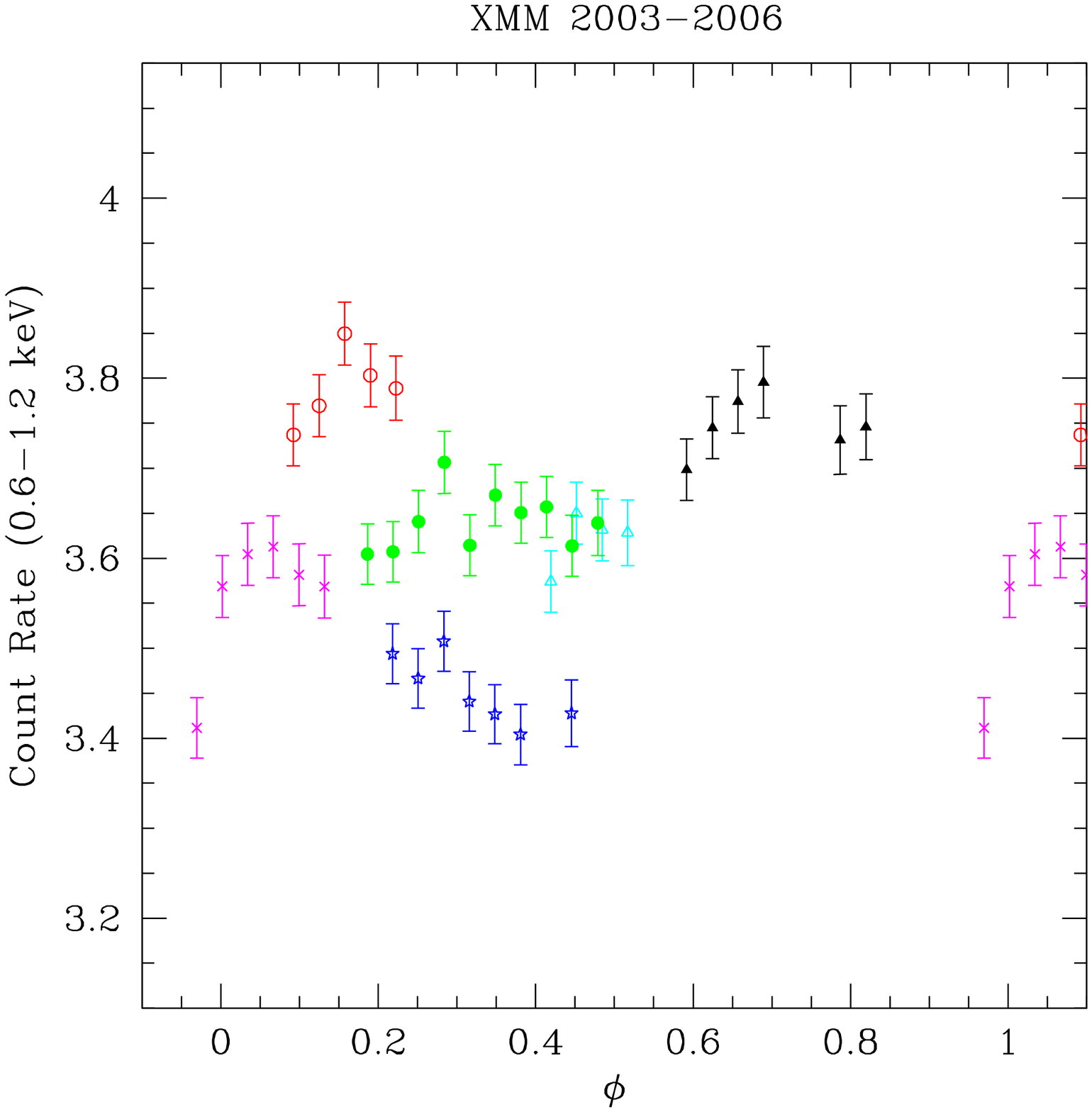}
\\
\includegraphics[width=5.5cm]{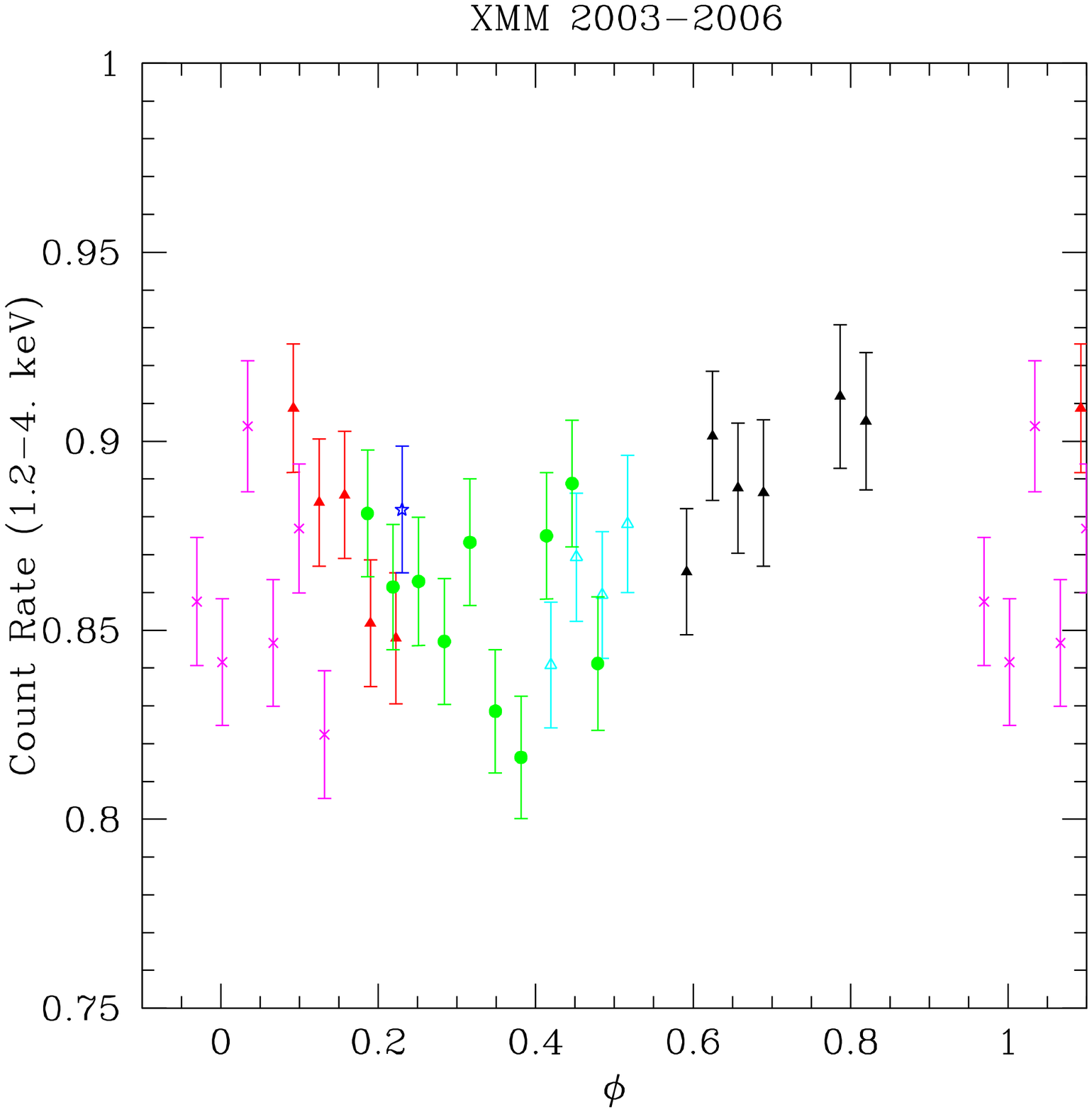}
\includegraphics[width=5.5cm]{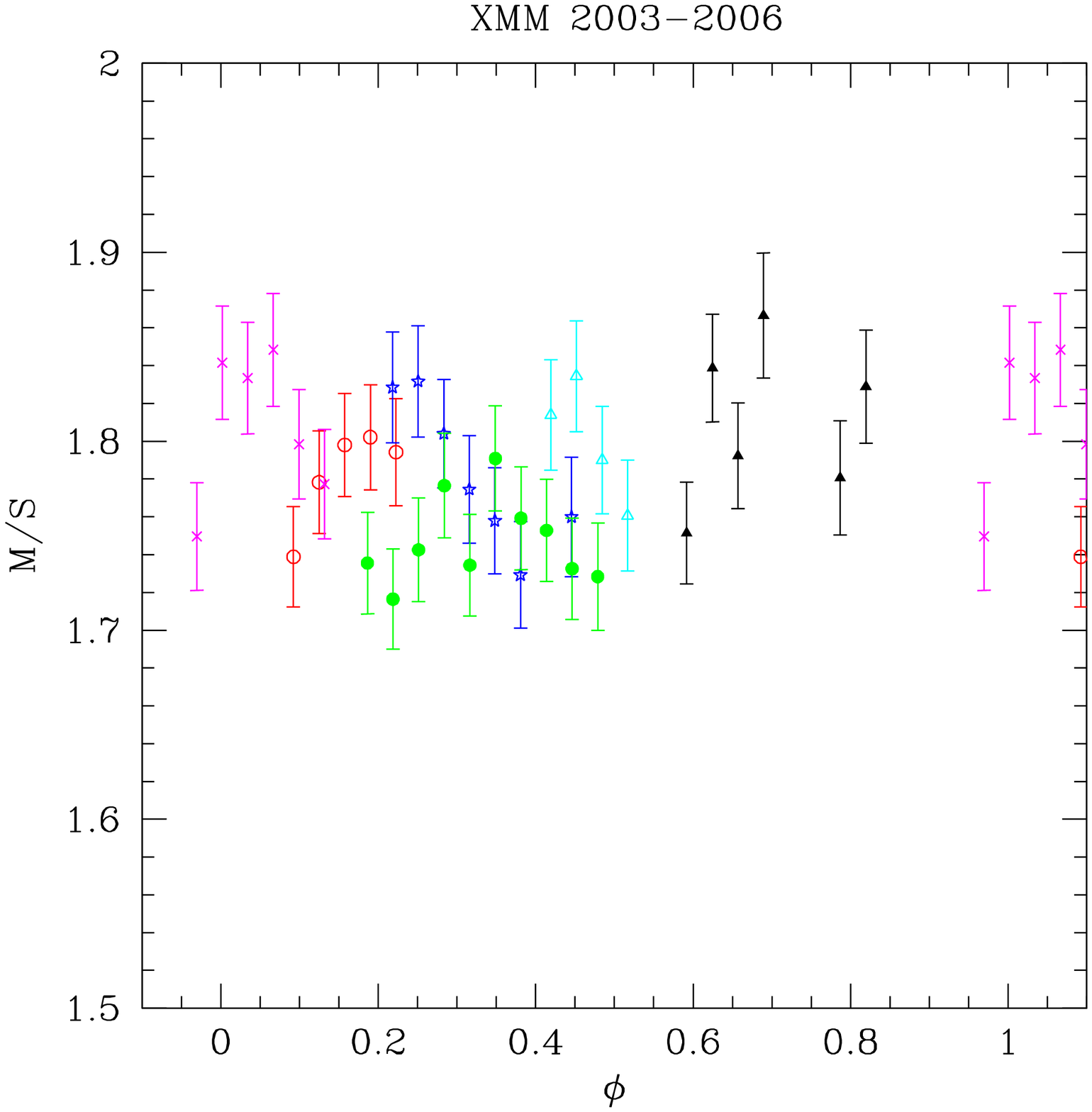}
\includegraphics[width=5.5cm]{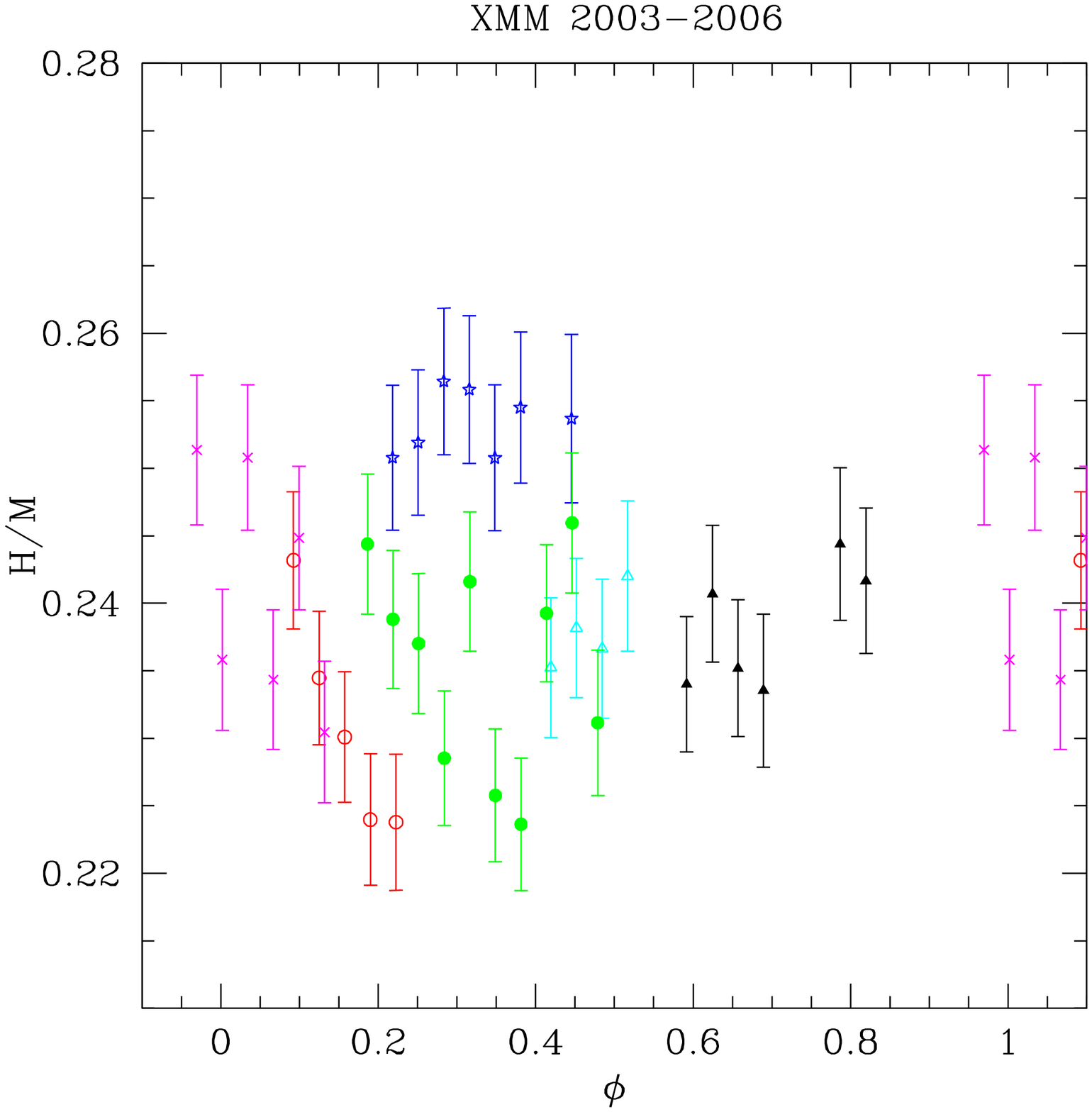}
\caption{Count rates in several energy bands of \zp, and associated hardness ratios for \xmm\ data from 2003--2006. The phases were calculated using $P=1.780938$\,d and $T_0$=2\,450\,000 \citep{how14}. To facilitate the comparisons, the ratio between upper and lower limits of the y-axes are all equal.}
\label{xmm3}
\end{figure*}

\begin{figure*}
\includegraphics[width=5.5cm]{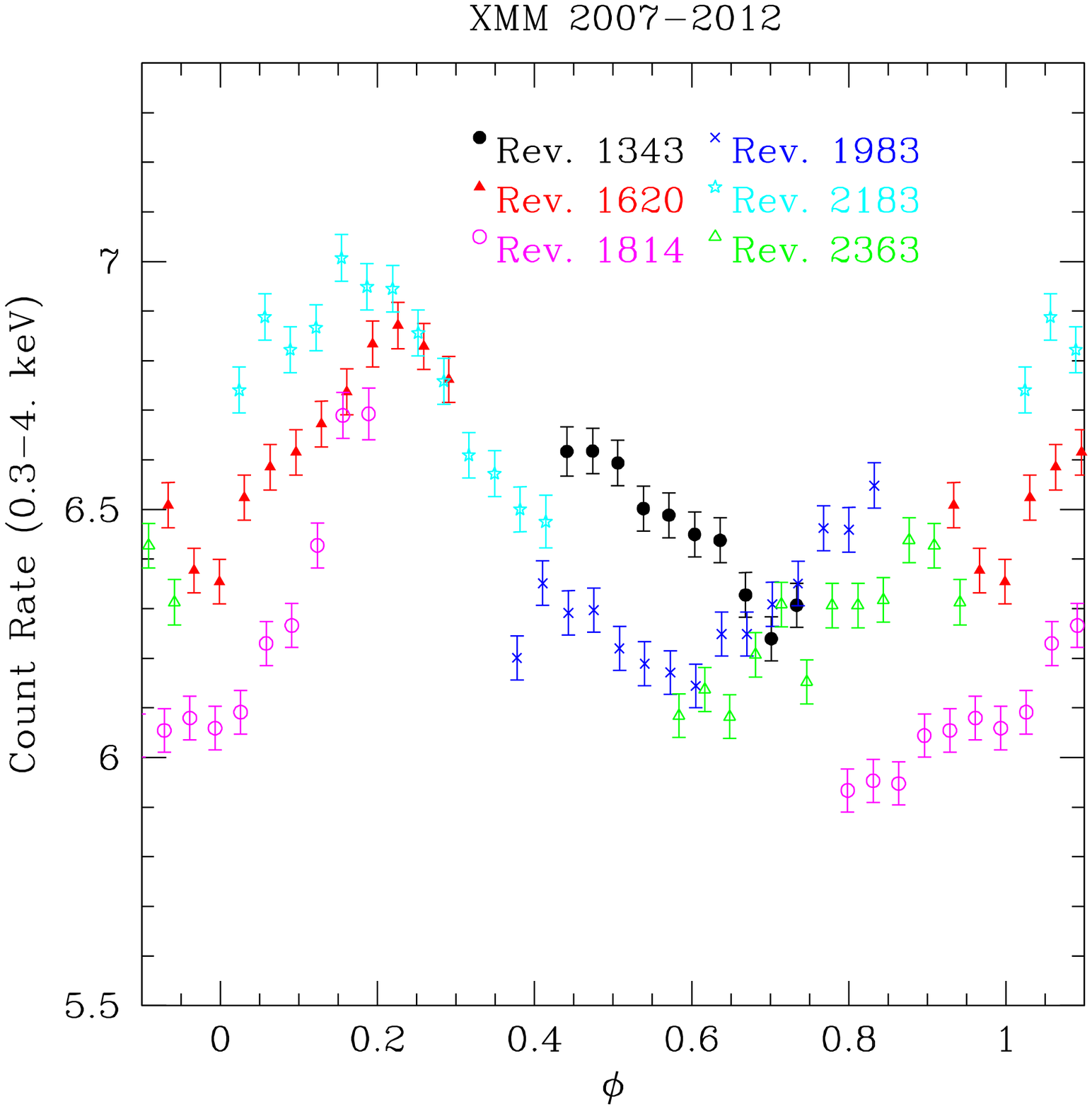}
\includegraphics[width=5.5cm]{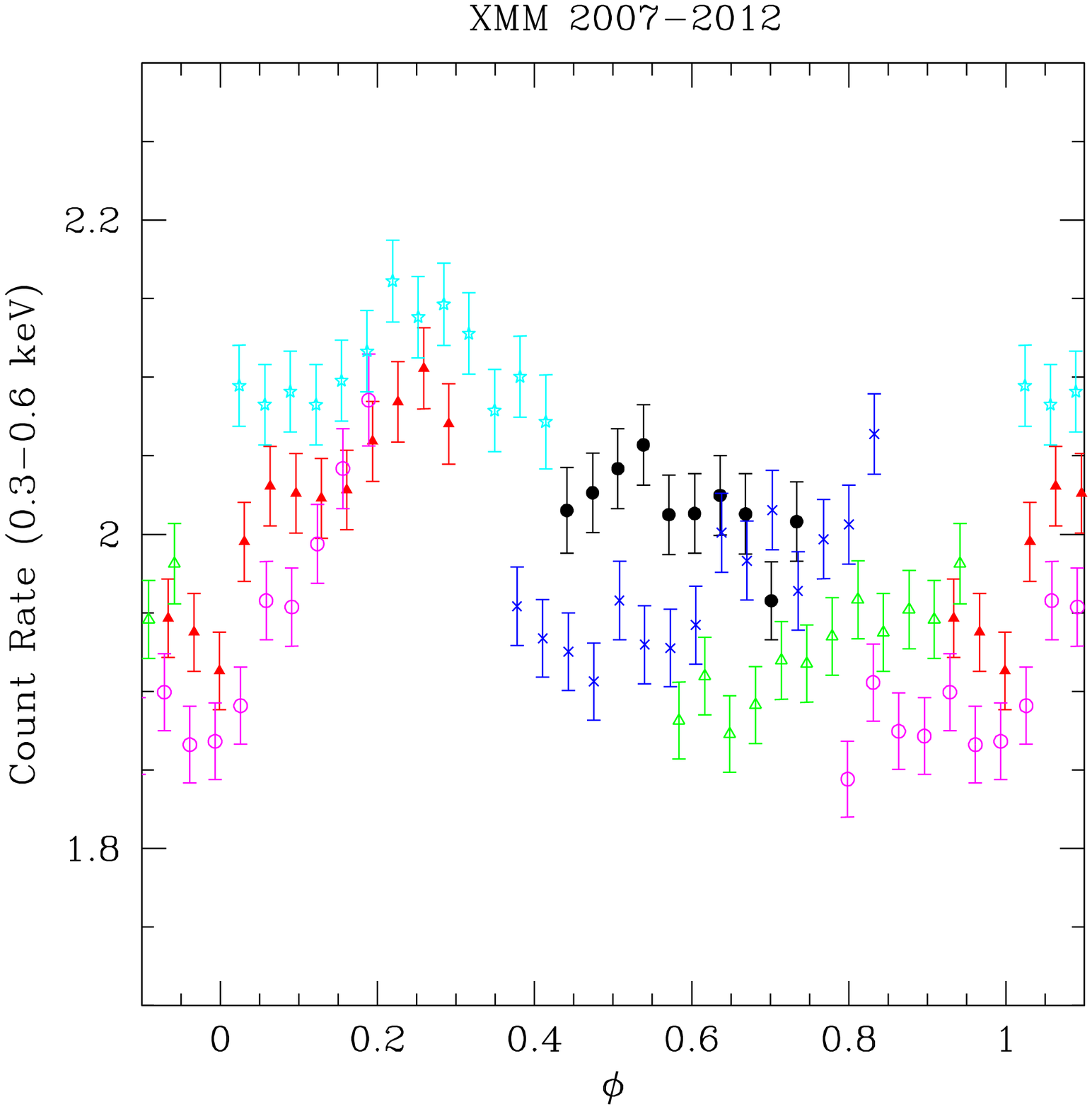}
\includegraphics[width=5.5cm]{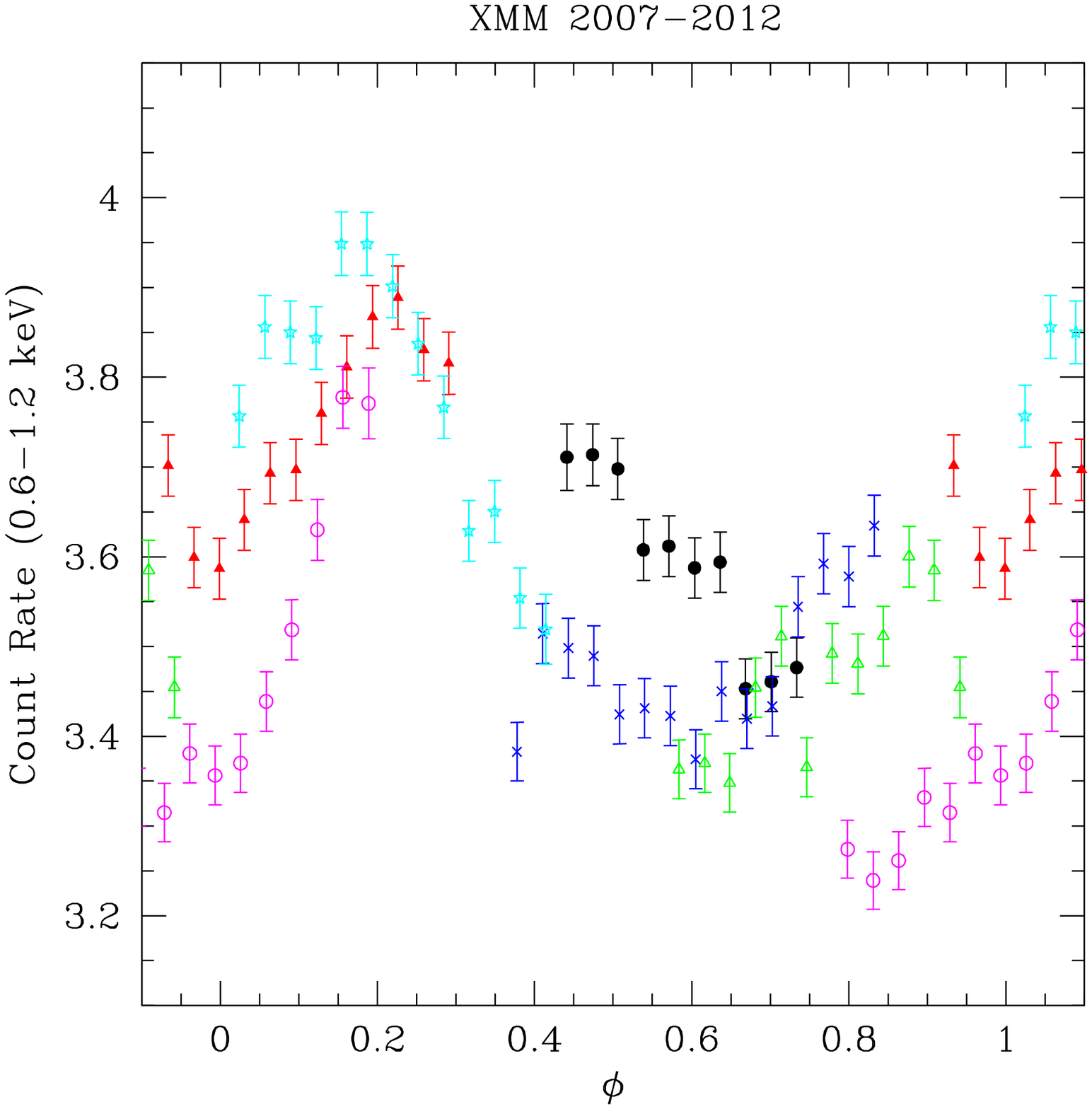}
\\
\includegraphics[width=5.5cm]{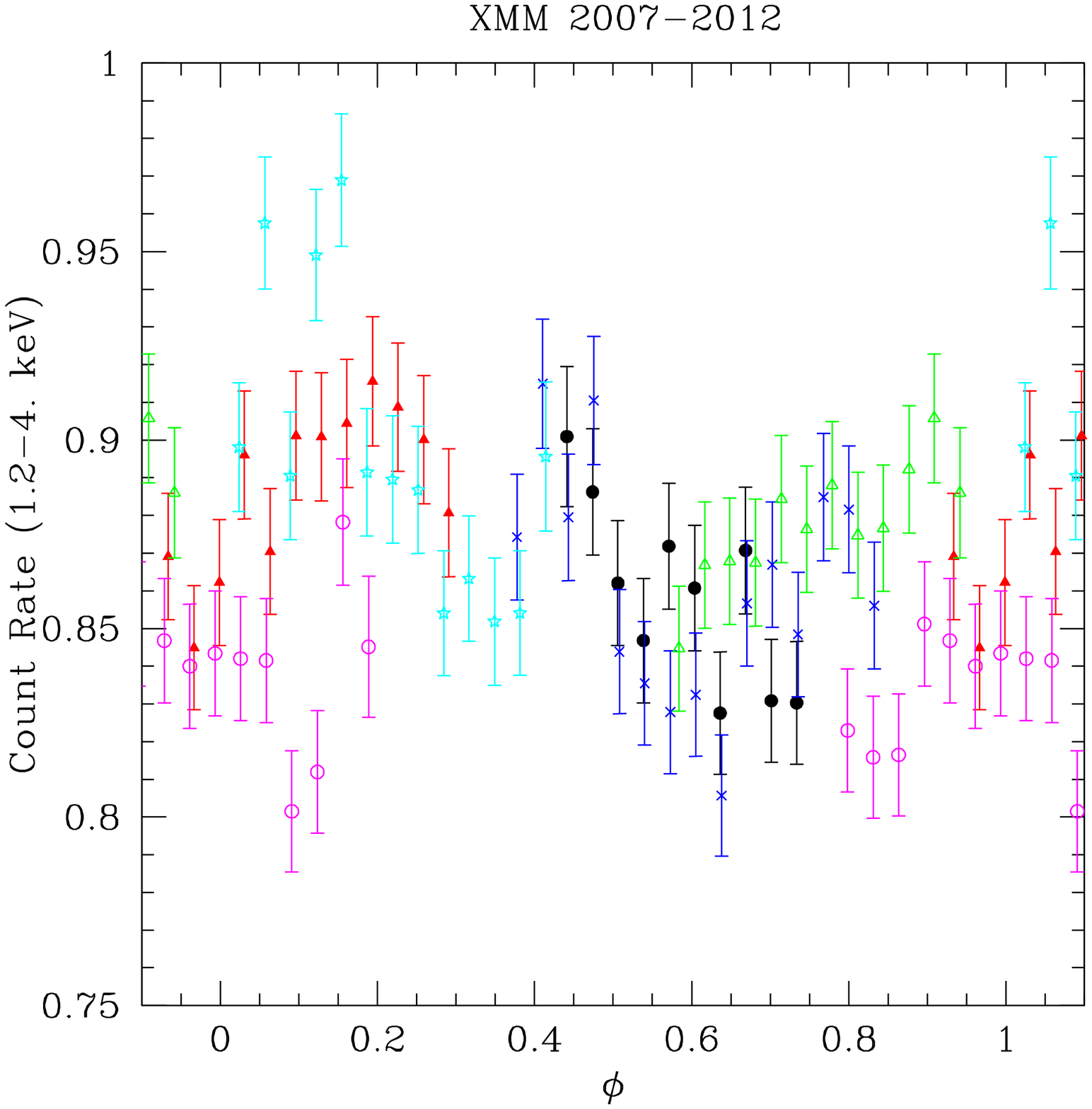}
\includegraphics[width=5.5cm]{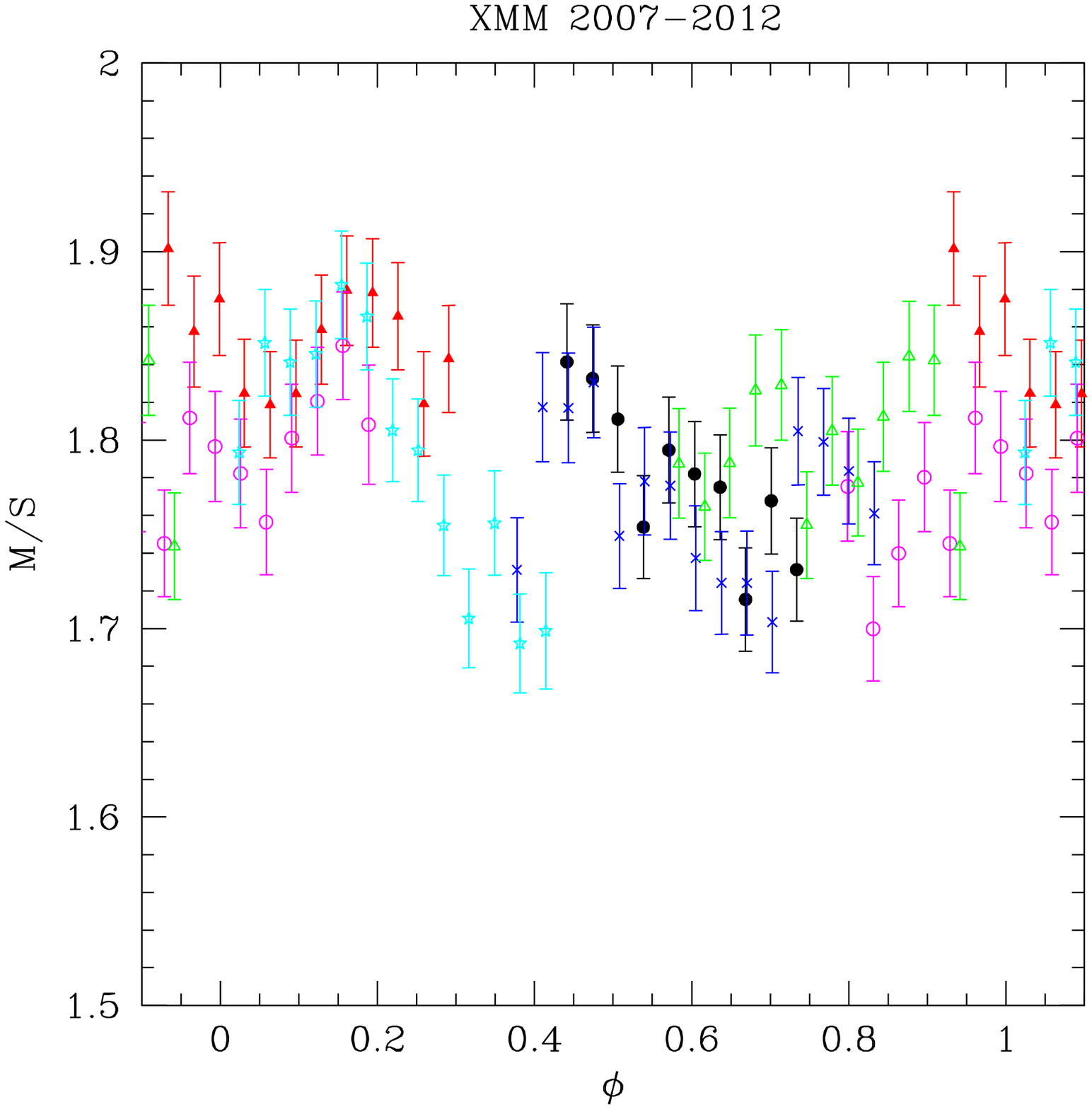}
\includegraphics[width=5.5cm]{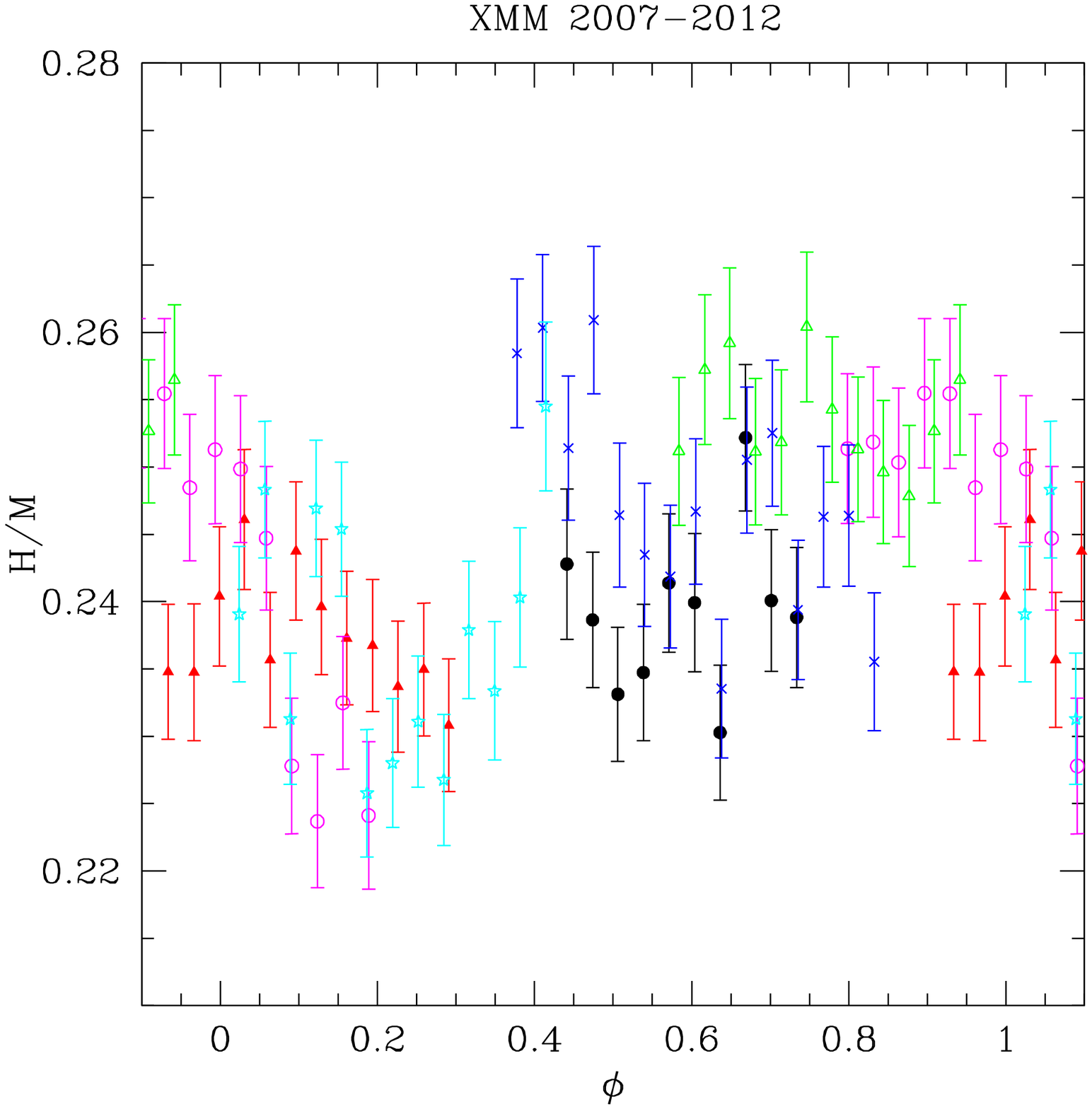}
\caption{Same as Figure \ref{xmm3} but for \xmm\ data from 2007--2012.}
\label{xmm1}
\end{figure*}

\begin{figure*}
\includegraphics[width=5.5cm]{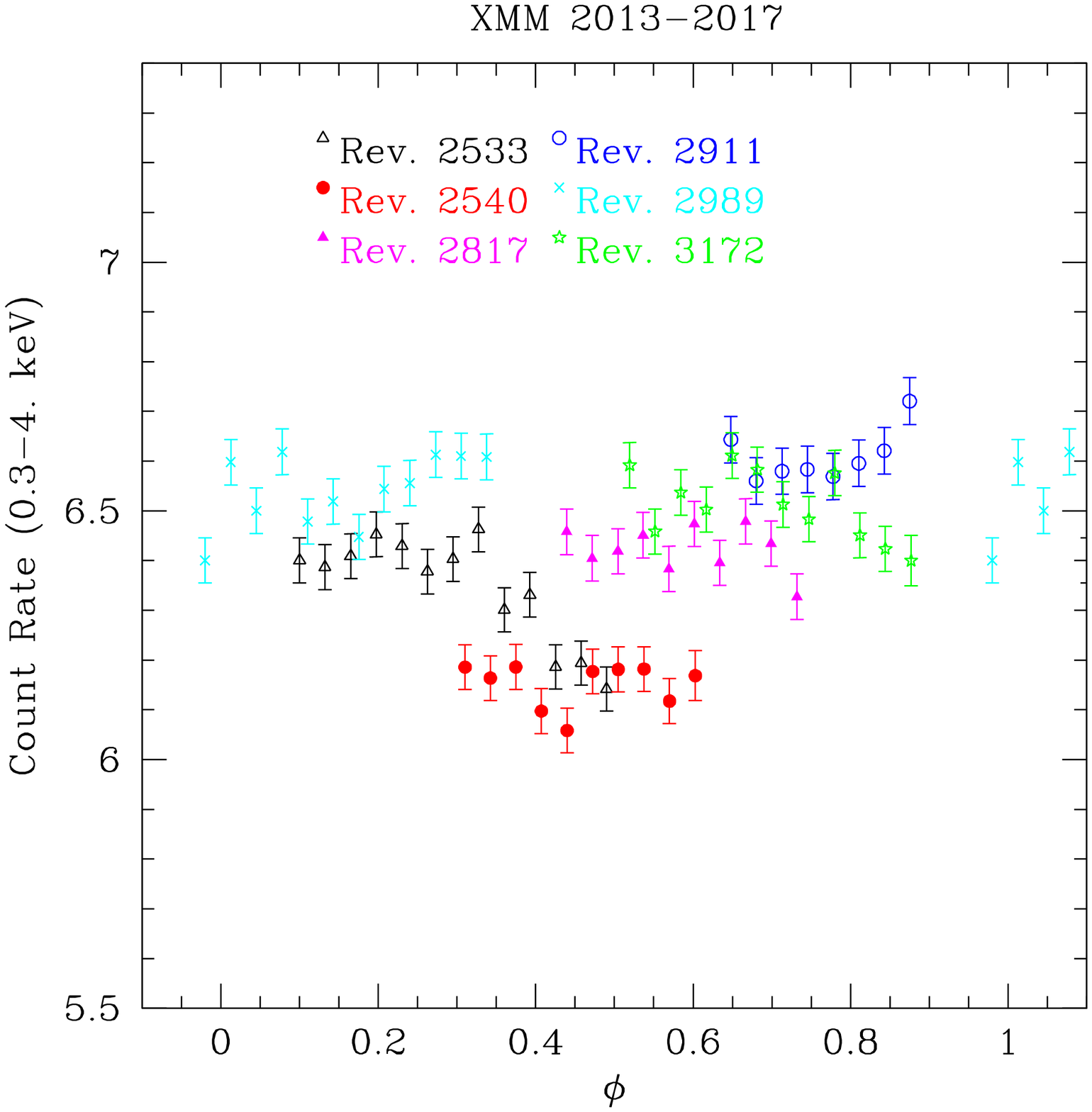}
\includegraphics[width=5.5cm]{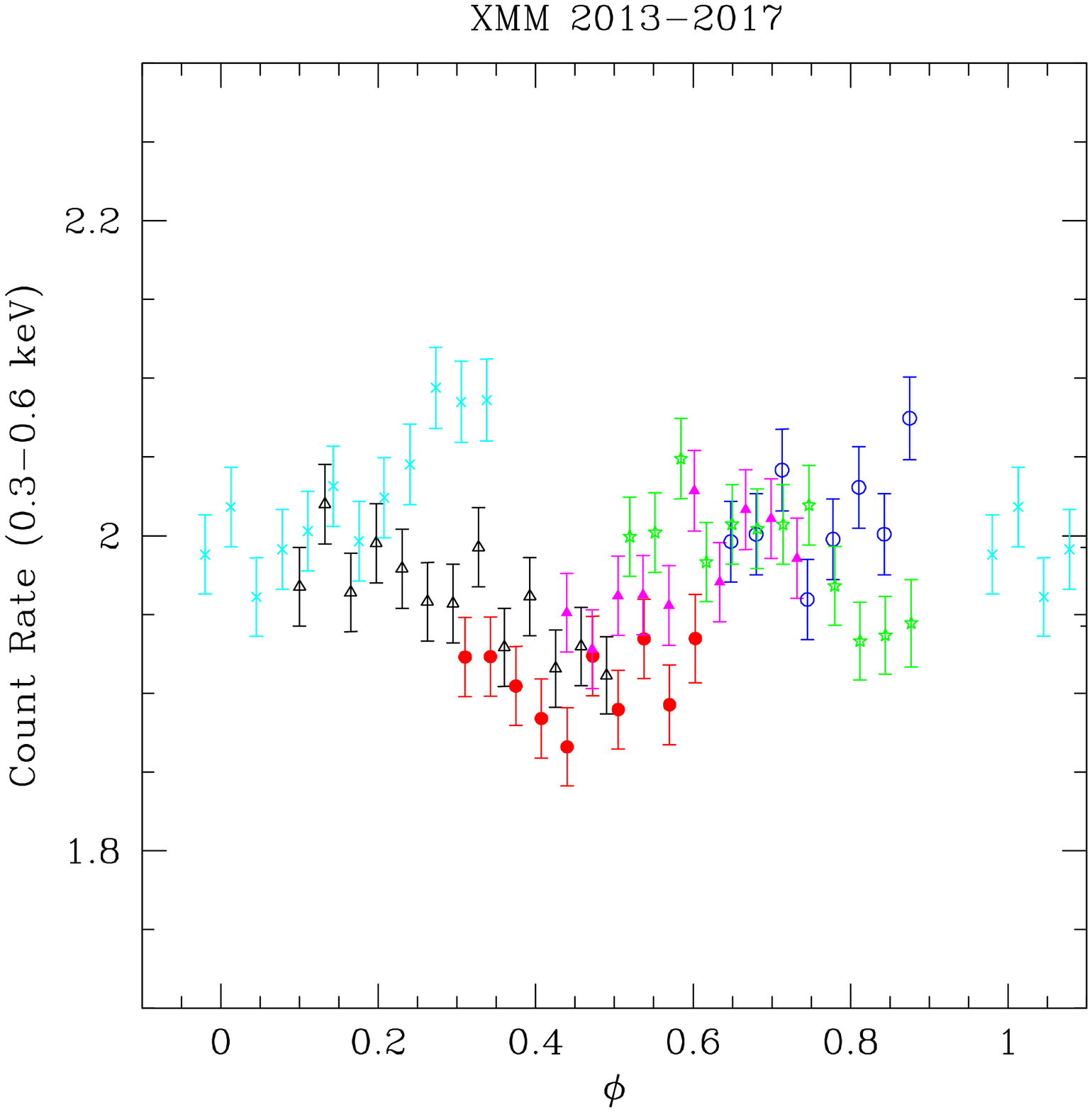}
\includegraphics[width=5.5cm]{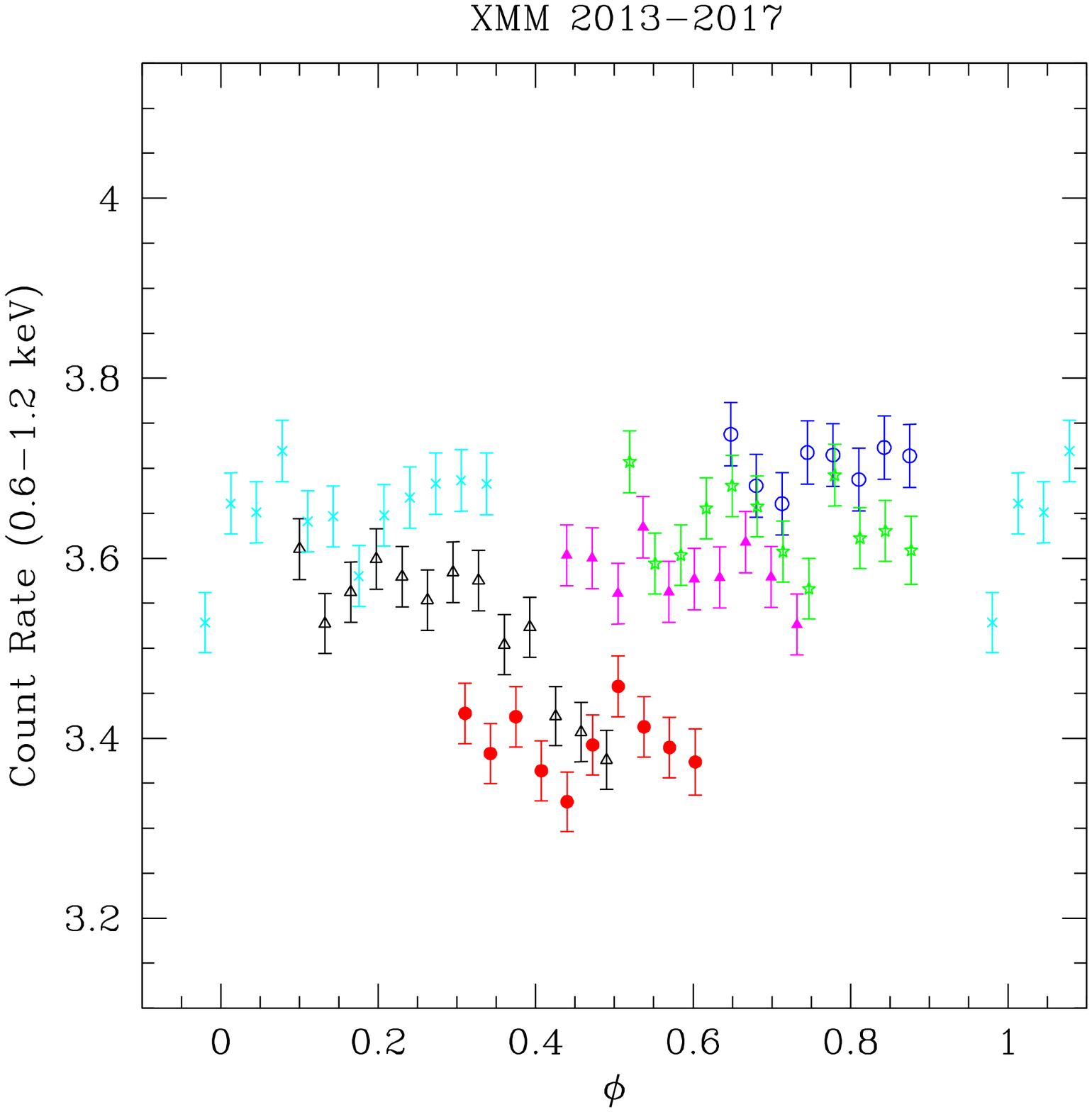}
\\
\includegraphics[width=5.5cm]{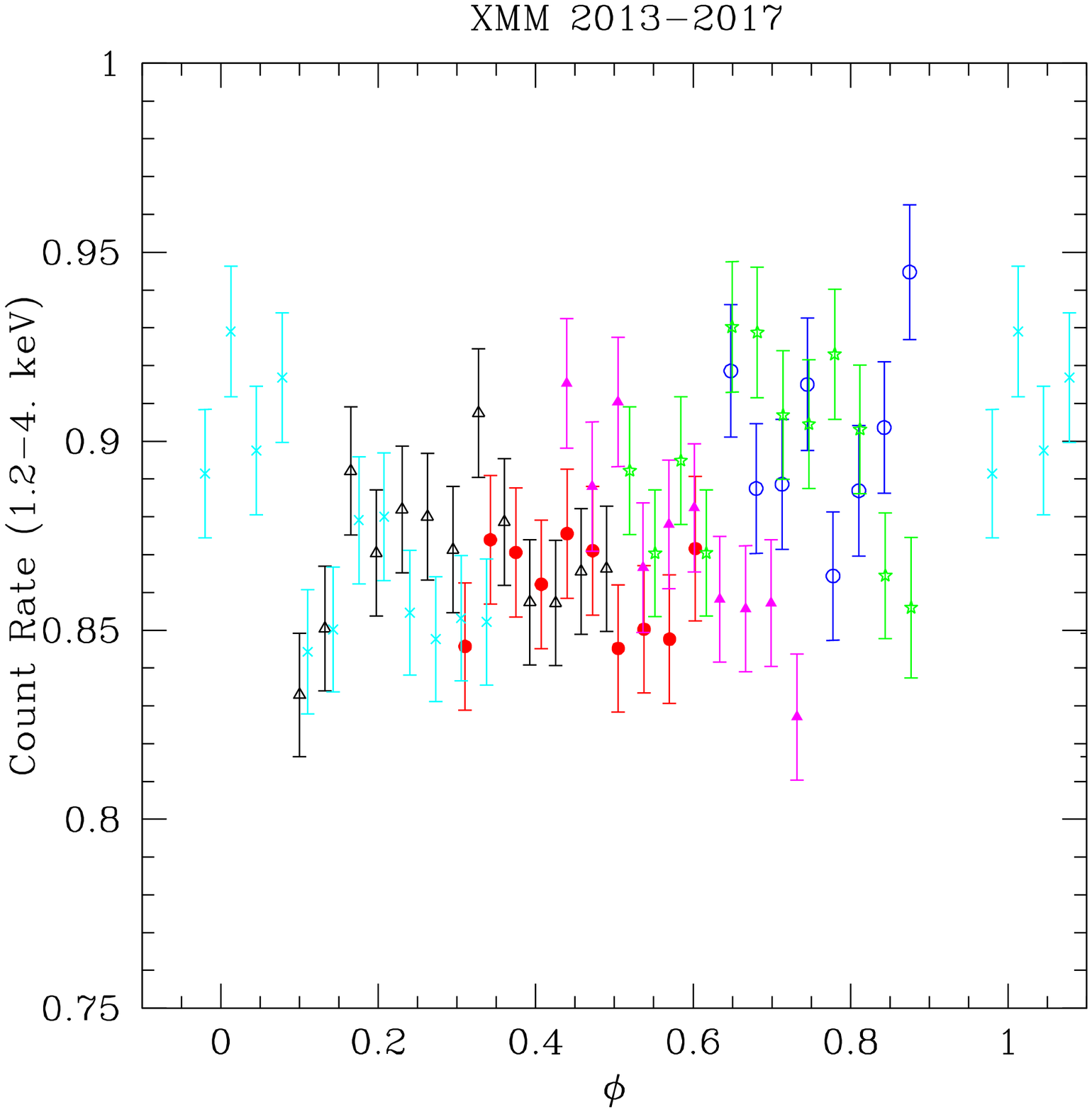}
\includegraphics[width=5.5cm]{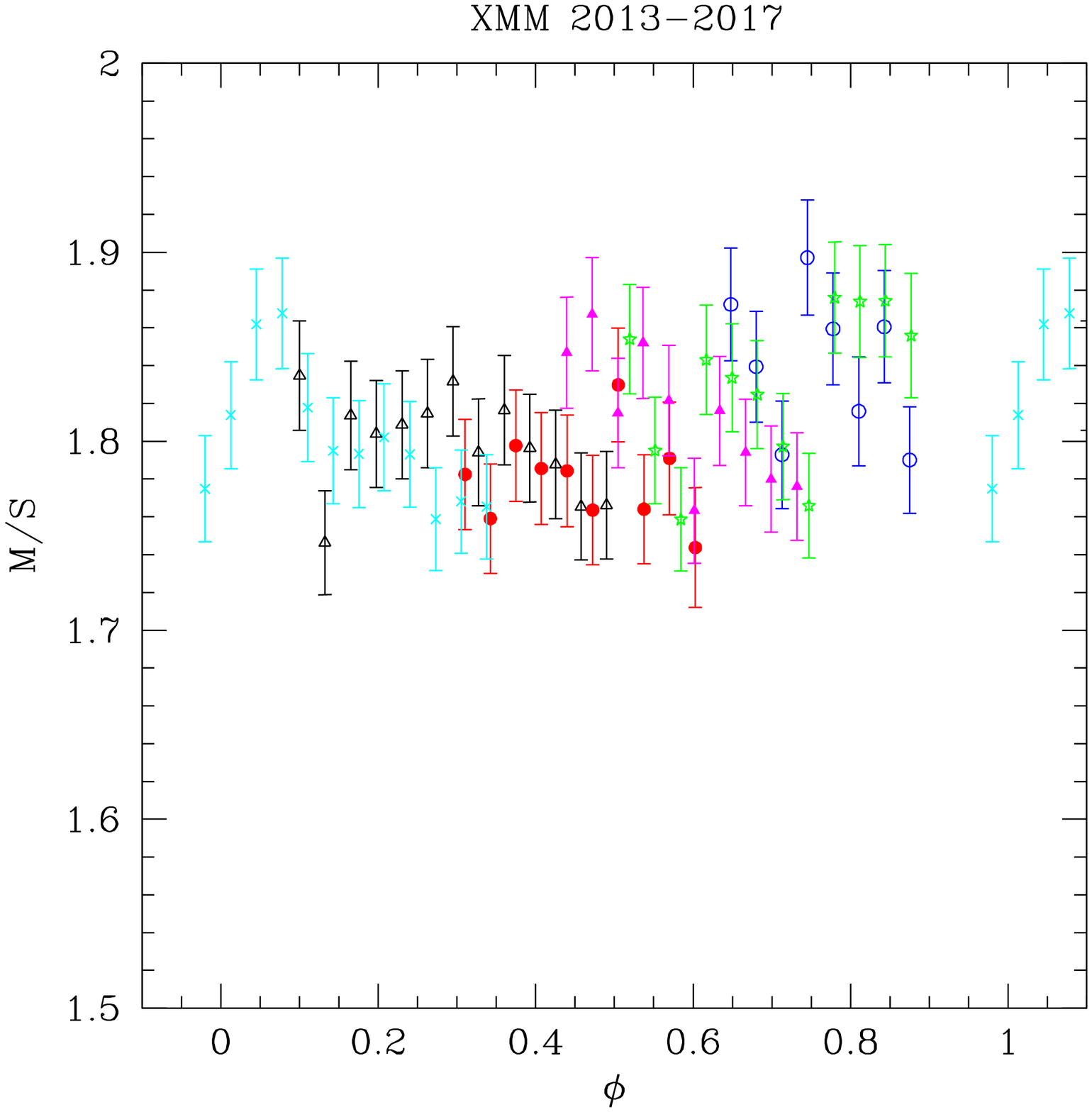}
\includegraphics[width=5.5cm]{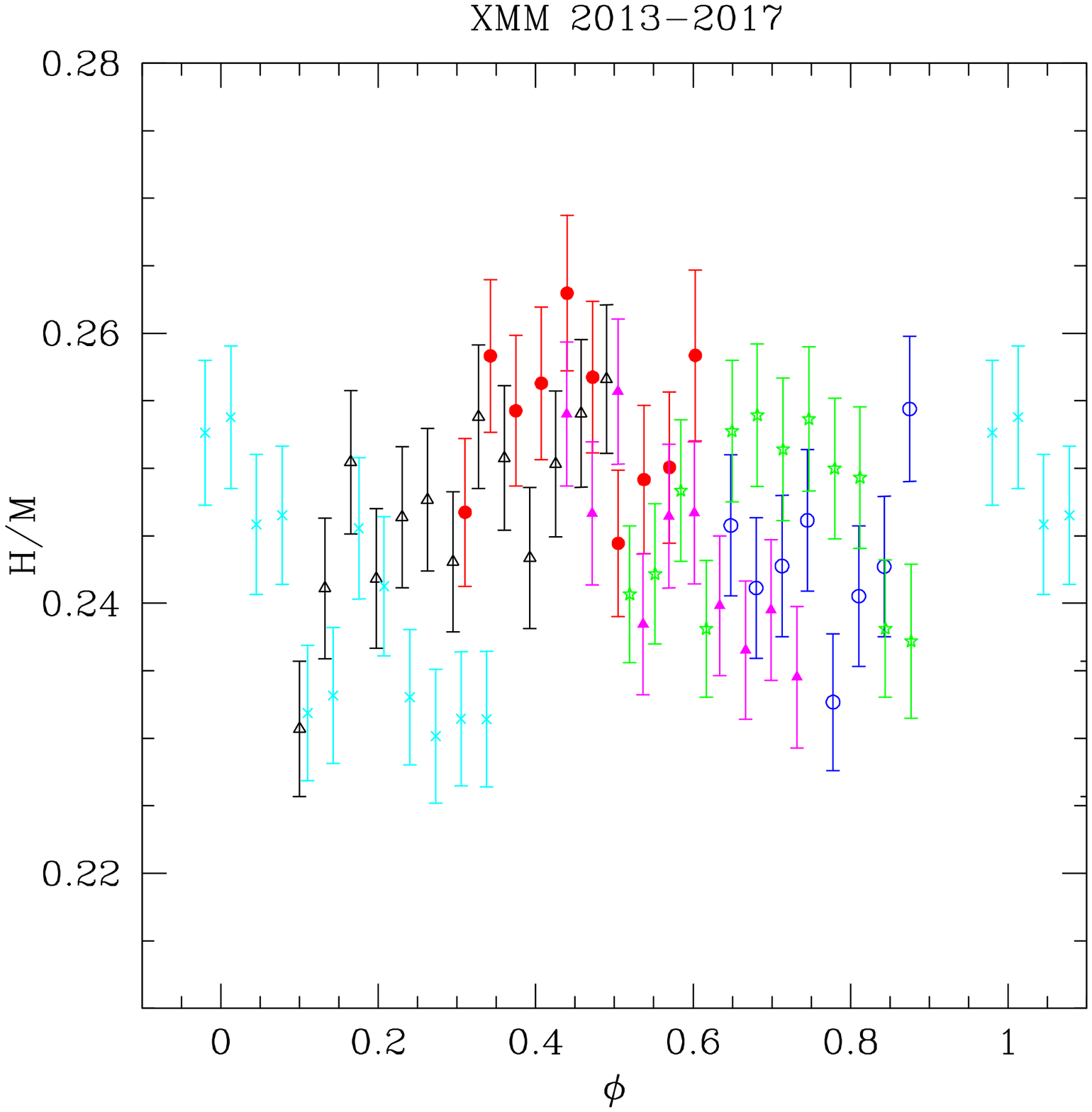}
\caption{Same as Figure \ref{xmm3} but for \xmm\ data from 2013--2017.}
\label{xmm2}
\end{figure*}

\begin{figure*}
\includegraphics[width=6cm]{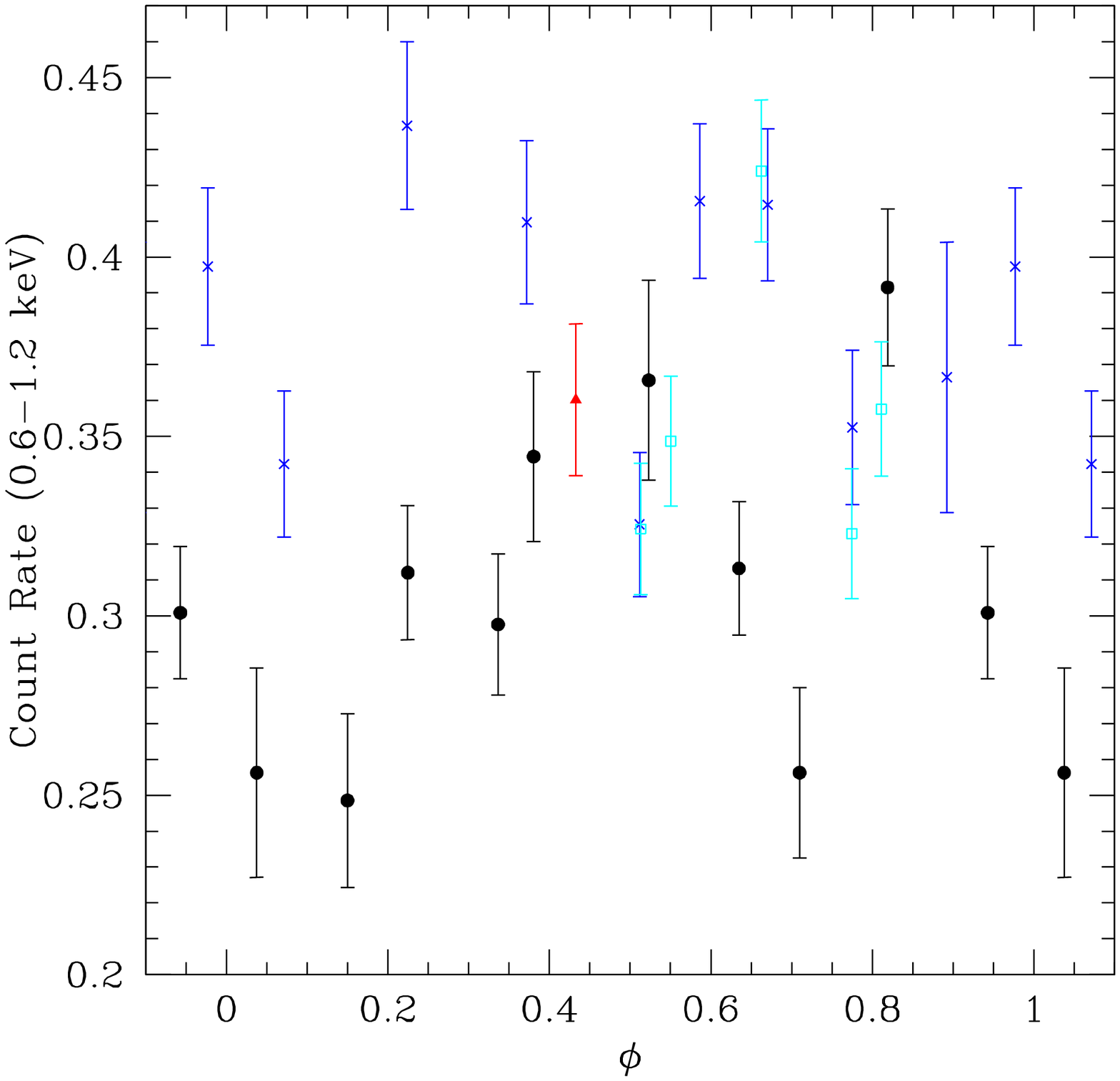}
\includegraphics[width=6cm]{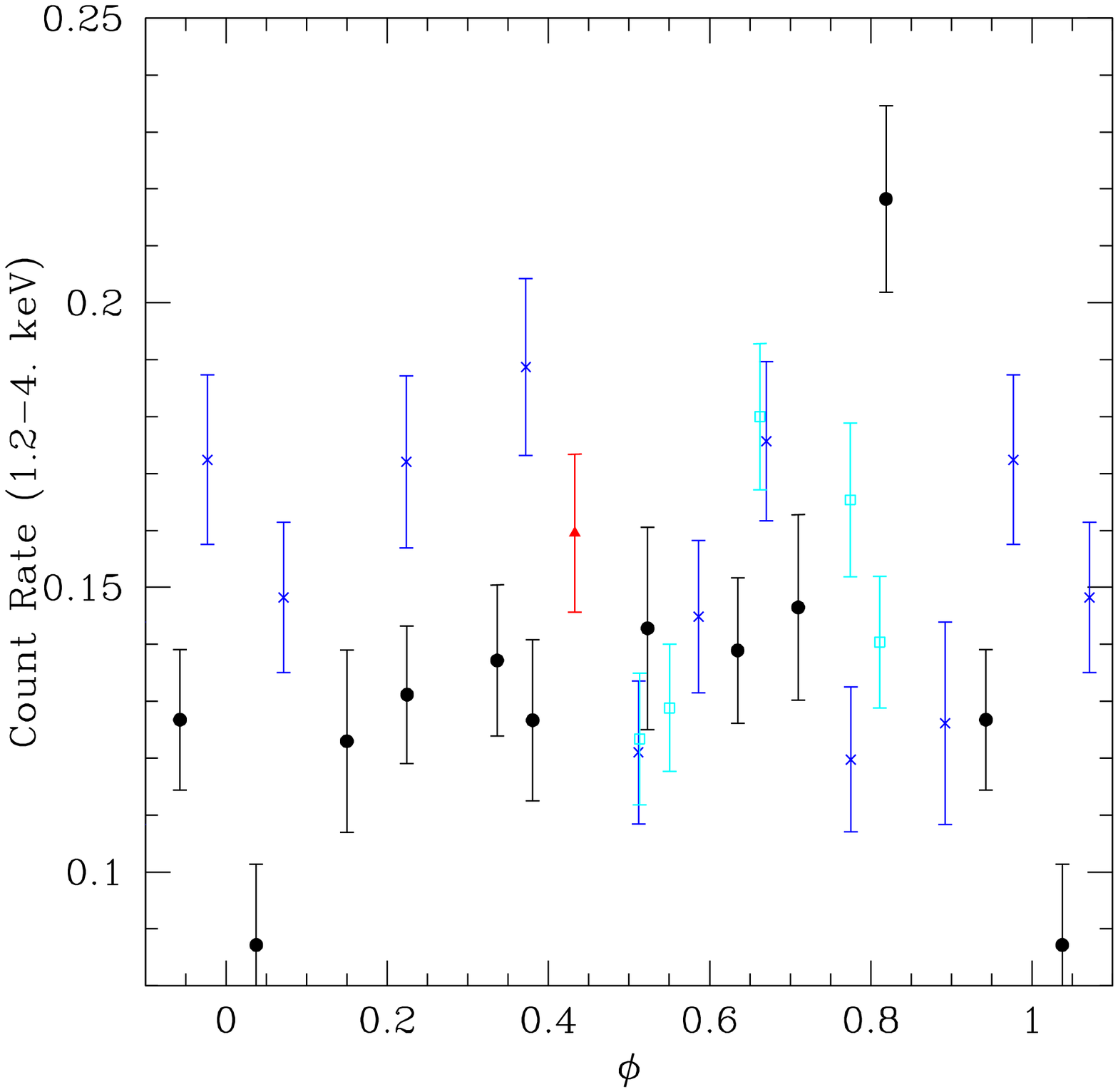}
\includegraphics[width=6cm]{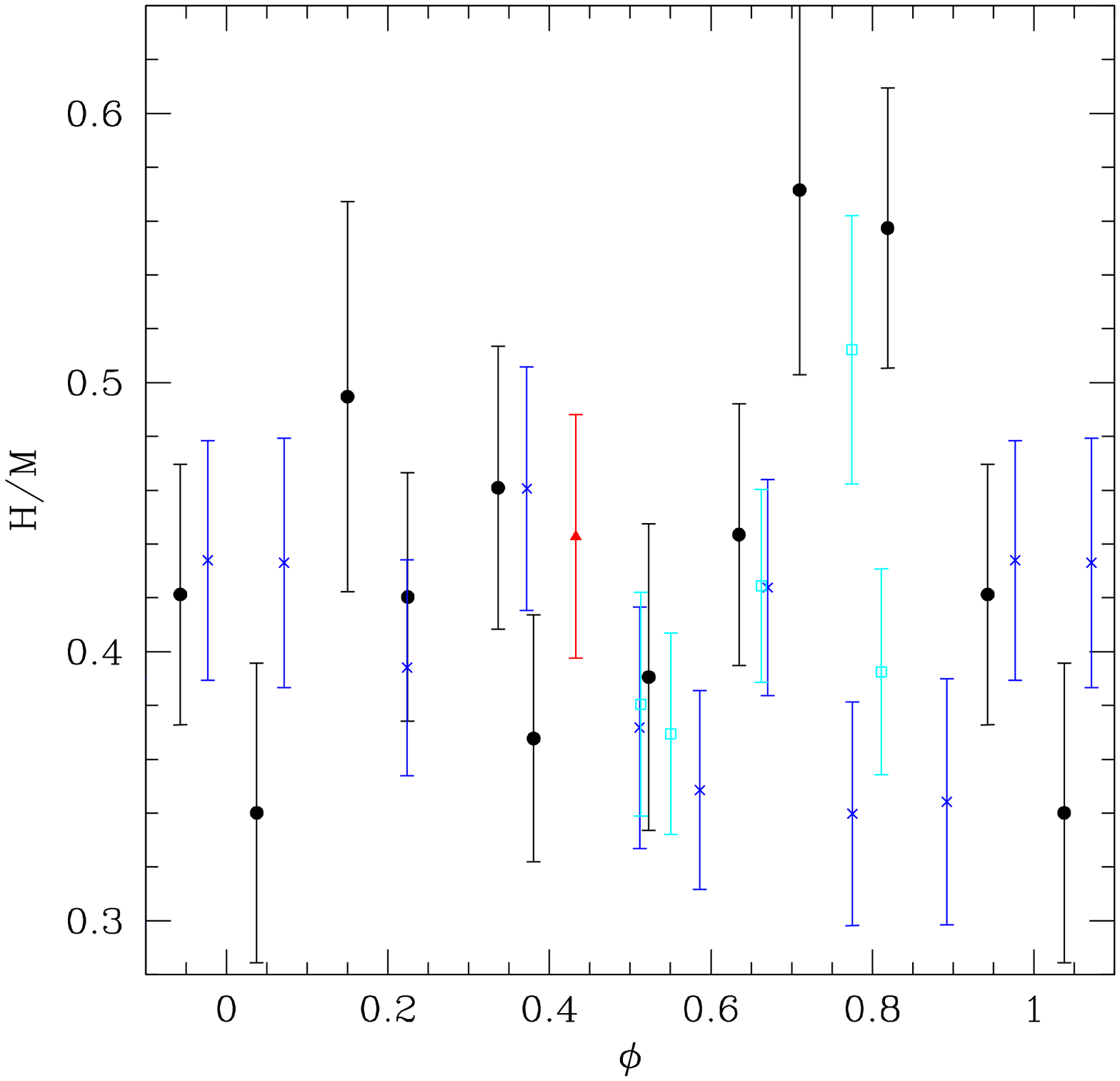}
\caption{Same as Figure \ref{xmm3} but for \swift\ data; those from Oct. 2016, Dec. 2016, Jan. 2017, and April 2017 are shown as red triangles, black dots, blue crosses, and cyan squares, respectively. }
\label{swlc}
\end{figure*}

\begin{figure*}
\includegraphics[width=6cm]{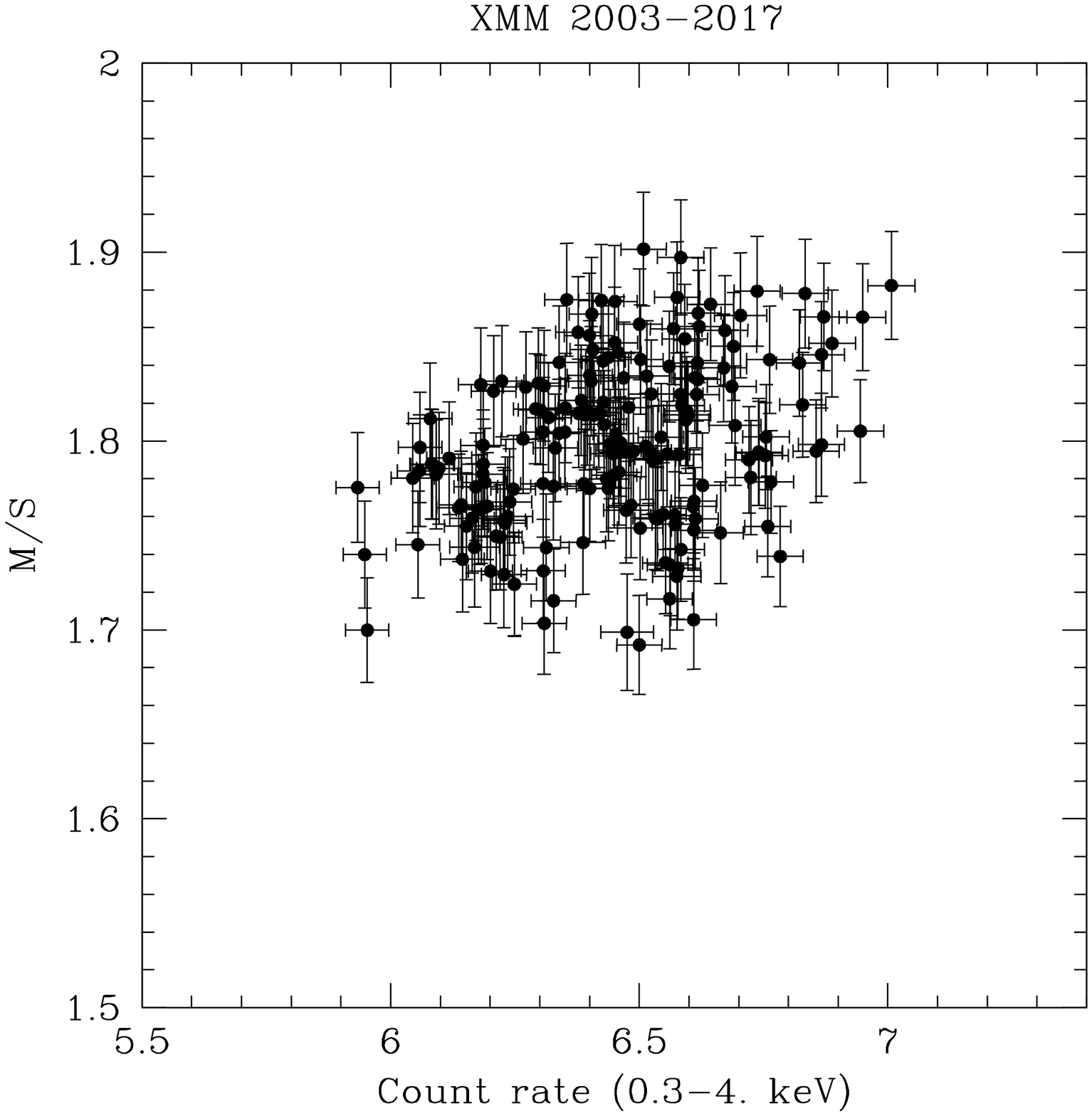}
\includegraphics[width=6cm]{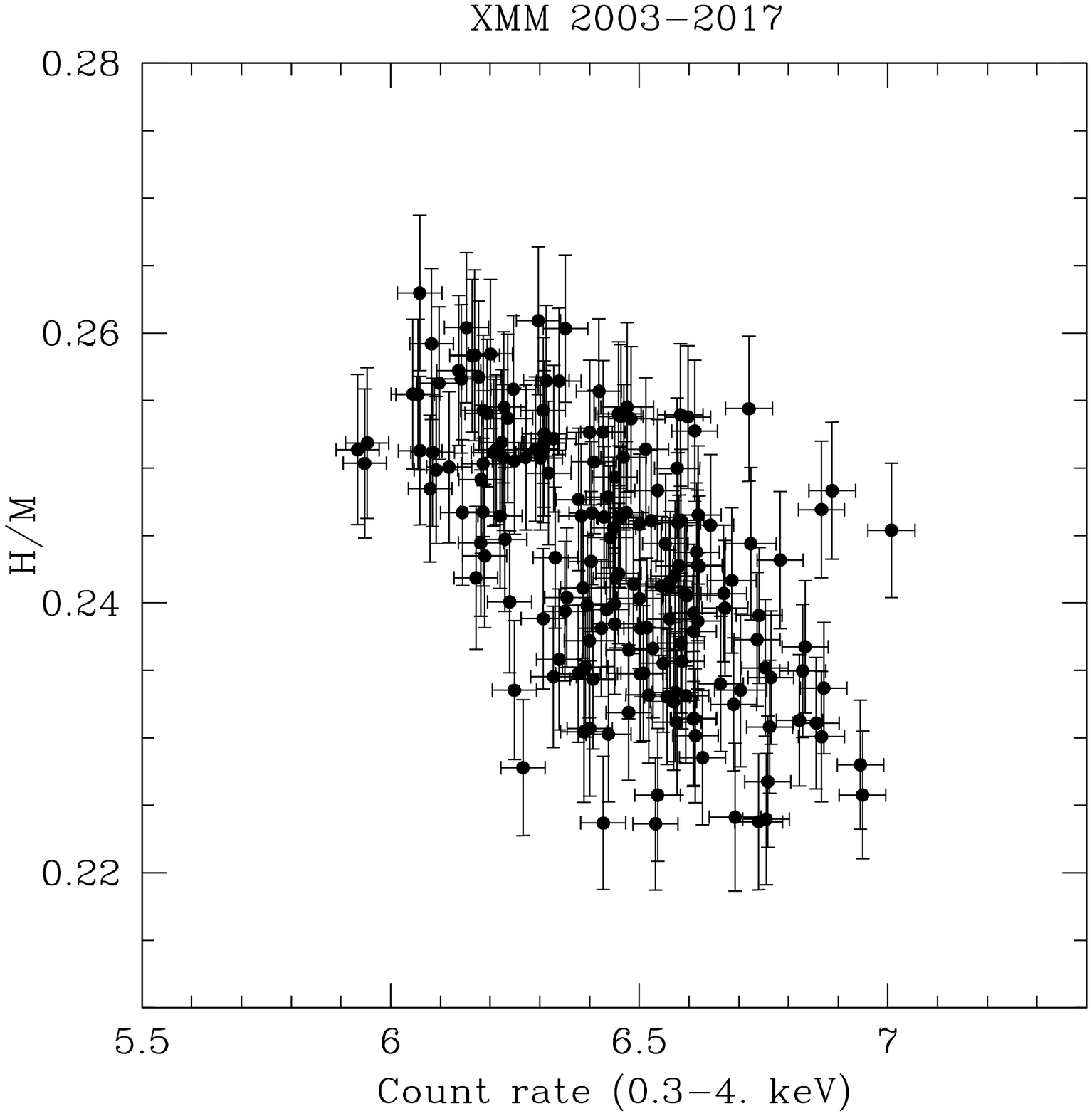}
\includegraphics[width=6cm]{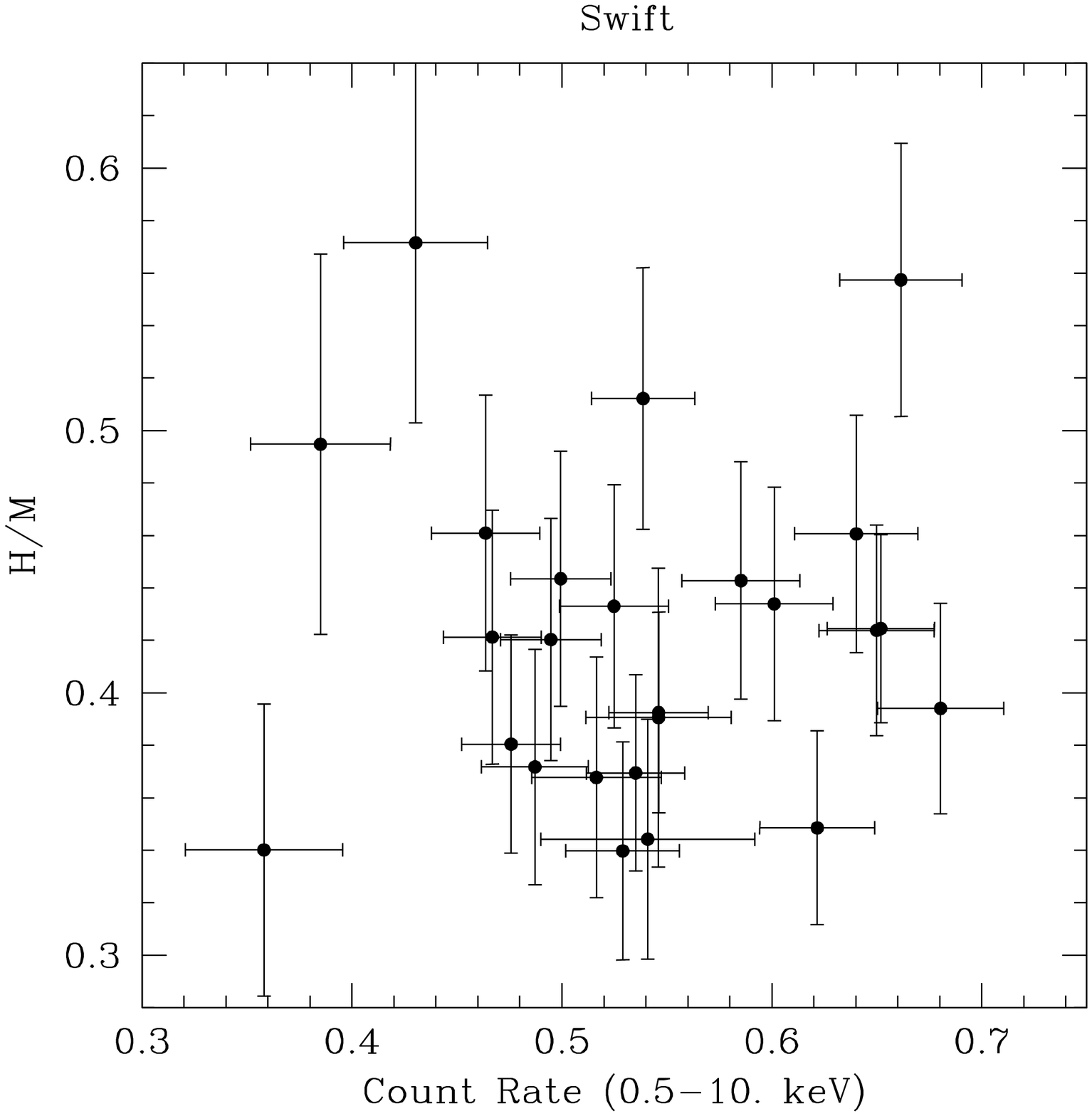}
\caption{Hardness ratios as a function of the source brightness for all \xmm\ data and for all \swift\ data.}
\label{xmmhr}
\end{figure*}

\setcounter{figure}{8}
\begin{figure*}
\includegraphics[width=4.5cm]{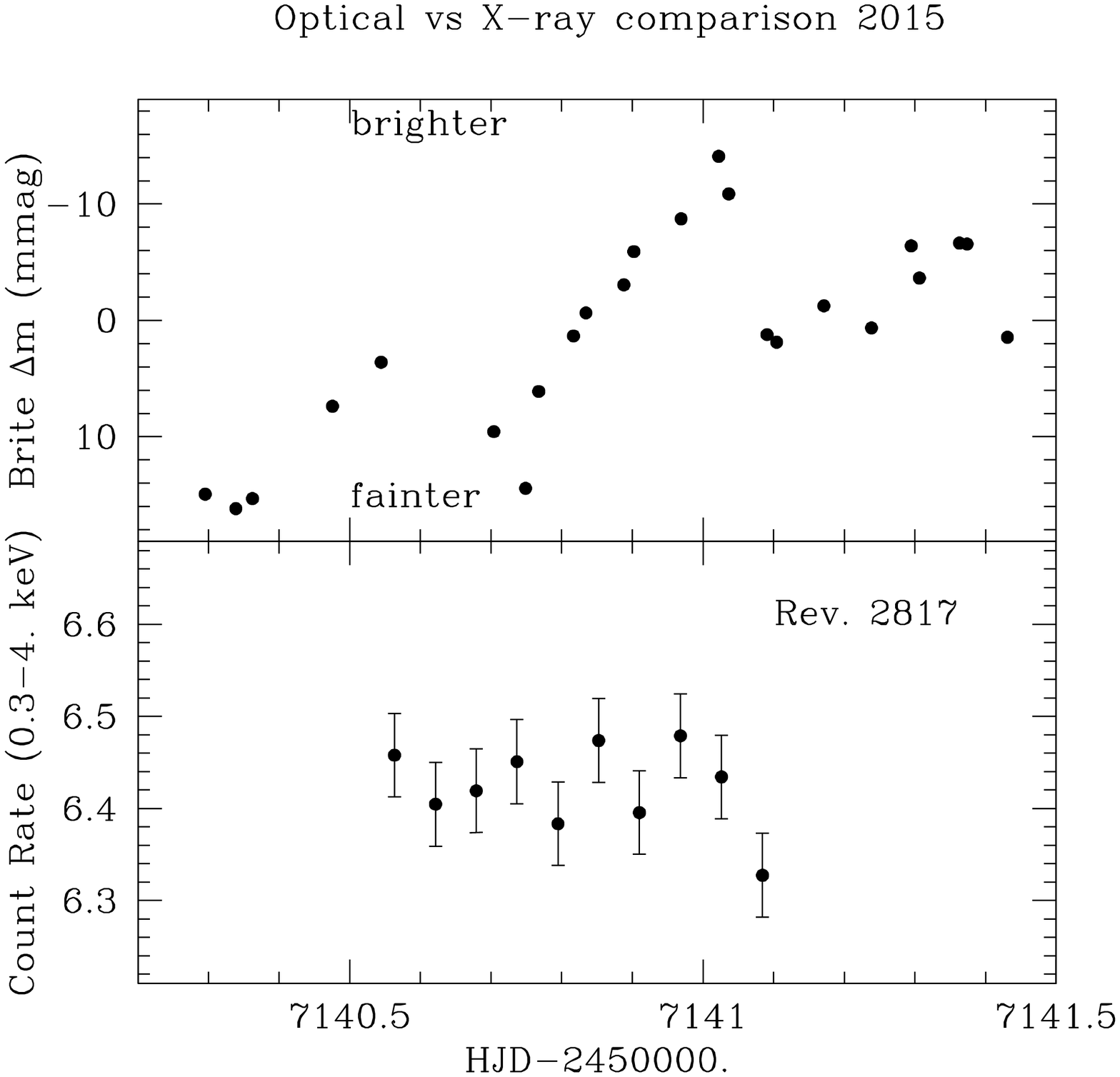}
\includegraphics[width=4.5cm]{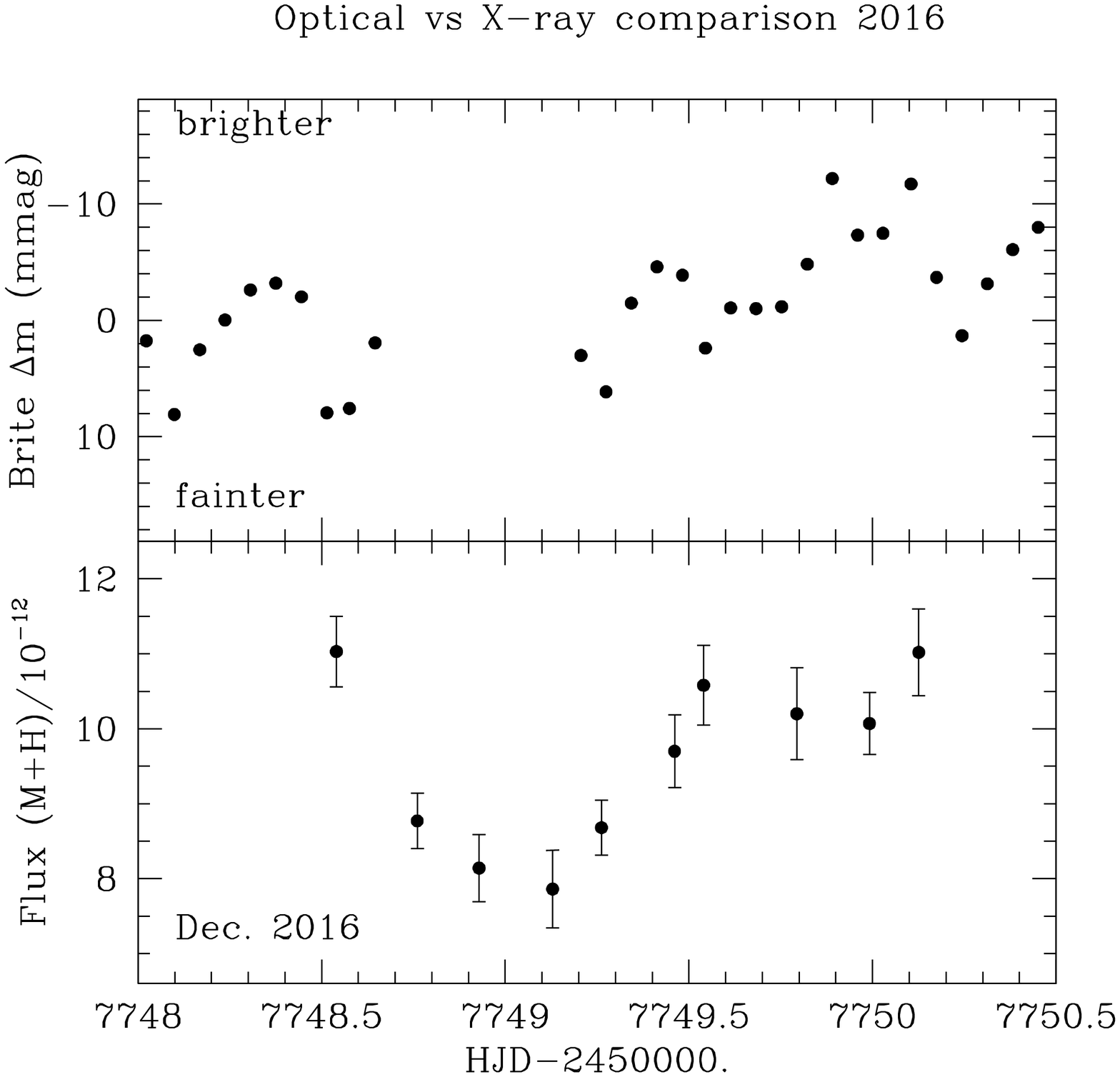}
\includegraphics[width=4.5cm]{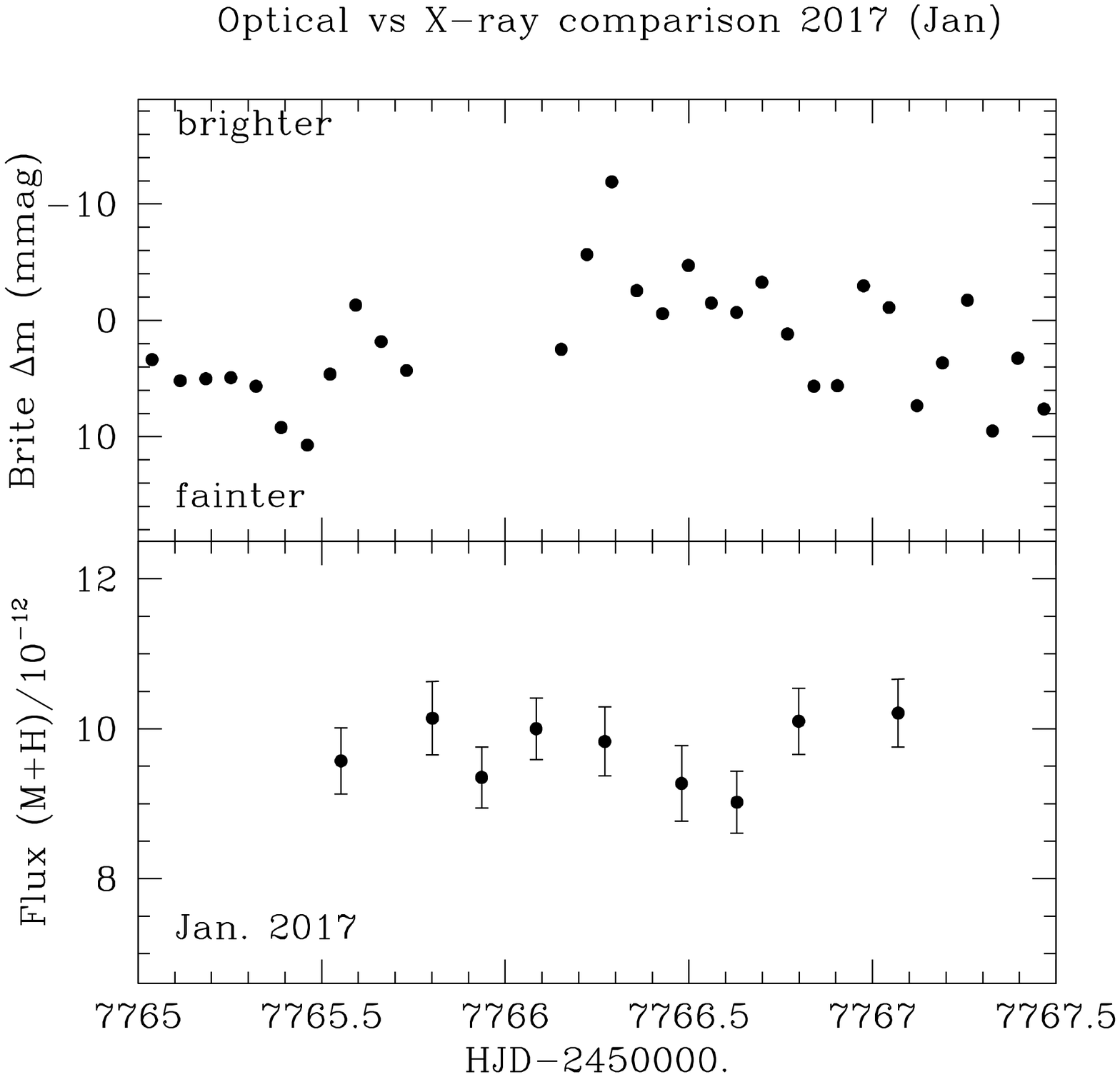}
\includegraphics[width=4.5cm]{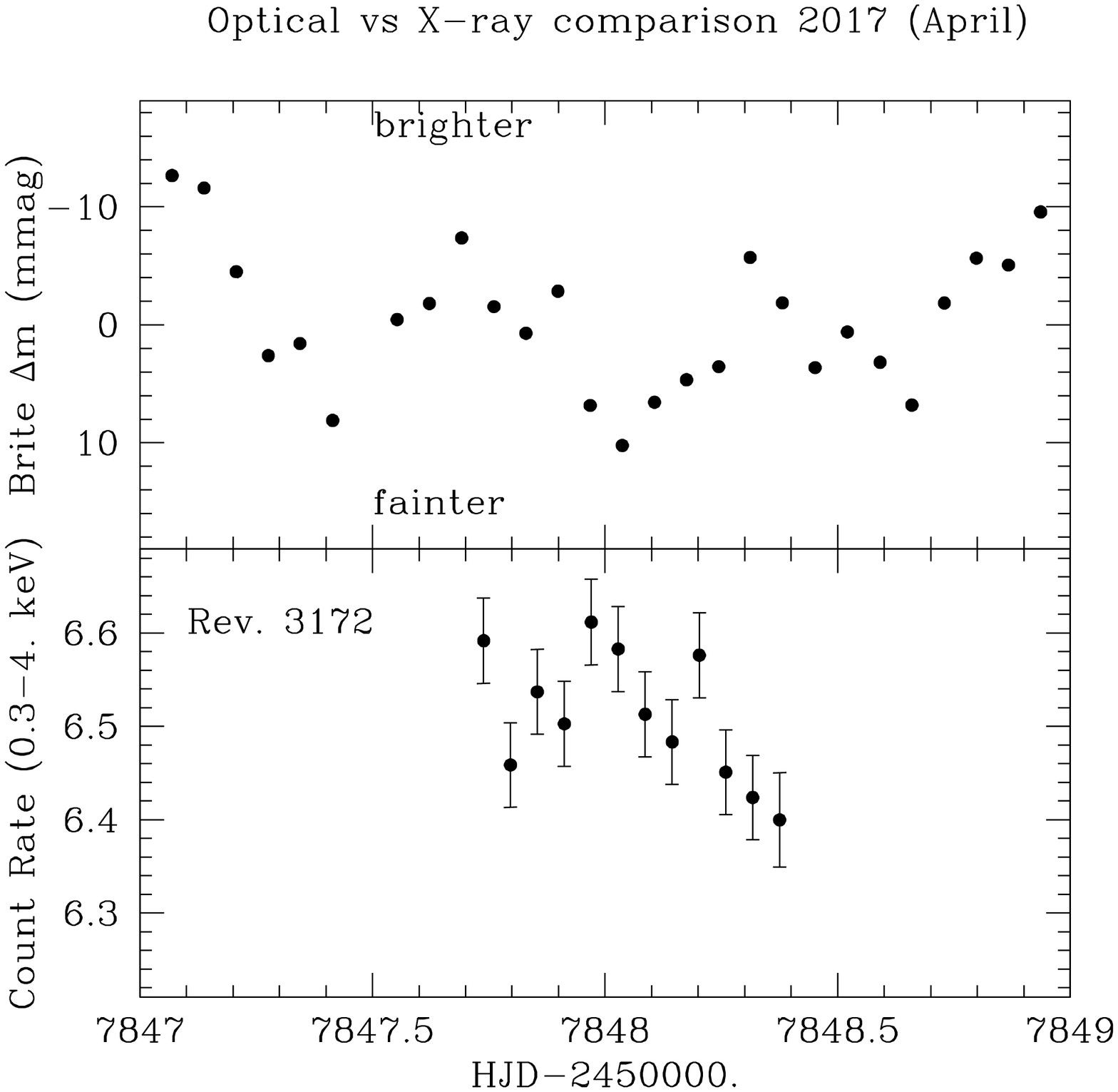}
\caption{Comparison of the X-ray light curves with the simultaneous $BRITE$ photometry in April 2015 (combined light curve from the two $BRITE$ filters, for details see Ramiaramanantsoa et al. 2017a, submitted), and December 2016--April 2017 (only one satellite, BRITE-Lem, in the blue filter, see Ramiaramanantsoa et al. 2017b, in prep). For \swift, fluxes from the spectral fits (Table \ref{swiftfit}) are used since they are more reliable.}
\label{britea}
\end{figure*}

\begin{figure*}
\includegraphics[width=4.5cm]{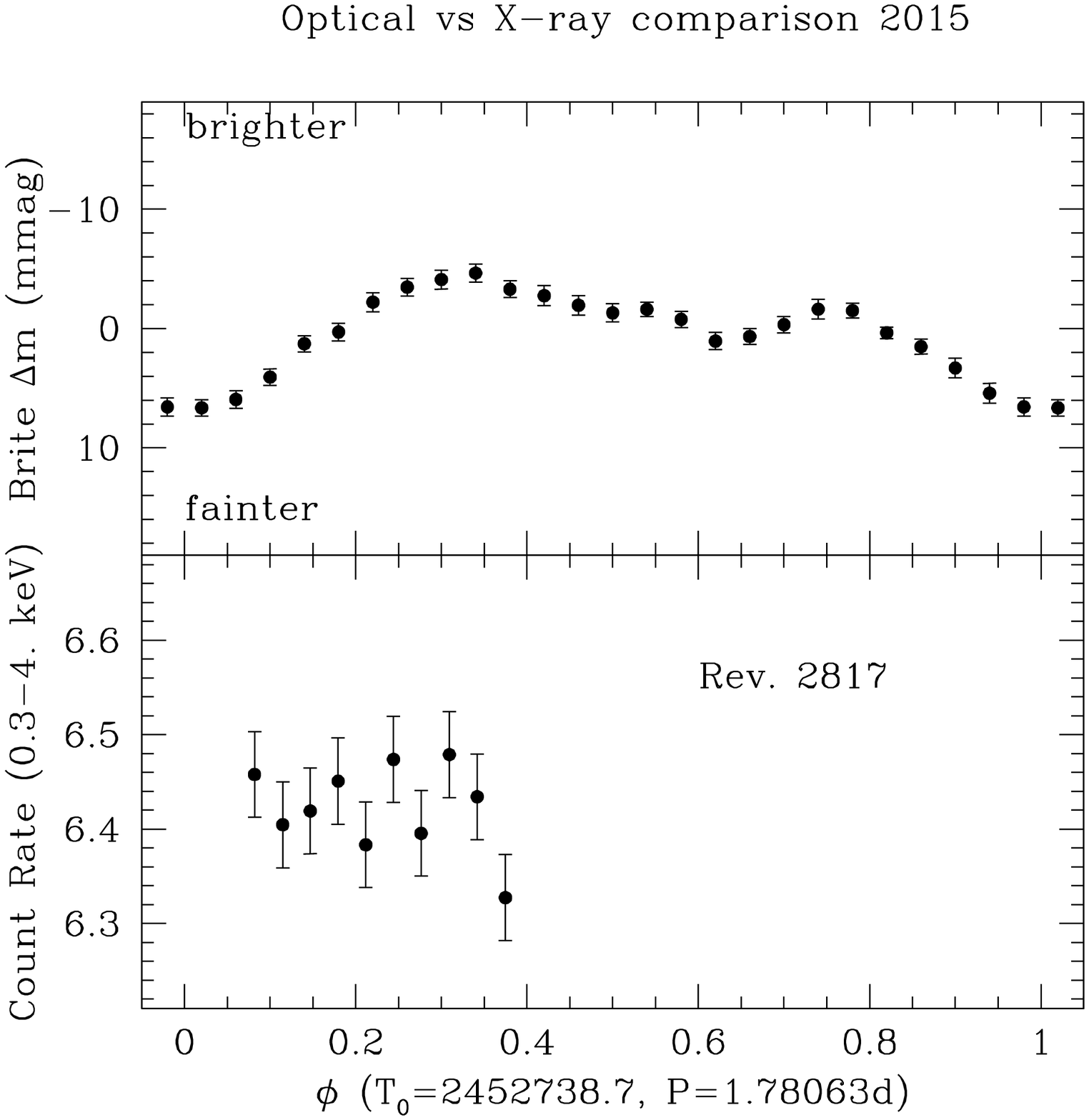}
\includegraphics[width=4.5cm]{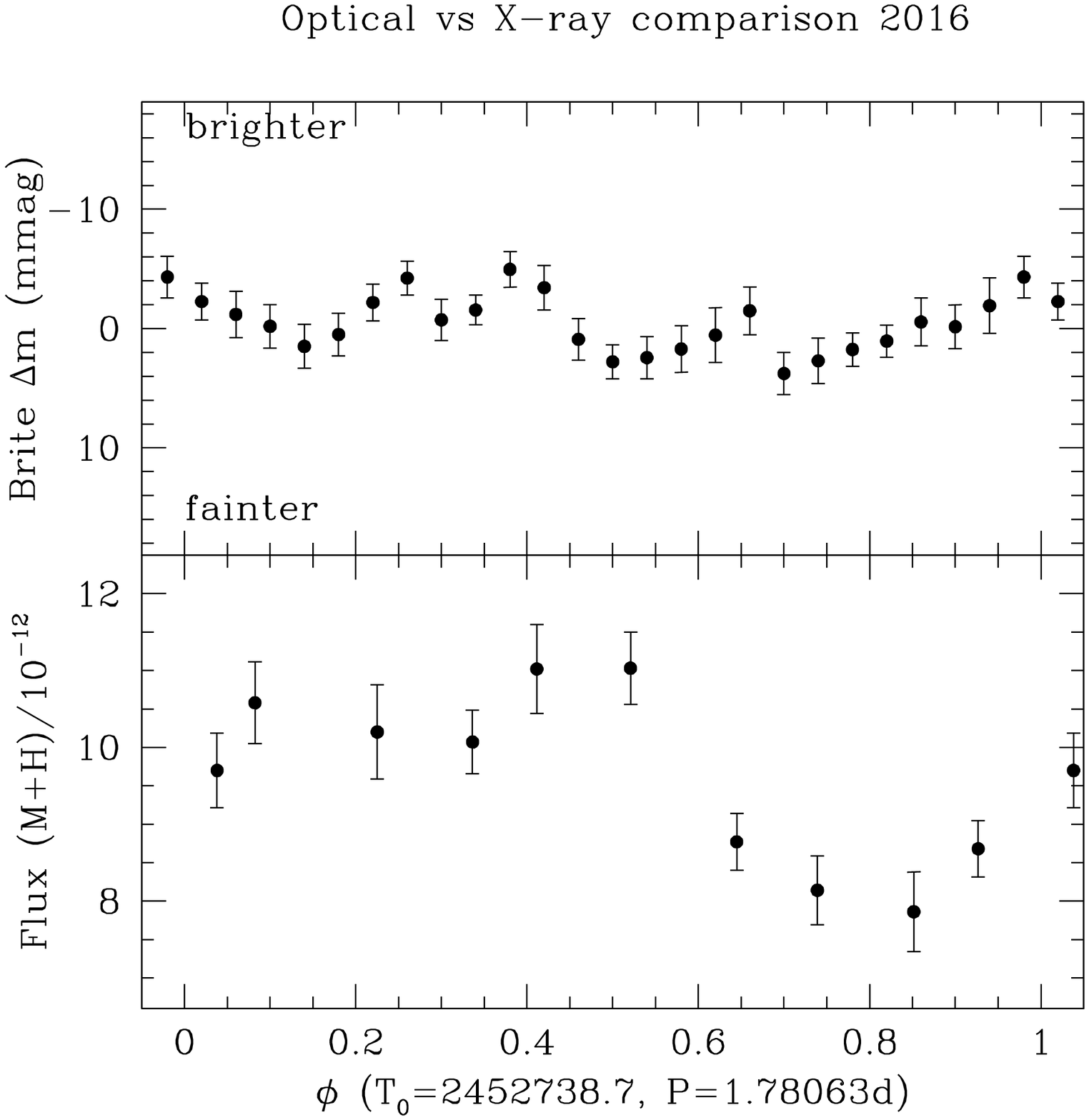}
\includegraphics[width=4.5cm]{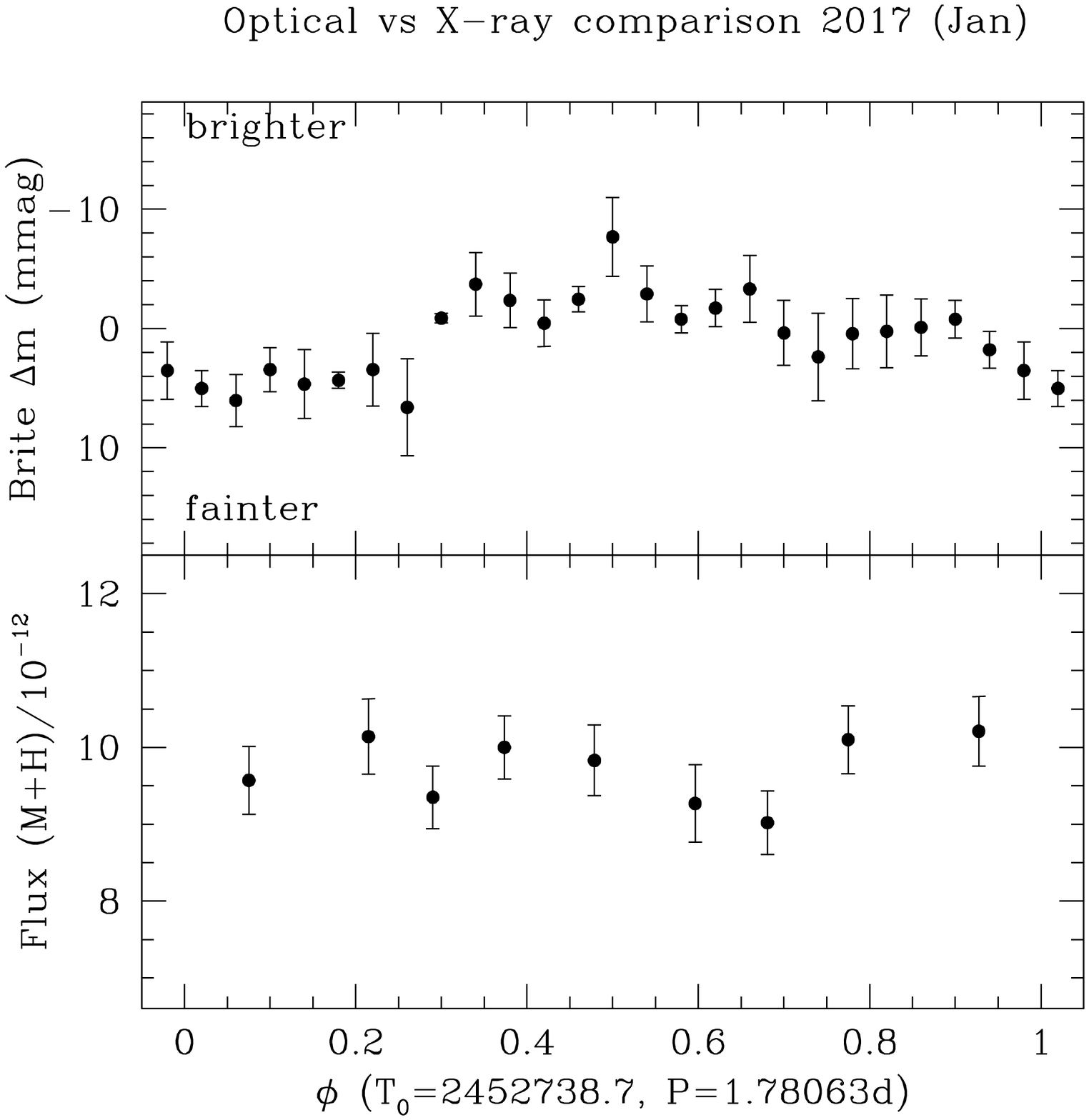}
\includegraphics[width=4.5cm]{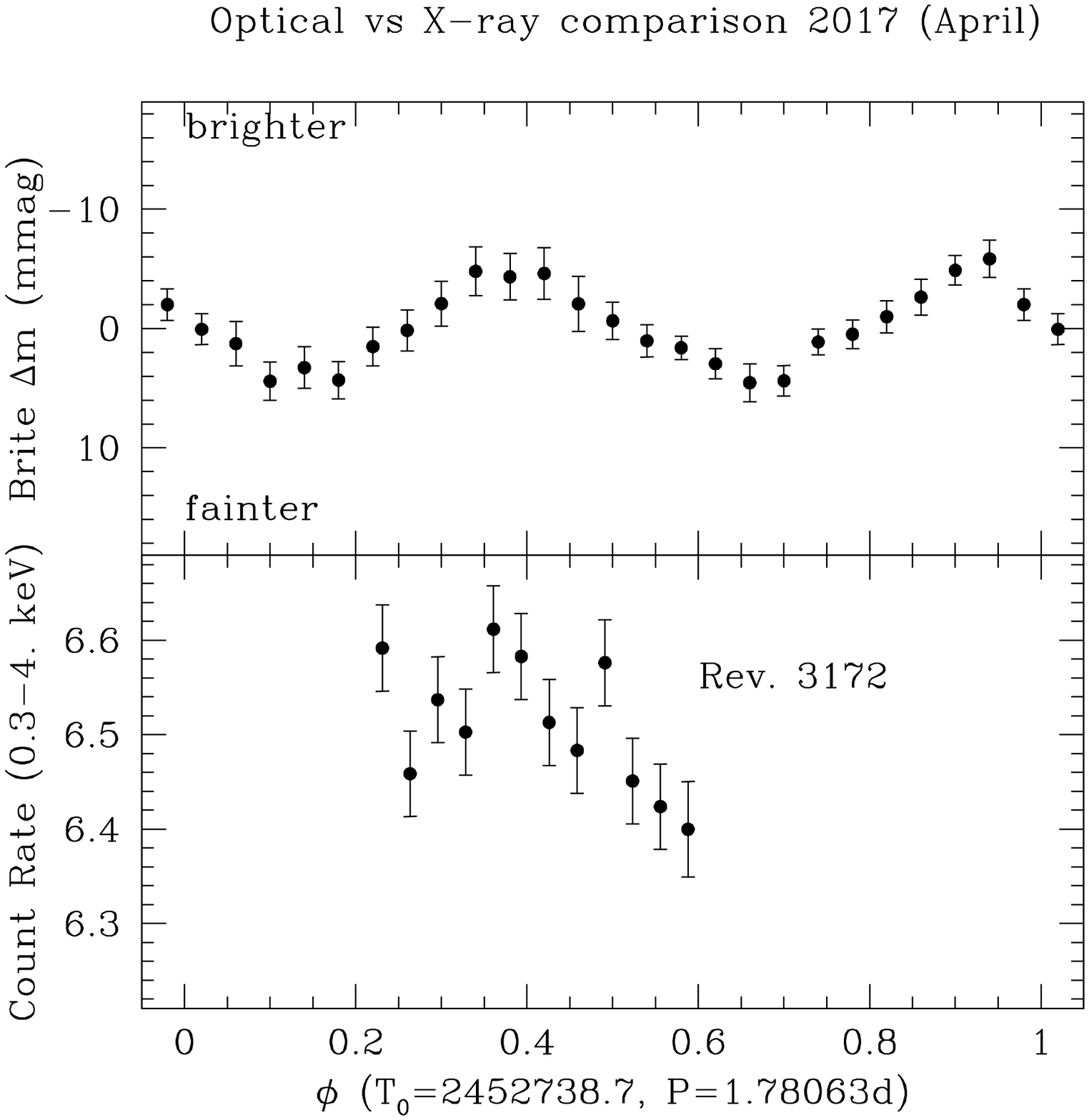}
\caption{Same as Figure \ref{britea} but using epochally averaged BRITE light curves valid for the date of the X-ray observations (cycles associated with ``Part V'' data for April 2015, ``Part VI'' for Dec 2016, ``Part VII'' for Jan 2017, and ``Part IX'' for April 2017, for details see Ramiaramanantsoa et al. 2017a, submitted). The ephemerides are those adopted by Ramiaramanantsoa et al. (2017a, submitted). }
\label{briteb}
\end{figure*}

\begin{figure*}
\includegraphics[width=6cm]{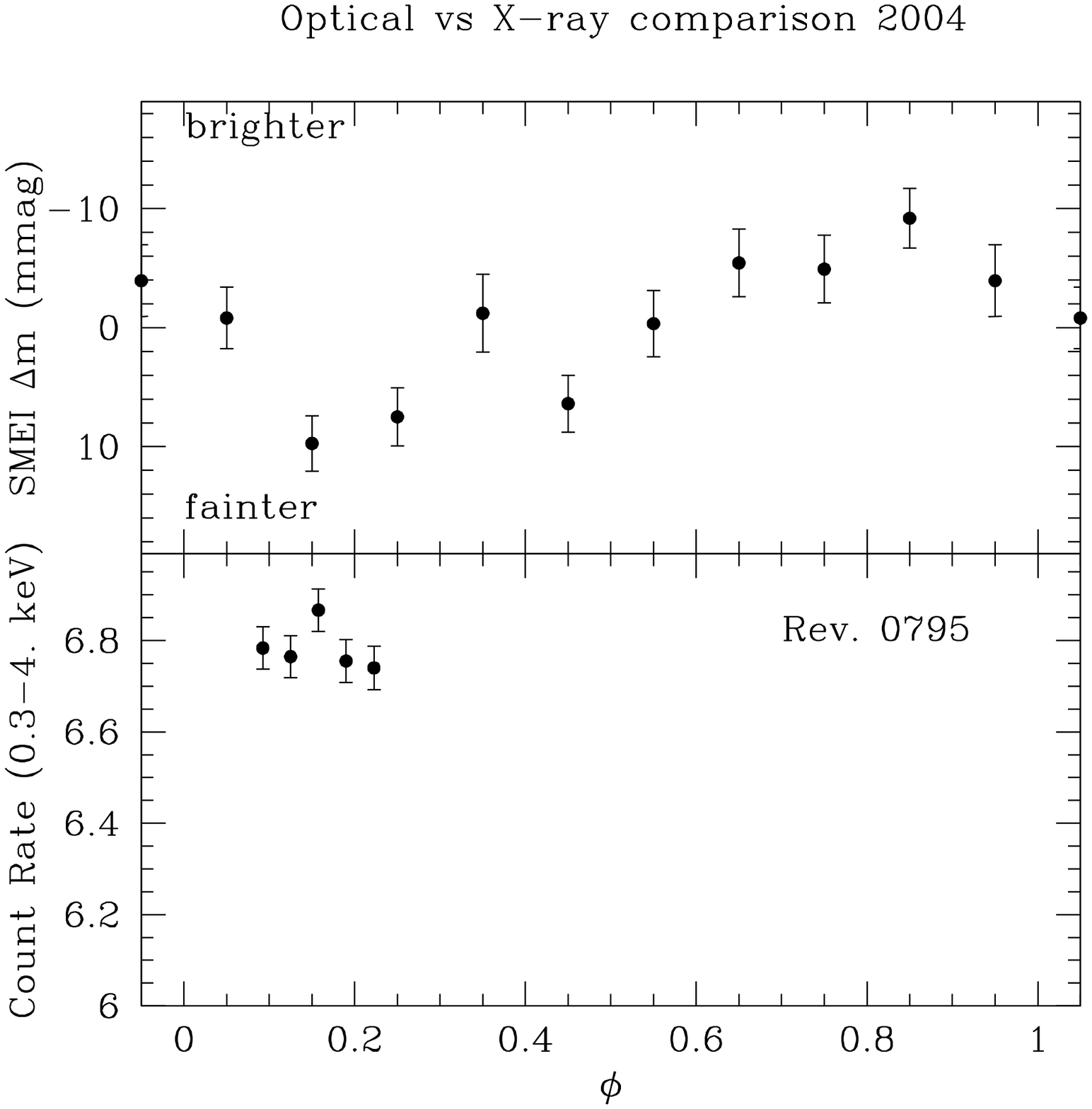}
\includegraphics[width=6cm]{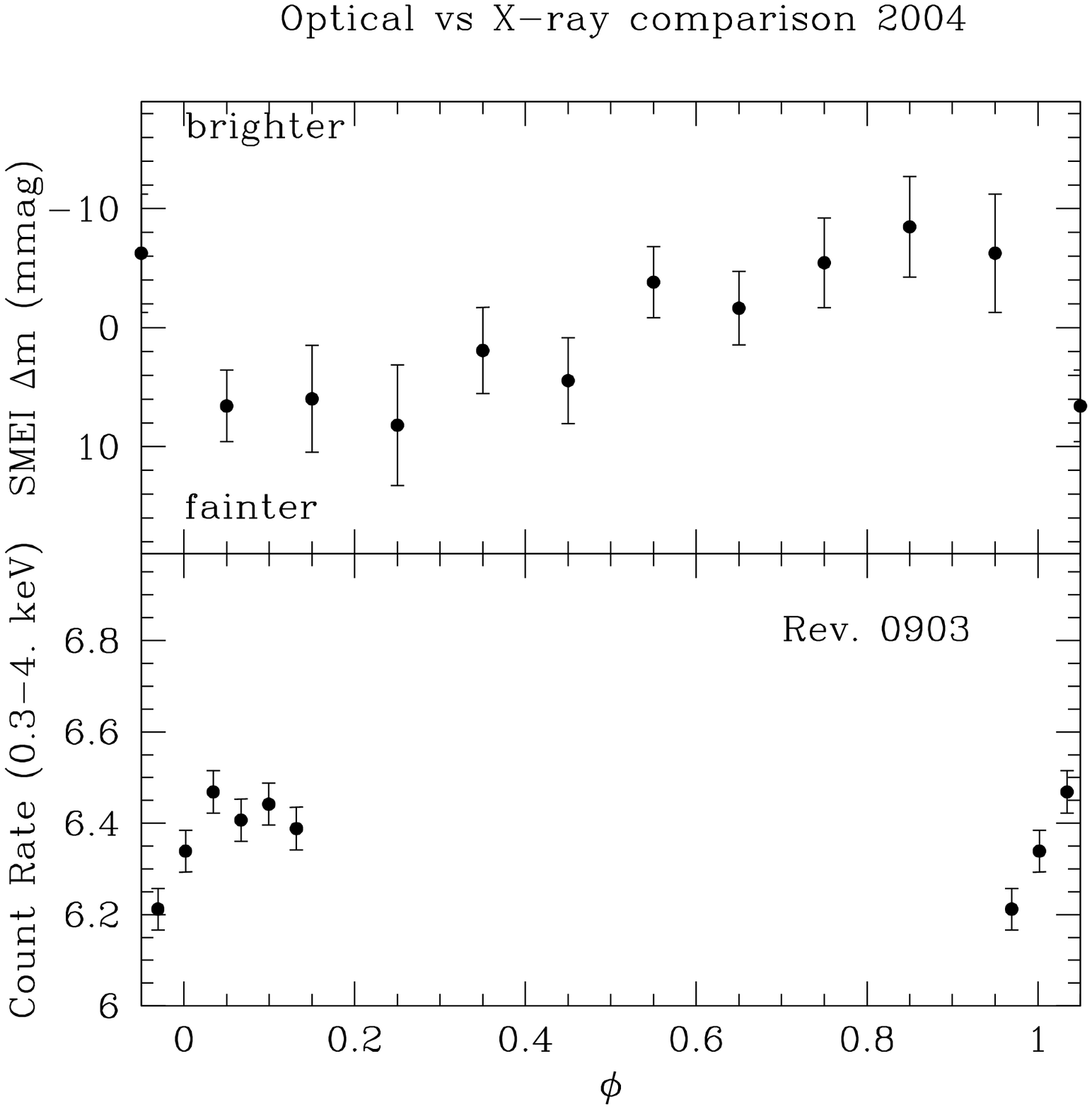}
\\
\includegraphics[width=6cm]{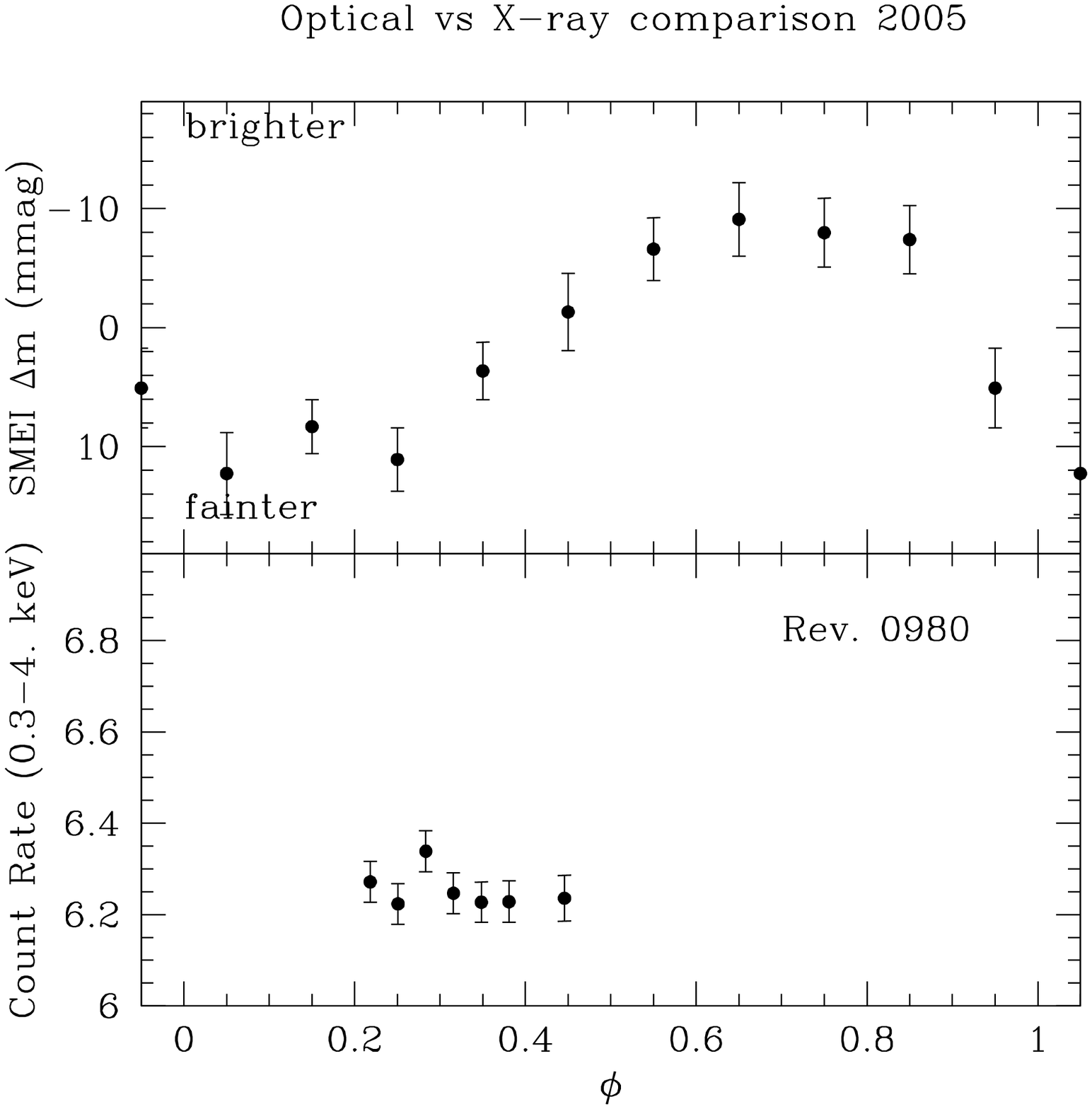}
\includegraphics[width=6cm]{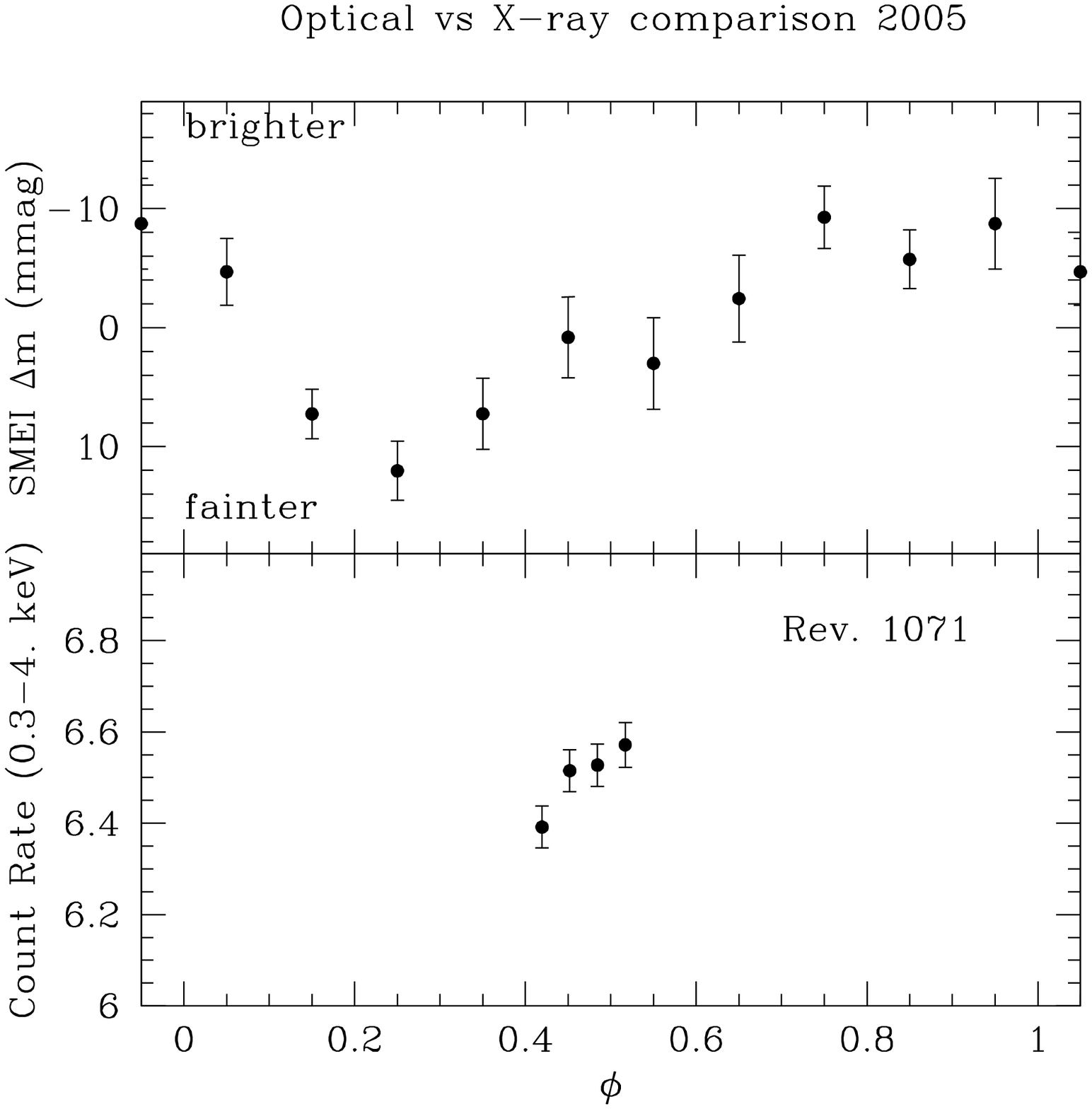}
\includegraphics[width=6cm]{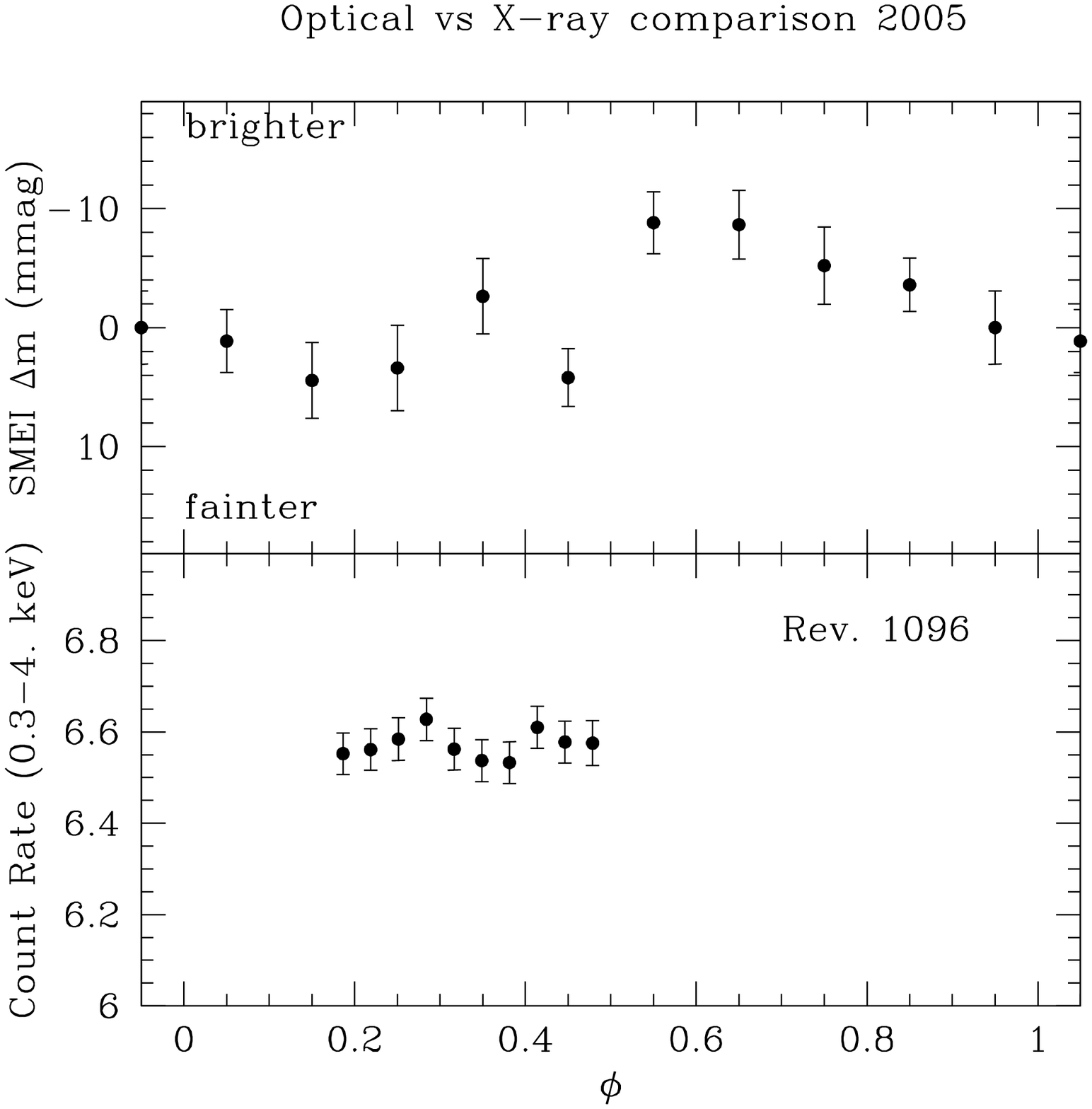}
\caption{Comparison of the X-ray light curves with the average 1.78\,d photometric cycle recorded with \emph{SMEI} $\pm15$\,d around the X-ray observing date. The ephemerides are those of \citet{how14}: $P=1.780938$\,d and $T_0$=2\,450\,000. }
\label{smei}
\end{figure*}

\end{document}